\documentclass[a4paper,11pt]{article}
\usepackage[top=3.5cm,bottom=3.5cm,left=2.5cm,right=2.5cm,a4paper]{geometry}

\usepackage{amsfonts,latexsym,amsmath,amssymb,amsthm}
\usepackage[colorlinks]{hyperref}
\hypersetup{
    colorlinks=true, 
    linkcolor={blue!50!black},
    citecolor={green!60!black},
    urlcolor= {red!80!black}
}
\usepackage{tikz,pgfplots,siunitx,multirow}
\usepackage{algorithmic,setspace,enumitem}
\usepackage{graphicx}
\usepackage[ruled,vlined]{algorithm2e}
\usepackage{empheq}
\usepackage[bottom]{footmisc} 
\usepgfplotslibrary{groupplots}
\usetikzlibrary{positioning,arrows,fit,patterns}
\usetikzlibrary{decorations.text}
\pgfplotsset{compat=newest}

\usepackage{subfig}
\DeclareCaptionLabelFormat{opening}{\textbf{#2)}} 
\usepackage{natbib} 
\setcitestyle{citesep={,}}

\makeatletter
\renewcommand*\env@matrix[1][\arraystretch]{%
  \edef\arraystretch{#1}%
  \hskip -\arraycolsep
  \let\@ifnextchar\new@ifnextchar
  \array{*\c@MaxMatrixCols c}}
\makeatother
\newcommand*\samethanks[1][\value{footnote}]{\footnotemark[#1]}

\newcommand{\myblue} {blue!80!white}

\newtheorem {remark}     {Remark}

\newcommand{\forward}     {\mathcal{F}}
\newcommand{\R}           {\mathbb{R}}
\newcommand{\Real}        {\mathrm{Re}}
\newcommand{\Cx}          {\mathbb{C}}

\newcommand{\divergence}  {\nabla \cdot}
\newcommand{\ii}          {\mathrm{i}}
\newcommand{\dd}          {\mathrm{d}}
\newcommand{\bx}          {\boldsymbol{x}}
\newcommand{\misfitL}     {\mathcal{J}_{L2}}
\newcommand{\misfitG}     {\mathcal{J}_\mathfrak{r}}
\newcommand{\restrict}    {\mathcal{R}}
\newcommand{\pressure}    {p}
\newcommand{\n}           {\boldsymbol{\nu}} 
 
\newcommand{\velocity}    {\boldsymbol{v}}

\newcommand{\nrcv}        {n_{\text{rcv}}}
\newcommand{\nsrc}        {{n_{\text{src}}}}
\newcommand{\nsrcSim}     {n_{\text{src}}^\text{sim}}
\newcommand{\nsrcObs}     {n_{\text{src}}^\text{obs}}
\newcommand{\nstack}      {n_{\text{stack}}}
\newcommand{\npt}         {n_{\text{pt}}}
\newcommand{\titlebar}    {}
\newcommand{\lagrangian} {\mathcal{L}}
\newcommand{\woperator}  {\mathcal{A}}
\newcommand{\sol}        {U}
\newcommand{\adj}        {\gamma}
\newcommand{\tildsol}    {\tilde{\sol}}
\newcommand{\tildadj}    {\tilde{\adj}}
\newcommand{\srcall}     {F}
\newcommand{\misfitall}  {\mathcal{J}_{\bullet}}

\newcommand{\update}[1]{{\color{black}#1}}

\newlength{\modelwidth} \newlength{\modelheight} \newlength{\jumpvert}
\newcommand{\modelfile} {}

\title{Full Reciprocity-Gap Waveform Inversion in the frequency 
       domain, enabling sparse-source acquisition}

\author{
Florian Faucher\thanks{Faculty of Mathematics, University of Vienna, Oskar-Morgenstern-Platz 1,
                       A-1090 Vienna, Austria.
                      (\href{mailto:florian.faucher@univie.ac.at}
                      {\texttt{florian.faucher@univie.ac.at}}).} 
\and
Giovanni Alessandrini\thanks{
                     Dipartimento di Matematica e Geoscienze, 
                     Universit\`{a} di Trieste, Italy.}
\and
H\'el\`ene Barucq\thanks{Inria Project-Team Magique-3D, E2S--UPPA, CNRS, Pau, France}
\and
Maarten V. de Hoop\thanks{Department of Computational and
                  Applied Mathematics and Department of Earth Science, Rice
                  University, Houston TX 77005, USA.}
\and
Romina Gaburro\thanks{
              Department of Mathematics and Statistics, 
              Health Research Institute (HRI), University of Limerick,  
              Limerick, Ireland.}
\and
Eva Sincich\samethanks[2]
}
\date{\today}

\usepackage[capitalize,nameinlink]{cleveref}
\crefname{section}   {Section}   {Sections}
\crefname{subsection}{Subsection}{Subsections}
\Crefname{section}   {Section}   {Sections}
\Crefname{subsection}{Subsection}{Subsections}
\Crefname{figure}    {Figure}    {Figures}

\crefformat     {equation}{equation~\textup{#2#1#3}}
\crefrangeformat{equation}{equations~\textup{#3#1#4--#5#2#6}}
\crefmultiformat{equation}{equations~\textup{#2#1#3}}{ and \textup{#2#1#3}}
{, \textup{#2#1#3}}{, and \textup{#2#1#3}}
\crefrangemultiformat{equation}{equations~\textup{#3#1#4--#5#2#6}}%
{ and \textup{#3#1#4--#5#2#6}}{, \textup{#3#1#4--#5#2#6}}{, and \textup{#3#1#4--#5#2#6}}
\Crefformat     {equation}{#2Equation~\textup{#1}#3}
\Crefrangeformat{equation}{Equations~\textup{#3#1#4--#5#2#6}}
\Crefmultiformat{equation}{Equations~\textup{#2#1#3}}{ and \textup{#2#1#3}}
{, \textup{#2#1#3}}{, and \textup{#2#1#3}}
\Crefrangemultiformat{equation}{Equations~\textup{#3#1#4--#5#2#6}}%
{ and \textup{#3#1#4--#5#2#6}}{, \textup{#3#1#4--#5#2#6}}{, and \textup{#3#1#4--#5#2#6}}

\crefdefaultlabelformat{#2\textup{#1}#3}
\crefname{prop}{Proposition}{Propositions}
\Crefname{prop}{Proposition}{Propositions}
\crefname{defi}{Definition}{Definitions}
\Crefname{defi}{Definition}{Definitions}
\crefname{thm}{Theorem}{Theorems}
\Crefname{thm}{Theorem}{Theorems}
\crefname{rmk}{Remark}{Remarks}
\Crefname{rmk}{Remark}{Remarks}

\begin{document}
\maketitle 

\begin{abstract}

  The quantitative reconstruction of sub-surface Earth 
  properties from the propagation of waves follows an 
  iterative minimization of a misfit functional.
  In marine seismic exploration, the observed data 
  usually consist of measurements of the pressure 
  field but dual-sensor devices also provide the normal 
  velocity.
  Consequently, a reciprocity-based misfit functional 
  is specifically designed, and defines the \emph{Full Reciprocity-gap 
  Waveform Inversion} (\emph{FRgWI}) method.
  This misfit functional provides 
  additional features compared to the more traditional 
  least-squares approaches with, in particular, that the 
  observational and computational acquisitions can be 
  different.
  Therefore, the positions and wavelets of the sources 
  from which the measurements are acquired are not needed 
  in the reconstruction procedure and, in fact, the 
  numerical acquisition (for the simulations) can be 
  arbitrarily chosen.
  Based on three-dimensional experiments, FRgWI is 
  shown to behave better than Full Waveform Inversion 
  (FWI) in the same context.
  Then, it allows for arbitrary numerical acquisitions 
  in two ways: when few measurements are given, 
  a dense numerical acquisition (compared to the observational 
  one) can be used to compensate. 
  On the other hand, with a dense observational acquisition,
  a sparse computational one is shown to be sufficient, for 
  instance with multiple-point sources, hence reducing the numerical 
  cost.
  FRgWI displays accurate reconstructions in both situations 
  and appears more robust with respect to cross-talk than the 
  least-squares shot-stacking.

\end{abstract}

\section{Introduction}

The Full Waveform Inversion (FWI) method has been 
extensively developed in the last decades for
quantitative recovery of sub-surface Earth media in 
seismic exploration. 
The concept of FWI is to minimize with respect to 
the Earth parameters a misfit criterion defined 
from the simulations of wave propagation and the 
measured seismograms (i.e., the `full waveform').
The method was originally introduced in the work of 
\cite{Bamberger1977,Bamberger1979} for one-dimensional wave 
equation, and was extended in the work of
\cite{Lailly1983} and \cite{Tarantola1984,Tarantola1987a}.
The method was first used with time-domain wave propagation,
and the frequency-domain formulation of FWI, which requires 
a Fourier transform of the original time-dependent seismic 
traces, was established by \cite{Pratt1996,Pratt1998}.

In marine seismic, the data usually consist of 
measurements of the pressure field from hydrophones, 
but new devices, the \emph{dual-sensors}, have been 
recently deployed and also give access to the vertical 
velocity, see \cite{Carlson2007} and \cite{Tenghamn2007}.
This additional information has shown advantages 
to reduce noise for image processing, 
cf.~\cite{Whitmore2010} and \cite{Ronholt2015}, 
and has motivated new analysis, such as in the work of 
\cite{Alessandrini2018,Alessandrini2019}, and 
specific numerical methodology by \cite{Faucher2017,
Zhong2018} for the seismic inverse problem. 
In our work, we implement a misfit functional dedicated 
to the dual-sensors data and demonstrate its efficiency 
for the recovery of sub-surface physical parameters.

The FWI relies on an iterative minimization of the 
misfit functional which, in the traditional 
least-squares approach, is simply the $L^2$ difference 
between the observations and the simulations.
In addition to the numerical challenges induced by the large 
scale domain, one main difficulty of FWI (which is a nonlinear 
and ill-posed inverse problem) is that the misfit functional 
suffers from local minima, in particular when the low frequency 
data are missing and/or in absence of a priori information, 
as it has been observed by, e.g., 
\cite{Gauthier1986,LuoSchuster1991,Bunks1995} and \cite{Faucher2020Geo}.
In order to mitigate the `cycle-skipping' effect, 
selection of increasing frequency content in the data 
is commonly employed, cf.~\cite{Bunks1995,Sirgue2004}.
In our work, we follow the frequency domain formulation, 
where this approach is natural, and further employ increasing 
\emph{sequential} frequency, as advocated by \cite{Faucher2020Geo}.

Several alternatives to the least-squares have been studied
to enhance the convexity of the misfit functional, such as
logarithmic function, mentioned by \cite{Tarantola1987},
that is particularly appropriate in complex-frequency FWI, 
see \cite{Shin2006,Shin2008} and \cite{Faucher2017}. 
Comparisons of misfit using the phase and amplitude of signals 
are carried out in the work of \cite{Shin2007a,Shin2007b} 
and \cite{Shin2007c}.
\cite{Fichtner2008} studies the use of a misfit based upon the 
phase, correlation, or the envelope; the later is also advocated 
by \cite{Bozdag2011}, and is shown to be efficient for the 
reconstruction of attenuation properties in global Earth seismology 
by \cite{Karaouglu2017}. 
\cite{Brossier2010} study the use of $L^1$ criterion. 
The optimal transport distance is considered in 
the work of \cite{Metivier2016} and \cite{Yang2018} and 
is shown to improve the misfit functional convexity.
In order to avoid the local minima, one can also rely on 
a specific model parametrization such as the 
migration-based travel time (MBTT) method which decomposes
the velocity model into a (smooth) background profile and 
a reflectivity in the data-space, see \cite{Clement2001} and 
\cite{Faucher2019IP}. This approach is shown to increase 
the size of the attraction basins for background velocity 
reconstruction by~\cite{Faucher2019IP}.

In the case where Cauchy data are available 
on the boundary, that is, both a field and its normal 
derivative (e.g., the pressure and the normal derivative
of the pressure, which actually relates to the normal 
velocity, see the appendix) 
the reconstruction can use a combination of the Dirichlet 
and Neumann traces, as in the work of \cite{KohnVogelius1985} 
and \cite{Colton2005} in inverse scattering, or \cite{Alessandrini2018}
for seismic.
This approach is labeled as ``\emph{reciprocity-gap}''
in the literature, see, e.g., \cite{deHoop2000,Colton2005}.
Lipschitz-stability can be obtained for the \emph{partial} 
Cauchy data inverse problem, in particular in the seismic context 
with surface measurements, as proved 
by~\cite{Alessandrini2018,Alessandrini2019}.
This result further holds for piecewise-linear parameters, 
employed for the numerical experiments of
\cite{Alessandrini2019} and \cite{Faucher2017}.
In the context of acoustic waves governed by the Euler's 
equations, we shall see that Cauchy data are in fact 
equivalent to the pressure field and the normal velocity, 
i.e., dual-sensors data.
In the first appendix, we further connect the boundary 
measurements to the volume properties of the media.
Hence, the minimization of the misfit functional 
defined from \emph{surface} data is connected to 
the recovery of the \emph{volume} medium parameters.
In our work, we employ the reciprocity-gap formulation 
and further combine all sources, defining the 
\emph{Full Reciprocity-gap Waveform Inversion} (FRgWI) 
framework for seismic imaging using dual-sensors data.

In the time-harmonic formulation, the reciprocity-gap
results in a misfit functional where observations 
and simulations are multiplied. 
While our work is conducted in the frequency
domain, it can similarly be carried out in the time 
domain where it relates to the family of correlation 
misfits.
Correlation-based misfit functionals have been 
studied in seismic tomography, for instance by
\cite{VanLeeuwen2010} and in the work of 
\cite{Choi2011} and \cite{Zhang2016}, however
using a single type of measurements (i.e., the 
acoustic pressure fields). 
The use of the velocity fields is studied in the 
time-domain by \cite{Zhong2019}, assuming that 
all directional velocities are available and 
convolving the same fields. 
A fundamental difference with these earlier studies, 
and in particular with \cite{Zhong2019}, is that in our work, 
we correlate \emph{different} fields (i.e. pressure with 
velocity). 
This is the essence of reciprocity-gap and it is crucial 
in order to relate surface measurements to global model 
reconstruction using Green's identity (see the first appendix).
In addition, our framework is naturally designed for the 
normal velocity, in accordance with the dual-sensors 
devices.
In the work of \cite{Menke2003} and \cite{Bodin2014}
in seismology, observations and simulations are also combined, 
but using a cross-convolution formula made of the vertical  
and horizontal components of the waves.
More generally, approaches based upon a specific filtering 
of the data, such as in the work of~\cite{Warner2016} and 
\cite{Guasch2019}, also rely on a minimization where the 
fields are \emph{not} directly the quantities compared. 

The main feature of FRgWI is to allow different 
observational and numerical acquisitions, 
cf.~\cite{Faucher2017,Alessandrini2019,Zhong2019}.
Minimal information regarding the observational 
sources is required: the source function and the
source positions are not needed to conduct the 
reconstruction.
This is due to the definition of the misfit
functional which does not compare an observation 
with a simulation directly, but products of an 
observation with a simulation.
This flexibility in the choice of numerical probing
sources opens up many perspectives and, in particular,
to use denser or sparser computational acquisitions 
than the given observational one.
The use of sparse acquisition relates to the 
shot-stacking approach for data decimation, 
which sums several single point-source data 
using the linearity of the wave equation.
It has early been employed in seismic to reduce the 
computational cost by \cite{Mora1987}. 
It is based upon the redundancy of information in 
the data, but has a major drawback as it is 
difficult to avoid the cross-talk between the 
encoded sources, cf. \cite{Krebs2009,Zhang2018}.
It has motivated several works to efficiently 
assemble the encoded sources in seismic, i.e., 
source blending \citep{Berkhout2008}. We
mention, for instance, the random combination 
of sources changing with iteration by \cite{Krebs2009},
the approach based upon compress sensing by \cite{Li2012},
while wavelet encoding is used by \cite{Zhang2018}; 
see also the references therein.
In our applications, we will see that FRgWI, 
by using arbitrary numerical sources, can 
work with multiple-point sources 
(for computational \emph{or} observational
acquisition) while being naturally robust 
to the cross-talk. 
In fact, it is not exactly cross-talk in this context 
because the key of FRgWI is that the measurements 
are \emph{not} modified, i.e., they are still tested 
\emph{independently}, one by one, with respect to the 
numerical simulations.
%

In this paper, we study the seismic inverse problem
associated with time-harmonic waves, using dual-sensors data. 
We first state the mathematical problem 
and define the two misfit functionals that we analyze for the 
iterative reconstruction: the traditional least-squares 
difference and the reciprocity-based functional.
Next, we detail the features provided by the reciprocity-gap 
formulation. 
We carry out three-dimensional reconstruction experiments:
first with a layered medium, where we compare the performance 
of reciprocity-gap and full waveform inversion.
We further demonstrate the efficiency of FRgWI
with simultaneous point-sources to reduce the computational
cost, and its robustness compared to traditional shot-stacking.
Finally, we carry out a larger-scale experiment including 
salt domes using the SEAM benchmark.
In particular, we highlight that FRgWI performs equally 
well in the case of acquisitions that are 
sparse for observations/dense for simulations and 
when acquisitions are
dense for observations/sparse for simulations.
\update{Our experiments are carried out using data acquired in the 
time-domain and, even though our reconstruction algorithm is conducted 
in the frequency-domain, which might present some scale limitations for 
field configurations, we believe that our method can be implemented with 
the time-domain wave equation.
This can be more appropriate for larger scales and it only requires a 
slight modification to our model. A discussion about it is carried out 
in the conclusions.}
\section{Time-harmonic seismic inverse problem with dual-sensors data}
\label{section:inverse_problem}

We work with time-harmonic wave propagation for 
the identification of the physical parameters in
a seismic context. 
The quantitative reconstruction is conducted
using an iterative minimization of a misfit
functional. 
We first give the misfit as the traditional 
$L^2$ difference and further design the 
reciprocity-gap version of the functional, which 
combines pressure and normal velocity data.

\subsection{Acoustic wave equation, forward problem from dual-sensors}

We consider a three-dimensional domain 
$\Omega \subset \R^3$ with boundary $\Gamma$.
The propagation of waves in acoustic media
is represented with the scalar pressure 
and vectorial velocity fields, that satisfy 
the Euler's equations 
\citep{Colton1996,Kirsch1996,deHoop2000},
\begin{subequations}
\label{eq:euler_main}
\begin{empheq}[left={(\text{Problem}~\ref{eq:euler_main})~~\empheqlbrace}]{alignat=2} 
   -\ii \omega \rho(\bx) \velocity(\bx) &= -\nabla \pressure(\bx), 
        &\text{in $\Omega$,}      \label{eq:euler_main_a} \\
   -\ii \omega \kappa(\bx)^{-1} \pressure(\bx)  &= -\divergence \velocity(\bx) + f(\bx),
        &\text{in $\Omega$,}                                           \\
    \pressure(\bx) & = 0,  &\text{on $\Gamma_1$ (Free Surface),}    
    \label{eq:euler_main_bc_fs}   \\
    \partial_{\n} \pressure(\bx) - \dfrac{\ii \omega}{c(\bx)} \pressure(\bx) 
                   & = 0, &\text{on $\Gamma_2$ (ABC).}
    \label{eq:euler_main_bc_abc}   
\end{empheq} \end{subequations}
The frequency is denoted by $\omega$,
$f$ is the (scalar) interior (harmonic) source term
and $\partial_{\n}$ denotes the normal derivative.
The propagation is characterized by the two physical 
parameters of the medium: the density $\rho$ and the 
bulk modulus $\kappa$. 
In addition, the velocity $c$ is defined such that 
\begin{equation}
 c(\bx) = \sqrt{\dfrac{\kappa(\bx)}{\rho(\bx)}}.
\end{equation}

For boundary conditions, we follow a geophysical 
context where the boundary is separated into two: 
$\Gamma = \Gamma_1 \cup \Gamma_2$. 
We denote by $\Gamma_1$ the interface between the 
air and the acoustic medium, where a \emph{free 
surface} boundary condition holds,~\cref{eq:euler_main_bc_fs}. 
On the other part of the boundary, $\Gamma_2$, 
we implement an \emph{Absorbing Boundary Conditions} 
(ABC), \cite{Engquist1977},~\cref{eq:euler_main_bc_abc},
to ensure that waves reaching the boundary are not 
reflected back into the domain. 
It corresponds to the fact that the domain $\Omega$ 
is a numerical restriction of the Earth.

The \emph{dual-sensor} devices have been recently
introduced in marine seismic exploration, and 
allow the recording of both the pressure field and 
the vertical velocity, see~\cite{Carlson2007} and 
\cite{Tenghamn2007}.
Consequently we define the \emph{forward problem}
(which maps the parameters to the data) 
$\forward_\omega^{(f)}$ at frequency $\omega$ for 
a source $f$ such that
\begin{equation} \label{eq:forward_problem}
 \forward_\omega^{(f)}(m) = 
 \Big\{ \pressure^{(f)}_\omega\mid_\Sigma \, , \, \,
        v_{\n,\omega}^{(f)}\mid_\Sigma \Big\}.
\end{equation}
The model parameters are referred to by 
$m= (\kappa, \, \rho)$, the normal velocity
by $v_{\n}$ and $\Sigma$ corresponds 
with the (discrete) set of receivers location: 
\begin{equation}
 \pressure\mid_\Sigma = \big\{ \, \pressure(\bx_1), \, \ldots, \, \pressure(\bx_{\nrcv}) \, \big\},
\end{equation}
where $\bx_i$ is the position of the $i^\text{th}$
receiver for a total of $\nrcv$ receivers.
For notation, we introduce the restriction 
operator $\restrict$, which reduces the fields 
to the set $\Sigma$, such that 
\begin{equation}
  \restrict \big( \pressure\big) = \pressure\mid_\Sigma~, \quad \quad 
  \restrict \big( v_{\n}   \big) = v_{\n}\mid_\Sigma.
\end{equation}

\subsection{Misfit functionals}

The inverse problem aims the \emph{quantitative}
reconstruction of the subsurface medium 
parameters ($\kappa$ and $\rho$) from data measured 
at the receivers location. 
Using pressure and vertical velocity measurements, 
we denote the data at frequency $\omega$ for a 
source $f$ by 
\begin{equation}
  d_\omega^{(f)} = \Big\{ \, d_{\omega,p}^{(f)} \, , \, \, \, 
                         d_{\omega,v}^{(f)} \, \Big\},
\end{equation}
where $d_{\omega,p}^{(f)}$ and $d_{\omega,v}^{(f)}$ 
are vectors of $\Cx^{\nrcv}$ and respectively refer 
to the pressure and normal velocity records, 
following~\cref{eq:forward_problem}.
We further denote by $d_{\omega}^{(f)}(\bx_i)$ the 
data recorded for the source $f$ at the $i^{\text{th}}$ 
receiver.

In the following, we omit the frequency index 
and space dependency for the sake of clarity
(in the experiments, we use increasing 
sequential frequency, as suggested
by \cite{Faucher2020Geo}). 
We introduce two misfit functionals
which evaluate the difference between the 
observations and simulations. 

Firstly, the functional $\misfitL$, which follows 
the traditional least-squares, is
\begin{equation} \label{eq:misfit_classic}
  \misfitL(m) = \dfrac{1}{2} \sum_{i=1}^\nsrc 
                \Big\Vert \restrict \big( \pressure^{(f_i)} \big) 
                       - d_{\pressure}^{(f_i)} \Big\Vert^2_2
              + \dfrac{\eta}{2} 
                \Big\Vert \restrict \big( v_{\n}^{(f_i)} \big) 
                       - d_{v}^{(f_i)} \Big\Vert^2_2,
\end{equation}
where $\eta$ is a scaling factor to adjust 
between the amplitudes of the pressure and 
velocity.
In our applications, it is taken such that 
$\Vert d_{\pressure} \Vert = \eta \Vert d_{v} \Vert$.

Secondly, we define an alternative misfit 
functional based upon the \emph{reciprocity-gap}, 
motivated by Green's identity:
\begin{equation} \label{eq:misfit_green}
\begin{aligned}
  \misfitG(m) & = \dfrac{1}{2} \sum_{i=1}^{\nsrcObs} \sum_{j=1}^{\nsrcSim}
                  \Big\Vert d_{v}^{(f_i) \, T} \restrict \big( \pressure^{(g_j)}      \big)
               -          d_{\pressure}^{(f_i) \, T} \restrict \big( v_{\n}^{(g_j)} \big)
                \Big\Vert^2_2, \\
              & = \dfrac{1}{2} \sum_{i=1}^{\nsrcObs} \sum_{j=1}^{\nsrcSim}
                  \Bigg\Vert 
                    \sum_{k=1}^{\nrcv}  \bigg( 
                      d_{v}^{(f_i)}(\bx_k)         \, \pressure^{(g_j)}(\bx_k)
                 -    d_{\pressure}^{(f_i)}(\bx_k) \,    v_{\n}^{(g_j)}(\bx_k) \bigg)
                \Bigg\Vert^2_2, \\
\end{aligned}
\end{equation}
where $^T$ denotes the transpose. 
The misfit functional is motivated by 
the Green's identity and has been 
introduced in the context of inverse 
scattering, from Cauchy data,
as mentioned in the introduction, 
cf.~\cite{KohnVogelius1985,Colton2005},
and used with partial seismic data by~\cite{Alessandrini2019}.
In the first appendix, 
we justify the formulation of the misfit functional 
$\misfitG$ using variational formulation of 
Problem~\ref{eq:euler_main}, and note 
that the mathematical foundation of the 
reciprocity-gap formulation relies \emph{naturally} 
on the \emph{normal} velocity.
Therefore, it perfectly matches the dual-sensors data, 
and does not necessitate the specific directional components
($v_x$, $v_y$ or $v_z$).

\subsection{Iterative reconstruction procedure}

The reconstruction procedure follows an iterative 
minimization of the selected misfit functional.
For least-squares formulation such as~\cref{eq:misfit_classic},
it is traditionally referred to as the Full Waveform 
Inversion (FWI) method in seismic (as one makes use 
of the full data seismograms) see the review of 
\cite{Virieux2009} and the references therein. 
Consequently, we shall refer to the minimization 
of~\cref{eq:misfit_green} as 
\emph{Full Reciprocity-gap Waveform Inversion}: \emph{FRgWI}.
In any case, the iterative minimization follows 
successive updates of the physical models, such 
that, at iteration $l$, the new model is given by
$m_{l+1}  = m_l - \alpha_l s_l$, 
where $\alpha$ is the scalar step size typically 
computed via a line search method \citep{Nocedal2006}, 
and $s$ is the search direction.

The search direction depends on the gradient of 
the cost function, which is computed using the 
adjoint state method. 
The method has its foundation in the body of 
work of \cite{Lions1971}, and is reviewed 
for seismic application by \cite{Plessix2006}.
Application with complex fields is further 
described by~\cite{Barucq2018,Faucher2019IP}.
The adjoint-state method for reciprocity-gap 
functional is briefly reviewed in the second 
appendix, 
see also~\cite{Alessandrini2019}.
In our implementation, the search direction 
only depends on the gradient of the misfit 
functional, and we employ the Limited-BFGS 
algorithm, see~\cite{Nocedal1980} and \cite{Nocedal2006}.
We review the steps for the iterative minimization
in \cref{algo:FWI}.

\begin{algorithm}[ht!]
\SetAlgoLined
 Initialization: starting models $m_1 = (\kappa_1, \, \rho_1)$, 
                 and measurements $d_\pressure, \, d_v$. \\
 \For{$\omega_i = \omega_1, \, \ldots, \, \omega_{N_\omega}$}{
   \For{$j = 1, \, \ldots, \, n_\text{iter}$}{
   set $l:= (i-1)n_\text{iter} + j$ \;
   solve Problem~\ref{eq:euler_main} using current 
     models $m_k$ and frequency $\omega_i$\;
   compute the misfit functional,~\cref{eq:misfit_classic} or~\cref{eq:misfit_green}\;
   compute the gradient of the misfit functional using the adjoint-state method\;
   compute the search direction, $s_l$, using Limited-BFGS\;
   compute the step length, $\alpha_l$, using line search method\;
   update the model: $m_{l+1}  = m_l - \alpha_l s_l$.
  }
 }
 \caption{Iterative reconstruction procedure: for the minimization 
          of~\cref{eq:misfit_classic}, the computational acquisition
          (source positions and wavelet) must follow the one employed
          to generate the measurements. For the minimization 
          of~\cref{eq:misfit_green}, the user can prescribe a computational
          acquisition by means of the positions and wavelet of the 
          probing sources $g_j$ in~\cref{eq:misfit_green}.}
 \label{algo:FWI}
\end{algorithm}

\subsection{Discretization with Hybridizable Discontinuous Galerkin method}

In the numerical implementation, both the pressure and 
velocity fields must be computed to feed the misfit 
functional.
We discretize Problem~\ref{eq:euler_main} using 
the \emph{Hybridizable Discontinuous Galerkin} 
(HDG) method, which is specifically designed for 
first-order problems, and avoid oversize linear 
systems.
Indeed, the global matrix using HDG is only 
composed of the degrees of freedom associated 
with the \emph{trace of the pressure field}, 
that is, only the degrees of freedom on the 
faces of the elements of the discretization 
mesh, see~\cite{Cockburn2009,Griesmaier2011} and \cite{Bonnasse2017}. 
Then, local (small) systems are solved to calculate
the volume solutions of both the pressure and the 
velocity fields, computed with similar accuracy.
We refer to \cite{Faucher2020adjoint} for the precise implementation.

With more traditional discretization methods such as 
Continuous Galerkin, Finite Differences of Internal 
Penalty Discontinuous Galerkin methods, one has to 
create a linear system which size is the total number
of degrees of freedom for all unknowns (i.e., the 
pressure and the three components of the velocity), 
possibly leading to cumbersome systems. 
Alternatively, one can solve only for the scalar pressure 
field, and post-process the solution to obtain the velocity 
but the computed velocity looses one order of accuracy 
compared to the discretized pressure field due 
to the derivative in~\cref{eq:euler_main_a}.
In the HDG method, we obtain both the pressure 
and velocity fields with the same accuracy, while the 
global linear system only contains the degrees of 
freedom of a scalar unknown \citep{Faucher2020adjoint}.
In addition, only the degrees of freedom on the faces
of the elements are taken into account, hence removing
all interior ones. 
Consequently, the HDG method has been shown to be more 
efficient (i.e., less memory consumption for the matrix 
factorization due to the smaller global matrix) 
compared to other approaches by~\cite{Kirby2012} 
and \cite{Bonnasse2017}, for high order discretization.

In the context of large scale seismic applications, 
the use of HDG is crucial (particularly in the 
frequency domain) to efficiently compute the pressure 
and the velocity, because it eventually leads to linear 
systems which sizes are \emph{not} larger than for 
second-order wave problems.
Then, For the resolution of the subsequent linear system, 
we use a direct solver \citep{Amestoy2001,XLiu2018}
for the multiple right-hand sides feature, in particular,
the solver \textsc{Mumps}, designed for sparse 
matrices \citep{Amestoy2006}.
Therefore, the matrix factorization of the forward 
problem is reused for the backward one that serves
to compute the gradient of the misfit functional 
depicted in the second appendix.

\section{Features of FRgWI and data acquisition}
\label{section:features}

Following \cite{Faucher2017} and \cite{Alessandrini2019},
the fundamental feature of the reciprocity-gap misfit 
functional~\cref{eq:misfit_green} compared 
to~\cref{eq:misfit_classic} is that the 
set of computational sources is separated 
from the observational ones (respectively $g_j$
and $f_i$ in~\cref{eq:misfit_green}).
It implies that 
\begin{enumerate}
  \item \emph{we do not have to know the observational source 
        positions (the location of the $f_i$ in~\cref{eq:misfit_green}) 
        for the reconstruction algorithm;}
  \item \emph{we do not have to know the observational source signatures (wavelet)};
\end{enumerate}
Consequently, and contrary to least-squares misfit 
such as~\cref{eq:misfit_classic}, 
minimal information is required regarding the 
observational acquisition (only the positions of the receivers). 
The recovery of the source wavelet in parallel to the 
iterative minimization procedure usually performs well
in seismic exploration, using the method prescribed by
\cite{Pratt1999a}, see \cref{rk:source-reconstruction}.
However, incorrect knowledge of the position of the
sources is a strong difficulty which can lead to 
the failure of the reconstruction procedure. 
Namely, FRgWI increases robustness by being free of 
such considerations.
In addition, 
\begin{enumerate}\setcounter{enumi}{2}
  \item \emph{the set of computational sources can 
        be arbitrarily taken, hence can differ from 
        the observational one (respectively 
        $\nsrcObs$ and $\nsrcSim$ in~\cref{eq:misfit_green})}.
\end{enumerate}
Therefore, it provides high flexibility regarding 
the choice of computational sources. 
It is important to notice that whatever computational
sources are selected, the observed data are still 
\emph{independently} tested one by one against
the simulations, i.e., the measurements are \emph{not}
modified in any way.

\subsection{Data acquisition}

In our experiments, we consider the source term for 
the wave propagation ($f$ or $g$) to be a Ricker 
function in time and a delta-Dirac function in space. 
In the frequency domain, it translates to a 
delta-Dirac function $\delta$ (with value 
equal to the discrete 
Fourier transform of the time-domain signal).
We refer to a \emph{point}-source when the source 
is localized at the position $\bx_k$: 
\begin{equation} \label{eq:single_ploint_source}
  f_{k}  = \delta(\bx - \bx_k) \, , 
  \qquad \text{point source in $\bx_k$.}
\end{equation}
We further refer to \emph{multiple-point} source
when it is composed of $\npt$ point-sources, 
such that 
\begin{equation} \label{eq:multi_point_source}
  f_{j} = \sum_{k=1}^{\npt} \delta(\bx - \bx_{j,k}^{(f)}) \, ,
  \qquad \text{multiple-point source.}
\end{equation}

The observational setup uses a fixed 
lattice of receivers, and sources located slightly above,
as illustrated in \cref{fig:multi-source-splitting_A}. 
Furthermore, the position of the receivers remains the 
same for all sources. 
This configuration is consistent with the \emph{TopSeis}
acquisition system for marine seismic developed by CGG 
(Compagnie G\'en\'erale de G\'eophysique) and Lundin Norwary AS. 
This principle has been recently deployed on field, showing 
benefits for near-offset coverage, and it has also won 
the `Exploration Innovation Prize 2019'\footnote{We refer to 
  $\text{https://www.cgg.com/en/What-We-Do/Offshore/Products-and-Solutions/TopSeis}$
  and~$\text{https://expronews.com/2019/05/23/topseis-a-worthy-winner/}$.
  \label{footnote:topseis}}.
Namely, it consists in having two boats: one that moves and 
carries the sources, and one that carries the receivers and 
remains fixed.

For the reconstruction using FRgWI, we can employ
arbitrary numerical sources, which do not have 
to coincide with the observational acquisition. 
Here, we investigate the two following configurations.

\subsection{Dense point-source computational acquisition for sparse multiple-point measurements}

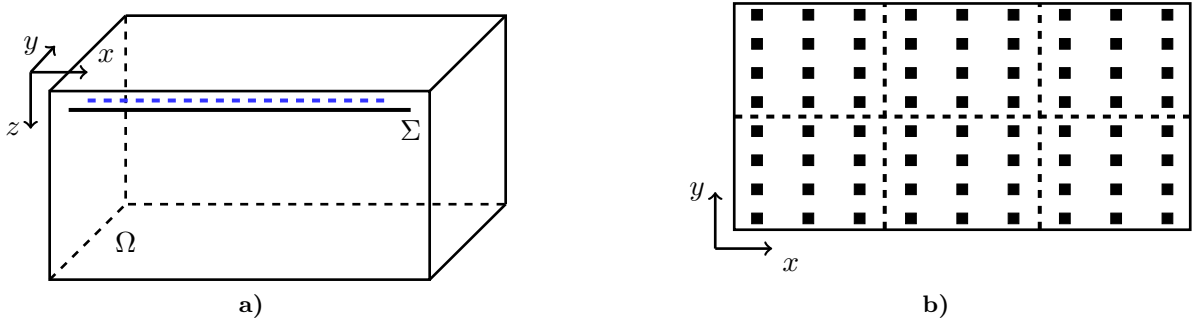
\begin{figure}[ht!] \centering
\subfloat[][]{\label{fig:multi-source-splitting_A} 
\begin{tikzpicture}[]
  \pgfmathsetmacro{\len}        {2.5}
  \pgfmathsetmacro{\wid}        {5}
  \draw[color=black,line width=1pt] (0,0) rectangle (\wid,\len);
  \pgfmathsetmacro{\gapx}        {1}
  \pgfmathsetmacro{\gapy}        {1}

  \coordinate (a1)at (0   , 0);
  \coordinate (a2)at (\wid, 0);
  \coordinate (a3)at (\wid, \len);
  \coordinate (a4)at (0   , \len);
  \coordinate (b1)at (\gapx     , \gapy);
  \coordinate (b2)at (\gapx+\wid, \gapy);
  \coordinate (b3)at (\gapx+\wid, \gapy+\len);
  \coordinate (b4)at (\gapx     , \gapy+\len);
  \draw[color=black,line width=1pt,dashed] (b4) -- (b1) -- (b2); 
  \draw[color=black,line width=1pt] (b2) -- (b3) -- (b4); 
  \draw[color=black,line width=1pt] (a4) -- (b4) ; 
  \draw[color=black,line width=1pt] (a2) -- (b2) ; 
  \draw[color=black,line width=1pt] (a3) -- (b3) ; 
  \draw[color=black,line width=1pt,dashed] (a1) -- (b1) ; 
        
  \coordinate (s1)at (0.10*\wid, 0.95*\len);
  \coordinate (s2)at (0.90*\wid, 0.95*\len);
  \draw[dashed,line width=1.5pt,\myblue] (s1) -- (s2);

  \coordinate (r1)at (0.05*\wid, 0.90*\len);
  \coordinate (r2)at (0.95*\wid, 0.90*\len);
  \draw[line width=1.5pt,black] (r1) -- (r2);

  \coordinate (x0)at (-0.25,\len + 0.25);
  \coordinate (xm)at ( 0.50,\len + 0.25);
  \coordinate (zm)at (-0.25,\len - 0.50);
  \coordinate (ym)at ( 0.07,\len + 0.60);
  
  \draw[line width=1pt,->] (x0) -- (xm);
  \draw[line width=1pt,->] (x0) -- (zm);
  \draw[line width=1pt,->] (x0) -- (ym);
  \node[anchor=south west,xshift=0] at (xm) {$x$};
  \node[anchor=east      ] at (zm) {$z$};
  \node[anchor=east,xshift=-0.2em] at (ym) {$y$};
  \node[xshift=1cm,yshift=0.5cm] at (0,0) {$\Omega$};
  \node[anchor=north,black] at (r2) {$\Sigma$};
    
\end{tikzpicture}
} \hfill
\subfloat[][]{\label{fig:multi-source-splitting_B} 
\begin{tikzpicture}[]
  \pgfmathsetmacro{\len}        {3.0}  
  \pgfmathsetmacro{\wid}        {6}  
  \pgfmathsetmacro{\step}       {3}
  \draw[color=black,line width=1pt] (0,0) rectangle (\wid,\len);
  \draw[dashed,line width=1.5pt] (0, 0.5*\len) -- (\wid, 0.5*\len);
  \draw[dashed,line width=1.5pt] (0.33*\wid,0) -- (0.33*\wid,\len);
  \draw[dashed,line width=1.5pt] (0.67*\wid,0) -- (0.67*\wid,\len);
  
  \pgfmathsetmacro{\kmax}       {7}
  \pgfmathsetmacro{\jmax}       {8}
  \foreach \k in {0, ..., \kmax} 
    \foreach \j in {0, ..., \jmax} 
  {
    \coordinate (s1)at (0.05*\wid + \j*0.90/\jmax*\wid, 0.05*\len + \k*0.90/\kmax*\len);
    \draw[mark=square*,mark size=2pt,mark options={color=black},only marks] 
          plot coordinates{(s1)} ;  ;
  }

  \coordinate (x0)at (-0.25,-0.25);
  \coordinate (xm)at ( 0.50,-0.25);
  \coordinate (ym)at (-0.25, 0.50);
  \draw[line width=1pt,->] (x0) -- (xm);
  \draw[line width=1pt,->] (x0) -- (ym);
  \node[anchor=north west,xshift=0] at (xm) {$x$};
  \node[anchor=east      ] at (ym) {$y$};
\end{tikzpicture}
}
\caption{\protect\subref{fig:multi-source-splitting_A}
         Domain of acquisition: all sources are positioned 
         at the same depth (blue dashed line) and the 
         receivers are slightly below (black line, $\Sigma$).
         The receivers remain fixed for all sources according
         to the Topseis acquisition.
        \protect\subref{fig:multi-source-splitting_B}
         $(x,y)$ source plane (all sources are at the same depth $z$):
         every black square corresponds with a point-source.
         The multiple-point sources are defined by taking 
         all points in a specified sub-domain,  
         indicated with the dashed line, i.e. using a
         \emph{structured} decomposition.
        }
\label{fig:multi-source-splitting}
\end{figure}

In the case where the observational acquisition is composed 
of few sources (e.g., to reduce the cost of field experiments), 
FRgWI can use a dense coverage of computational point-sources
to enhance the sensitivity.
The sources for the measurements can be multiple-points, e.g., if 
several air-guns are excited at the same time. 
In our experiments, the choice of positions for the 
multiple-points follows a structured decomposition:
adjacent sources are grouped, with a fixed
distance in $x$ and $y$ between each of them,
cf.~\cref{fig:multi-source-splitting_B}.
It appears more natural to employ such structured decision, 
e.g. the boat will carry the air-gun sources all together. 
In addition, note that we have observed in our experiments that 
this structured partition gives a better performance than using 
a random combination of sources.

\subsection{Sparse multiple-point computational acquisition for dense point-source measurements}

In the case where the observational acquisition is composed 
of many (point) sources (as it can happen in exploration 
seismic), FRgWI can instead use multiple-point sources in the 
computational acquisition to reduce the computational cost. 
It consists of a \emph{sparsification} of the observational
acquisition.

In order to compare with the traditional FWI approach, 
one can use the linearity of the wave equation and the 
multiple-point source can be assimilated to the well-known 
\emph{shot-stacking} approach, which rewrites the misfit 
functional \cref{eq:misfit_classic} such that 
\begin{equation} \label{eq:misfit_stack}
  \misfitL^{\text{stack}}(m) = 
    \sum_{j=1}^{\nstack}
    \dfrac{1}{2}
    \Big\Vert 
    \sum_{k = 1}^{\npt} \restrict \Big( \pressure^{(f_{j,k})} \Big) 
  - \sum_{k = 1}^{\npt} d_{\pressure}^{(f_{j,k})} \Big\Vert^2
  + \dfrac{\eta}{2} 
    \Big\Vert 
    \sum_{k = 1}^{\npt} \restrict \Big( v_{\n}^{(f_{j,k})} \Big) 
  - \sum_{k = 1}^{\npt} d_{v}^{(f_{j,k})} \Big\Vert^2,
\end{equation}
where we have a total of $\nstack$ multiple-point sources, 
each composed of $\npt$ points.

Therefore, we will investigate in the following numerical
sections different situations where one of the acquisitions
(observational or computational) is composed of point-sources 
while the other is made of multiple-point sources.
The main advantage of FRgWI is that it does not 
require any modification of the data, which are tested 
one by one against the computational acquisition
in~\cref{eq:misfit_green}. 
However, the shot-stacking version of FWI 
requires the summation of data in~\cref{eq:misfit_stack}, 
which results in the possible loss of information, 
i.e., cross-talk \citep{Zhang2018}.

\section{Numerical experiment 1: layered medium} 
\label{section:numerical_experiments_1}

In this section and the next, we carry out 
three-dimensional experiments of geophysical 
reconstruction to study the performance of FRgWI.
In this first test, we compare with the 
traditional least-squares functional, and we 
employ multiple-point sources to probe the 
robustness of arbitrary source positions;
we also test the combination of sparse and dense acquisitions.

\subsection{Velocity model}
\label{subsection:numerical:statoil}

\begin{figure}[ht!] \centering
\setlength{\modelwidth} {7.25cm}
\setlength{\modelheight}{4.00cm}
  \renewcommand{\modelfile}{vp_true}
  \begin{tikzpicture}
\pgfmathsetmacro{\xmin} {0.}
\pgfmathsetmacro{\xmax} {2.540}
\pgfmathsetmacro{\ymin} {0.}
\pgfmathsetmacro{\ymax} {1.440}
\pgfmathsetmacro{\zmin} {0.}
\pgfmathsetmacro{\zmax} {1.220}

\pgfmathsetmacro{\ycut} {1.14}
\pgfmathsetmacro{\zcut} {0.8}
\pgfmathsetmacro{\vmin} {1.5}
\pgfmathsetmacro{\vmax} {5.2}

\matrix[column sep=7mm, row sep=0mm] {
\begin{axis}[
  anchor=center,
  tick label style={font=\small},
  grid=both,minor tick num=1,
  xlabel={\small{$x$ (\si{\km})}},ylabel={\small{$y$ (\si{\km})}},
  zlabel={\small{depth (\si{\km})}}, 
  ztick pos=left,
  3d box,width=\modelwidth, 
  xmin=\xmin,ymin=\ymin,zmin=\zmin,xmax=\xmax,ymax=\ymax,zmax=\zmax,
  every axis x label/.style={at={(0.35, 0.00)},anchor=north},
  every axis y label/.style={at={(0.95, 0.02)},anchor=north},
  line width=.25pt]
  \addplot3[fill=white] graphics[points={
            (\xmin,\ymin,\zmax)  => (244,232-232)
            (\xmax,\ymin,\zmax)  => (0. ,232-150)
            (\xmin,\ymax,\zmax)  => (410,232-193)
            (\xmax,\ymax,\zmin)  => (165,232-0. )}
            ]{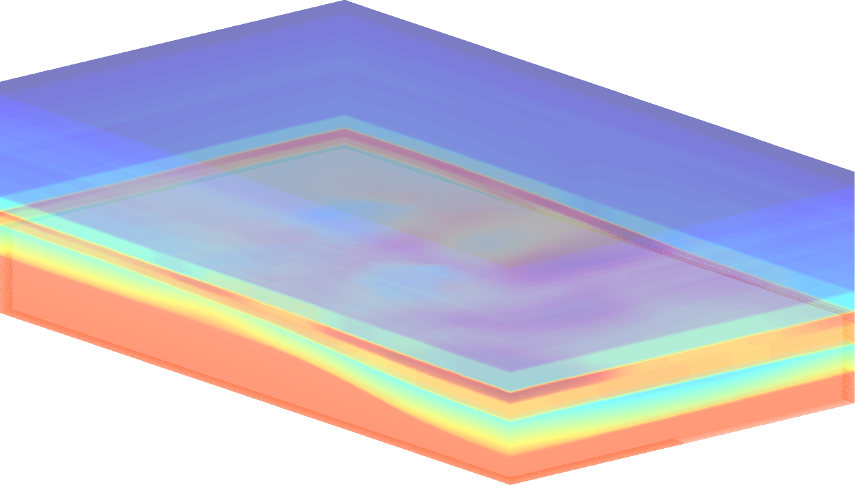};
  
  \addplot3[dashed,black,line width=1pt] 
      coordinates {(\xmin,\ycut,\zmin) (\xmin,\ycut,\zmax)
                   (\xmax,\ycut,\zmax) (\xmax,\ycut,\zmin) (\xmin,\ycut,\zmin)};
  \addplot3[dashed,black,line width=1pt] 
      coordinates {(\xmin,\ymin,\zcut) (\xmin,\ymax,\zcut) 
                   (\xmax,\ymax,\zcut) (\xmax,\ymin,\zcut) (\xmin,\ymin,\zcut)};
\end{axis}

&

\begin{axis}[%
anchor=center, width=\modelwidth, height=\modelheight,
xmin=\xmin, xmax=\xmax,ymin=\zmin, ymax=\zmax,hide axis,
tick label style={font=\small},y dir=reverse]
\addplot [forget plot] graphics 
         [xmin=\xmin,xmax=\xmax,ymin=\zmin,ymax=\zmax] {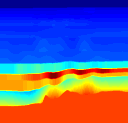};

\addplot[dashed,black,line width=1pt] 
      coordinates {(\xmin+.01,\zmin+.01) (\xmin+.01,\zmax-.01) 
                   (\xmax-.01,\zmax-.01) (\xmax-.01,\zmin+.01) (\xmin+.01,\zmin+.01)};
\end{axis}

\\

\begin{axis}[%
anchor=center, width=\modelwidth, height=\modelheight,
xmin=\xmin, xmax=\xmax,ymin=\ymin, ymax=\ymax,hide axis,
tick label style={font=\small}, y dir=reverse]
\addplot [forget plot] graphics 
         [xmin=\xmin,xmax=\xmax,ymin=\ymin,ymax=\ymax] {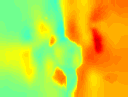};

\addplot[dashed,black,line width=1pt] 
      coordinates {(\xmin+.01,\ymin+.01) (\xmin+.01,\ymax-.01) 
                   (\xmax-.01,\ymax-.01) (\xmax-.01,\ymin+.01) (\xmin+.01,\ymin+.01)};
\end{axis}

&

\begin{axis}[anchor=south,yshift=.5cm,hide axis,height=.5cm,colorbar/width=.2cm,
     colormap/jet,colorbar horizontal,
     colorbar style={separate axis lines,
                     tick label style={font=\small},
                     title={\small{velocity (\si{\km\per\second})}}},
                     point meta min=\vmin, point meta max=\vmax,
                     height=.5\modelwidth]
\end{axis}
\\};

\draw[black,-stealth,->,dashed,line width=1pt] (-0.2,1.55)to[out=0,in=-180] ( 1.0 ,1.55);
\draw[black,-stealth,->,dashed,line width=1pt] (-4.5,1.0) to[out=-90,in=90] (-4.5,-0.5);

\end{tikzpicture}
  \caption{velocity model with layers of 
           size $2.55 \times 1.45 \times 1.22$ 
           \si{\km\cubed}. For visualization, we 
           extract sections at a fixed 
           depth $z = 800$ \si{\meter} 
           (bottom) and for $y = 1125$ 
           \si{\meter} (right).}
  \label{fig:statoil:true}
\end{figure}
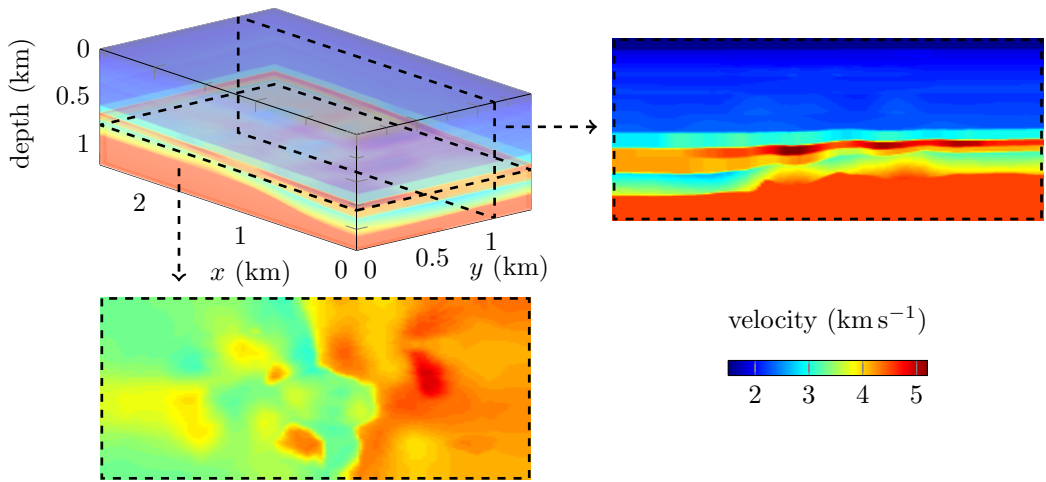

We consider a three-dimensional acoustic medium
(provided by Statoil) of size 
$2.55 \times 1.45 \times 1.22$ \si{\km\cubed}, in the 
$x$, $y$, and $z$ axes respectively (i.e., $z$
is the medium depth). 
We assume a constant density 
$\rho = 1000~\si{\kg\per\meter\cubed}$, and 
the subsurface velocity is pictured in 
\cref{fig:statoil:true}. 
It consists of different geophysical layers, 
with non-monotone variations, from \num{1500} 
to \num{5200} \si{\meter\per\second}. 
The first 500 \si{\meter} are mostly constant
with a velocity of about \num{1600}
\si{\meter\per\second}.

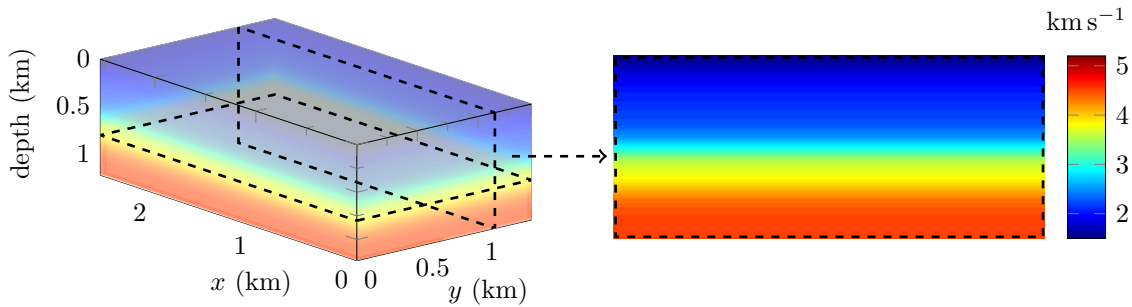
\begin{figure}[ht!] \centering
\setlength{\modelwidth} {7.25cm}
\setlength{\modelheight}{4.00cm}
  \renewcommand{\modelfile}{vp_start_bkg}
\begin{tikzpicture}
\pgfmathsetmacro{\xmin} {0.}
\pgfmathsetmacro{\xmax} {2.540}
\pgfmathsetmacro{\ymin} {0.}
\pgfmathsetmacro{\ymax} {1.440}
\pgfmathsetmacro{\zmin} {0.}
\pgfmathsetmacro{\zmax} {1.220}

\pgfmathsetmacro{\ycut} {1.14}
\pgfmathsetmacro{\zcut} {0.8}
\pgfmathsetmacro{\vmin} {1.5}
\pgfmathsetmacro{\vmax} {5.2}

\matrix[column sep=10mm] {
\begin{axis}[
  anchor=center,
  tick label style={font=\small},
  grid=both,minor tick num=1,
  xlabel={\small{$x$ (\si{\km})}},ylabel={\small{$y$ (\si{\km})}},
  zlabel={\small{depth (\si{\km})}}, 
  ztick pos=left,
  3d box,width=\modelwidth, 
  xmin=\xmin,ymin=\ymin,zmin=\zmin,xmax=\xmax,ymax=\ymax,zmax=\zmax,
  every axis x label/.style={at={(0.35, 0.00)},anchor=north},
  every axis y label/.style={at={(0.90,-0.03)},anchor=north},
  line width=.25pt]
  \addplot3[fill=white] graphics[points={
            (\xmin,\ymin,\zmax)  => (244,232-232)
            (\xmax,\ymin,\zmax)  => (0. ,232-150)
            (\xmin,\ymax,\zmax)  => (410,232-193)
            (\xmax,\ymax,\zmin)  => (165,232-0. )}
            ]{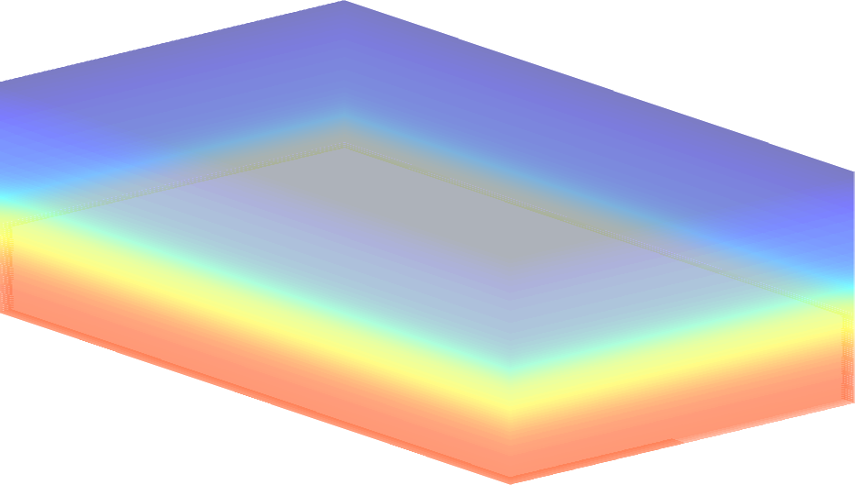};
  
  \addplot3[dashed,black,line width=1pt] 
      coordinates {(\xmin,\ycut,\zmin) (\xmin,\ycut,\zmax)
                   (\xmax,\ycut,\zmax) (\xmax,\ycut,\zmin) (\xmin,\ycut,\zmin)};
  \addplot3[dashed,black,line width=1pt] 
      coordinates {(\xmin,\ymin,\zcut) (\xmin,\ymax,\zcut) 
                   (\xmax,\ymax,\zcut) (\xmax,\ymin,\zcut) (\xmin,\ymin,\zcut)};
\end{axis}

&

\begin{axis}[%
yshift=-1mm,   
anchor=center, width=\modelwidth, height=\modelheight,
xmin=\xmin, xmax=\xmax,ymin=\zmin, ymax=\zmax,hide axis,
tick label style={font=\small},y dir=reverse,
colormap/jet, colorbar,
colorbar style={title={\small{\si{\km\per\second}}},
                tick label style={font=\small},
                xshift=-.0cm,
                },
                point meta min=\vmin, point meta max=\vmax]
\addplot [forget plot] graphics 
         [xmin=\xmin,xmax=\xmax,ymin=\zmin,ymax=\zmax] {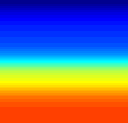};

\addplot[dashed,black,line width=1pt] 
      coordinates {(\xmin+.01,\zmin+.01) (\xmin+.01,\zmax-.01) 
                   (\xmax-.01,\zmax-.01) (\xmax-.01,\zmin+.01) (\xmin+.01,\zmin+.01)};
\end{axis}
\\};

\draw[black,-stealth,->,dashed,line width=1pt] (-0.75,0) to[out=0,in=-180] (0.50,0);

\end{tikzpicture}

  \caption{Starting model for the reconstruction of 
           the subsurface velocity of 
           \cref{fig:statoil:true}. It consists
           in a one-dimensional profile of increasing 
           speed with depth, with values from \num{1500}
           to about \num{4000} \si{\meter\per\second}.
           The horizontal section is for $y = 1125$ (right).}
  \label{fig:statoil:start}
\end{figure}

For the iterative reconstruction, we start
with a one-dimensional velocity profile 
that is pictured in \cref{fig:statoil:start}.
This initial guess does not encode any 
a priori information, and only has an increasing 
velocity with the depth.
Moreover, the range of values is lower compared
to the true medium of \cref{fig:statoil:true}.

\subsection{Time-domain data with noise}

We work with time-harmonic wave propagation but
follow a seismic context where measurements are
obtained in the time-domain. Furthermore, we 
incorporate noise in the synthetic seismograms to 
have a more realistic setup. 
The observational acquisition consists of 160 
point-sources, excited one by one.
They are positioned at the depth $z = 10$ \si{\meter},
and forms regular lattice such that sources are
every $160$ \si{\meter} along the $x$ axis, and 
every $150$ \si{\meter} along the $y$ axis.
There is a total of \num{1376} receivers, which
are located at a fixed depth of $100$ \si{\meter}, 
every $60$ \si{\meter} along the $x$ axis, 
and every $50$ \si{\meter} along the $y$ axis.
Note that the receivers remain at the same position for 
all sources, cf. \cref{fig:multi-source-splitting}.

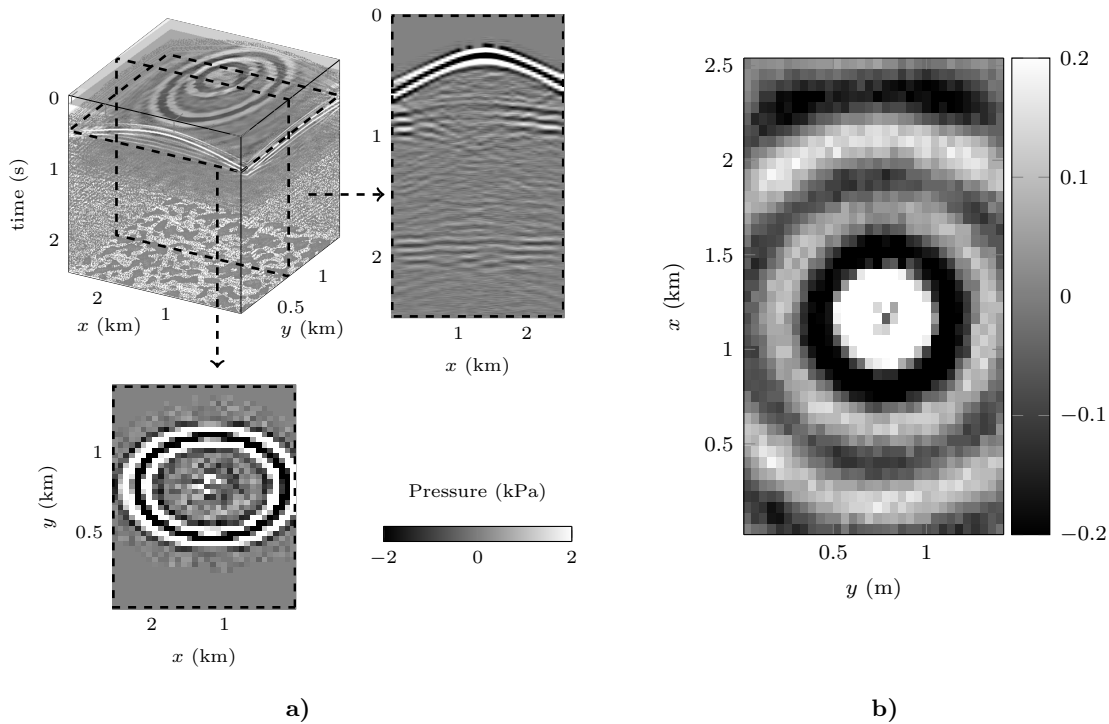
\begin{figure}[ht!] \centering
\subfloat[][]{\includegraphics[scale=1]{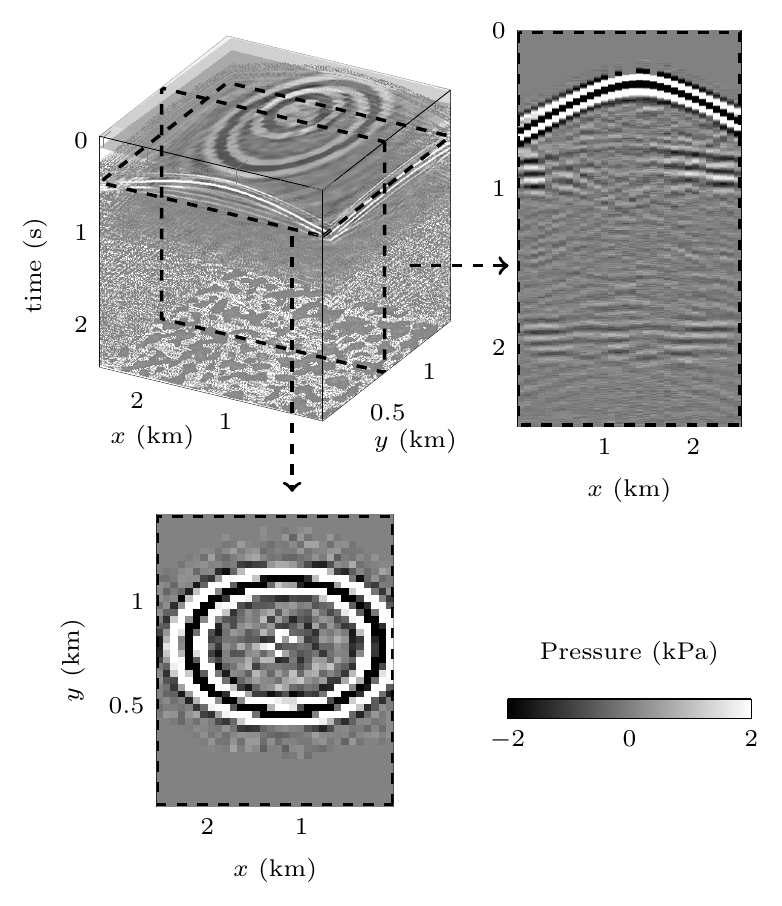}
              \label{fig:statoil:data_A}} \hspace*{0cm}
\renewcommand{\modelfile}{fourier-transform_pressure-noise_5hz_-1_1e2} 
\setlength{\modelwidth} {5.00cm}
\setlength{\modelheight}{7.90cm}
\subfloat[][]{\raisebox{1cm}{{
\begin{tikzpicture}

\pgfmathsetmacro{\zmin} {0.020}
\pgfmathsetmacro{\zmax} {2.540}
\pgfmathsetmacro{\xmin} {0.020}
\pgfmathsetmacro{\xmax} {1.415}
\pgfmathsetmacro{\cmin} {-0.2}
\pgfmathsetmacro{\cmax} {0.2}

\begin{axis}[%
width=\modelwidth,
height=\modelheight,
every node/.append style={font=\scriptsize},
        tick label style={font=\scriptsize},
             label style={font=\scriptsize},
axis on top, separate axis lines,
xmin=\xmin, xmax=\xmax, xlabel={\scriptsize $y$ (\si{\m})},
ymin=\zmin, ymax=\zmax, ylabel={\scriptsize $x$ (\si{\km})}, 
colormap/jet,colorbar,
colormap={blackwhite}{gray(0cm)=(0);gray(1cm)=(1)},
colorbar style={ 
xshift=-.2cm},point meta min=\cmin,point meta max=\cmax
]
\addplot [forget plot] graphics [xmin=\xmin,xmax=\xmax,ymin=\zmin,ymax=\zmax] {{\modelfile}.png};
\end{axis}
\end{tikzpicture}%
}}
              \label{fig:statoil:data_B}}
\caption{\protect\subref{fig:statoil:data_A} 
         Seismic traces in time including noise and
         \protect\subref{fig:statoil:data_B}
         corresponding real part Fourier transform at
         $5$ \si{\Hz} frequency.
         These are associated with a single-point source located in 
         $(x_{s},\, y_{s},\, z_{s}) = (\num{1160},\, 780,\, 10)~\si{\meter}$,
         for the velocity model of \cref{fig:statoil:true}.
         The measurements are obtained from \num{1376} receivers
         positioned at depth $z=100$ \si{\meter}.
        }
\label{fig:statoil:data}
\end{figure}

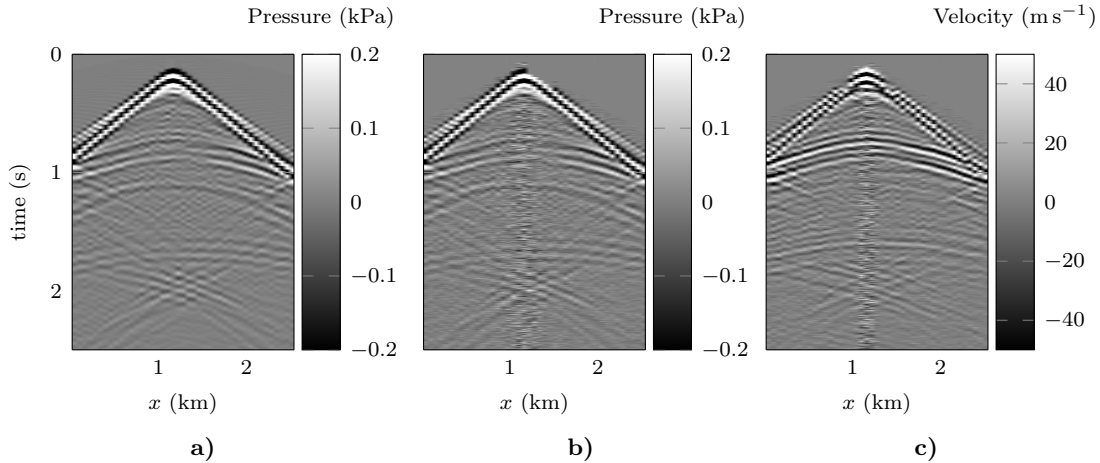
\begin{figure}[ht!] \centering
\setlength{\modelwidth} {4.50cm}
\setlength{\modelheight}{5.50cm}
  \renewcommand{\titlebar}{Pressure (\si{\kilo\pascal})}
  \pgfmathsetmacro{\xmin} {0.02} \pgfmathsetmacro{\xmax}{2.54}
  \pgfmathsetmacro{\cmin} {-0.2} \pgfmathsetmacro{\cmax} {0.2}
  \pgfmathsetmacro{\zmax} {2.5}
  \renewcommand{\modelfile}{trace-rcvY16_pressure_-2_2e3}
  \subfloat[][]{\begin{tikzpicture}

\pgfmathsetmacro{\zmin} {0.000}
\pgfmathsetmacro{\zmaxall} {4.000}

\begin{axis}[%
width=\modelwidth,
height=\modelheight,
every node/.append style={font=\scriptsize},
        tick label style={font=\scriptsize},
             label style={font=\scriptsize},
axis on top, separate axis lines,
xmin=\xmin, xmax=\xmax, xlabel={$x$  (\si{\kilo\meter})},
ymin=\zmin, ymax=\zmax, ylabel={time (\si{\second})}, y dir=reverse,
colormap/jet,colorbar,
colormap={blackwhite}{gray(0cm)=(0);gray(1cm)=(1)},
colorbar style={title={{\scriptsize{\titlebar}}},
xshift=-.2cm},point meta min=\cmin,point meta max=\cmax
]
\addplot [forget plot] graphics [xmin=\xmin,xmax=\xmax,ymin=\zmin,ymax=\zmaxall] {{\modelfile}.png};
\end{axis}
\end{tikzpicture}%
                \label{fig:statoil:data-2d_A}}       \hspace*{-.3cm}  
  \renewcommand{\modelfile}{trace-rcvY16_pressure-noise_-2_2e3}
  \subfloat[][]{\begin{tikzpicture}

\pgfmathsetmacro{\zmin}    {0.000}
\pgfmathsetmacro{\zmaxall} {4.000}

\begin{axis}[%
width=\modelwidth,
height=\modelheight,
every node/.append style={font=\scriptsize},
        tick label style={font=\scriptsize},
             label style={font=\scriptsize},
axis on top, separate axis lines,
xmin=\xmin, xmax=\xmax, xlabel={$x$  (\si{\kilo\meter})},
ymin=\zmin, ymax=\zmax, 
y dir=reverse, yticklabels={,,},
colormap/jet,colorbar,
colormap={blackwhite}{gray(0cm)=(0);gray(1cm)=(1)},
colorbar style={title={{\scriptsize{\titlebar}}},
xshift=-.2cm},point meta min=\cmin,point meta max=\cmax
]
\addplot [forget plot] graphics [xmin=\xmin,xmax=\xmax,ymin=\zmin,ymax=\zmaxall] {{\modelfile}.png};
\end{axis}
\end{tikzpicture}%
                \label{fig:statoil:data-2d_B}} \hspace*{-.8cm}  
  \pgfmathsetmacro{\cmin} {-50} \pgfmathsetmacro{\cmax} {50}
  \renewcommand{\modelfile}{trace-rcvY16_dpressure-noise_-5_5e1}
  \renewcommand{\titlebar}{Velocity (\si{\meter\per\second})}
  \subfloat[][]{\begin{tikzpicture}

\pgfmathsetmacro{\zmin}    {0.000}
\pgfmathsetmacro{\zmaxall} {4.000}

\begin{axis}[%
width=\modelwidth,
height=\modelheight,
every node/.append style={font=\scriptsize},
        tick label style={font=\scriptsize},
             label style={font=\scriptsize},
axis on top, separate axis lines,
xmin=\xmin, xmax=\xmax, xlabel={$x$  (\si{\kilo\meter})},
ymin=\zmin, ymax=\zmax, 
y dir=reverse, yticklabels={,,},
colormap/jet,colorbar,
colormap={blackwhite}{gray(0cm)=(0);gray(1cm)=(1)},
colorbar style={title={{\scriptsize{\titlebar}}},
xshift=-.2cm},point meta min=\cmin,point meta max=\cmax
]
\addplot [forget plot] graphics [xmin=\xmin,xmax=\xmax,ymin=\zmin,ymax=\zmaxall] {{\modelfile}.png};
\end{axis}
\end{tikzpicture}%
                \label{fig:statoil:data-2d_C}}
  \caption{Time-domain 
           \protect\subref{fig:statoil:data-2d_A} 
            pressure trace without noise, 
           \protect\subref{fig:statoil:data-2d_B} 
            pressure trace with 10\si{\deci\bel} signal-to-noise ratio and
           \protect\subref{fig:statoil:data-2d_C}
            normal velocity trace with 10\si{\deci\bel} signal-to-noise ratio.
            These correspond with a line of receivers at a fixed 
            $y = \num{695}$ \si{\meter}, for a single-point source 
            located in $(x_{s},\, y_{s},\, z_{s}) 
            = (\num{1160},\, 780,\, 10)~\si{\meter}$
            for the velocity model shown in \cref{fig:statoil:true}. 
            For our 
            experiment, we apply $10$ \si{\deci\bel} 
            signal-to-noise ratio to the synthetic data
            and employ the Fourier transform of the noisy
            traces. The available frequency ranges from 
            $5$ to $15$ \si{\Hz}, see \cref{fig:statoil:data}.}
  \label{fig:statoil:data-2d}
\end{figure}

We generate the time-domain 
seismograms\footnote{We have used the parallel time-domain code 
                     Hou10ni, see 
                     $\text{https://team.inria.fr/magique3d/software/hou10ni/}$;
                     it relies on Internal Penalty Discontinuous Galerkin 
                     discretization (while we use HDG).
                     Also, the meshes are different between the time-domain modeling
                     and the harmonic inversion.}
and incorporate noise in the resulting traces,
with a signal-to-noise ratio of 
\num{10} \si{\deci\bel}.
Then, we proceed with the discrete Fourier 
transform to feed \cref{algo:FWI}.
In \cref{fig:statoil:data}, 
we picture the noisy time-domain pressure
trace for a single-point source, and the corresponding 
Fourier-transform that we employ for the 
reconstruction.
We respect the seismic constraint that the
low-frequencies are not available from the 
time-domain data (because of noise and listening 
time) and we only work with data from $5$ to
$15$ \si{\Hz} frequency.
In \cref{fig:statoil:data-2d}, we picture
two-dimensional time-space sections of the 
traces for the pressure and normal velocity, 
and illustrate the effect of the added noise
(10 \si{\deci\bel} signal-to-noise ratio in our 
experiments).

\subsection{Performance comparison of the misfit functionals}

\begin{figure}[ht!] \centering
\setlength{\modelwidth} {7.25cm}
\setlength{\modelheight}{4.00cm}
  \renewcommand{\modelfile}{cp_p-v_15hz_gauss2}
\begin{tikzpicture}
\pgfmathsetmacro{\xmin} {0.}
\pgfmathsetmacro{\xmax} {2.540}
\pgfmathsetmacro{\ymin} {0.}
\pgfmathsetmacro{\ymax} {1.440}
\pgfmathsetmacro{\zmin} {0.}
\pgfmathsetmacro{\zmax} {1.220}

\pgfmathsetmacro{\ycut} {1.14}
\pgfmathsetmacro{\zcut} {0.8}
\pgfmathsetmacro{\vmin} {1.5}
\pgfmathsetmacro{\vmax} {5.2}

\matrix[column sep=7mm, row sep=0mm] {
\begin{axis}[
  anchor=center,
  tick label style={font=\small},
  grid=both,minor tick num=1,
  xlabel={\small{$x$ (\si{\km})}},ylabel={\small{$y$ (\si{\km})}},
  zlabel={\small{depth (\si{\km})}}, 
  ztick pos=left,
  3d box,width=\modelwidth, 
  xmin=\xmin,ymin=\ymin,zmin=\zmin,xmax=\xmax,ymax=\ymax,zmax=\zmax,
  every axis x label/.style={at={(0.35, 0.00)},anchor=north},
  every axis y label/.style={at={(0.95, 0.02)},anchor=north},
  line width=.25pt]
  \addplot3[fill=white] graphics[points={
            (\xmin,\ymin,\zmax)  => (244,232-232)
            (\xmax,\ymin,\zmax)  => (0. ,232-150)
            (\xmin,\ymax,\zmax)  => (410,232-193)
            (\xmax,\ymax,\zmin)  => (165,232-0. )}
            ]{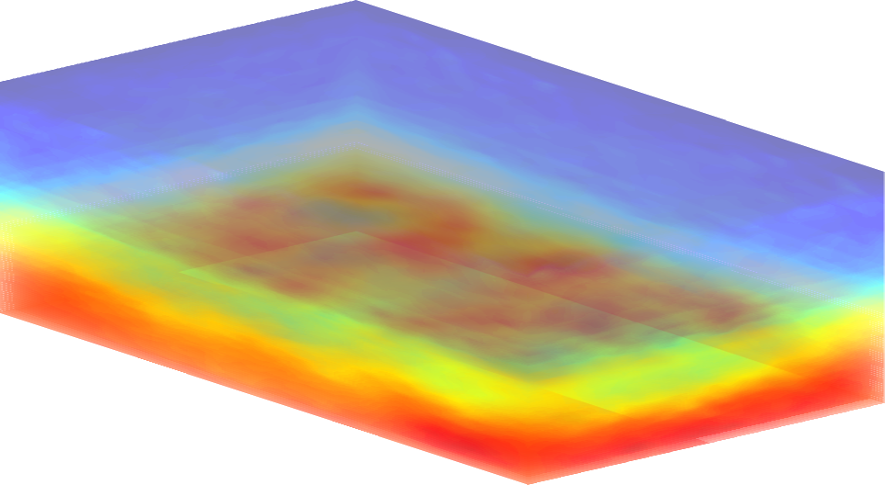};
  
  \addplot3[dashed,black,line width=1pt] 
      coordinates {(\xmin,\ycut,\zmin) (\xmin,\ycut,\zmax)
                   (\xmax,\ycut,\zmax) (\xmax,\ycut,\zmin) (\xmin,\ycut,\zmin)};
  \addplot3[dashed,black,line width=1pt] 
      coordinates {(\xmin,\ymin,\zcut) (\xmin,\ymax,\zcut) 
                   (\xmax,\ymax,\zcut) (\xmax,\ymin,\zcut) (\xmin,\ymin,\zcut)};
\end{axis}

&

\begin{axis}[%
anchor=center, width=\modelwidth, height=\modelheight,
xmin=\xmin, xmax=\xmax,ymin=\zmin, ymax=\zmax,hide axis,
tick label style={font=\small},y dir=reverse]
\addplot [forget plot] graphics 
         [xmin=\xmin,xmax=\xmax,ymin=\zmin,ymax=\zmax] {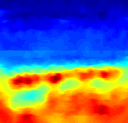};

\addplot[dashed,black,line width=1pt] 
      coordinates {(\xmin+.01,\zmin+.01) (\xmin+.01,\zmax-.01) 
                   (\xmax-.01,\zmax-.01) (\xmax-.01,\zmin+.01) (\xmin+.01,\zmin+.01)};
\end{axis}

\\

\begin{axis}[%
anchor=center, width=\modelwidth, height=\modelheight,
xmin=\xmin, xmax=\xmax,ymin=\ymin, ymax=\ymax,hide axis,
tick label style={font=\small}, y dir=reverse]
\addplot [forget plot] graphics 
         [xmin=\xmin,xmax=\xmax,ymin=\ymin,ymax=\ymax] {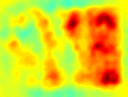};

\addplot[dashed,black,line width=1pt] 
      coordinates {(\xmin+.01,\ymin+.01) (\xmin+.01,\ymax-.01) 
                   (\xmax-.01,\ymax-.01) (\xmax-.01,\ymin+.01) (\xmin+.01,\ymin+.01)};
\end{axis}

&

\begin{axis}[anchor=south,yshift=.5cm,hide axis,height=.5cm,colorbar/width=.2cm,
     colormap/jet,colorbar horizontal,
     colorbar style={separate axis lines,
                     tick label style={font=\small},
                     title={\small{velocity (\si{\km\per\second})}}},
                     point meta min=\vmin, point meta max=\vmax,
                     height=.5\modelwidth]
\end{axis}
\\};

\draw[black,-stealth,->,dashed,line width=1pt] (-0.2,1.55)to[out=0,in=-180] ( 1.0 ,1.55);
\draw[black,-stealth,->,dashed,line width=1pt] (-4.5,1.0) to[out=-90,in=90] (-4.5,-0.5);

\end{tikzpicture}
  \caption{velocity reconstruction from the minimization of 
           $\misfitL$ in~\cref{eq:misfit_classic}, starting 
           with the model of \cref{fig:statoil:start}.
           The reconstruction uses $30$ iterations per frequency
           between $5$ and $15$ \si{\Hz}, with pressure and normal 
           velocity data.
           For visualization, we picture sections at a fixed 
           depth $z = 800$ \si{\meter} (bottom) and for 
           $y = 1125$ \si{\meter} (right).}
  \label{fig:statoil:fwi_15hz}
\end{figure}
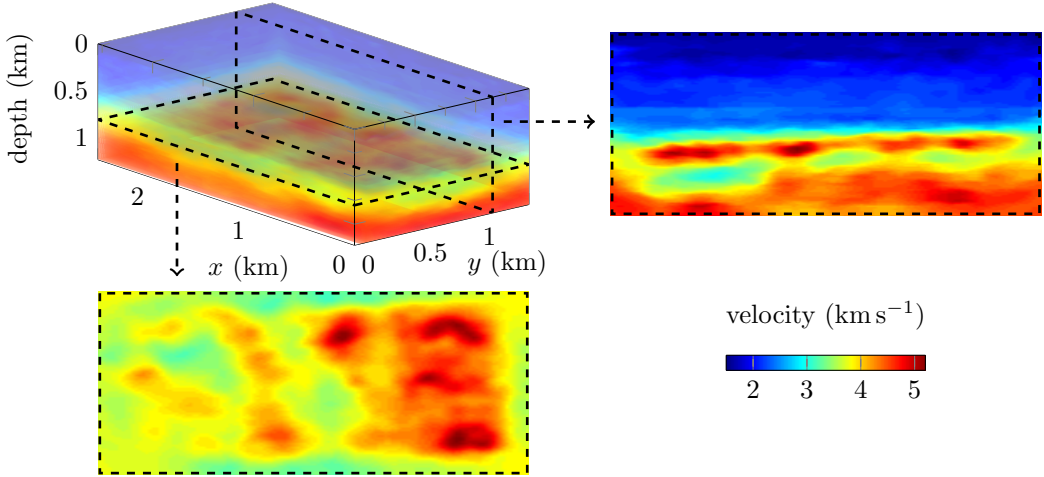

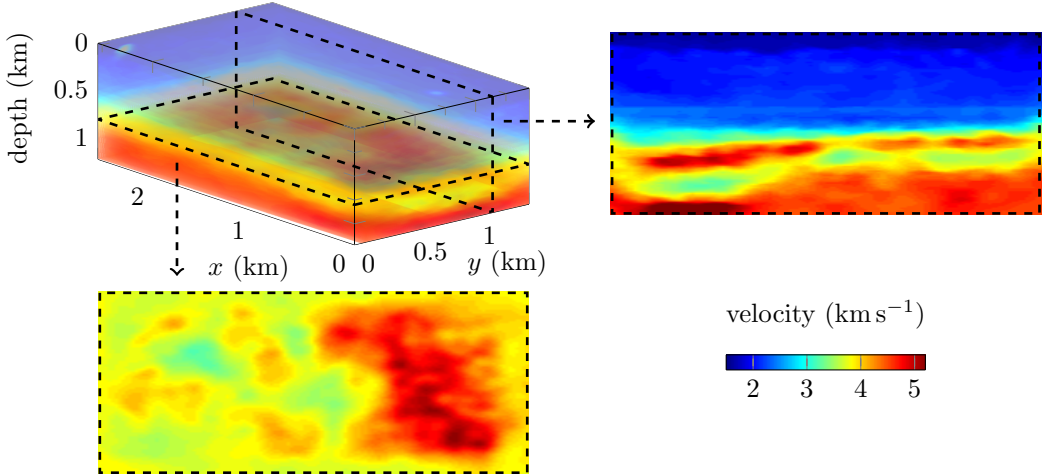
\begin{figure}[ht!] \centering
\setlength{\modelwidth} {7.25cm}
\setlength{\modelheight}{4.00cm}
  \renewcommand{\modelfile}{cp_reciprocity-same-src_nlcg_15hz_gauss2}
\begin{tikzpicture}
\pgfmathsetmacro{\xmin} {0.}
\pgfmathsetmacro{\xmax} {2.540}
\pgfmathsetmacro{\ymin} {0.}
\pgfmathsetmacro{\ymax} {1.440}
\pgfmathsetmacro{\zmin} {0.}
\pgfmathsetmacro{\zmax} {1.220}

\pgfmathsetmacro{\ycut} {1.14}
\pgfmathsetmacro{\zcut} {0.8}
\pgfmathsetmacro{\vmin} {1.5}
\pgfmathsetmacro{\vmax} {5.2}

\matrix[column sep=7mm, row sep=0mm] {
\begin{axis}[
  anchor=center,
  tick label style={font=\small},
  grid=both,minor tick num=1,
  xlabel={\small{$x$ (\si{\km})}},ylabel={\small{$y$ (\si{\km})}},
  zlabel={\small{depth (\si{\km})}}, 
  ztick pos=left,
  3d box,width=\modelwidth, 
  xmin=\xmin,ymin=\ymin,zmin=\zmin,xmax=\xmax,ymax=\ymax,zmax=\zmax,
  every axis x label/.style={at={(0.35, 0.00)},anchor=north},
  every axis y label/.style={at={(0.95, 0.02)},anchor=north},
  line width=.25pt]
  \addplot3[fill=white] graphics[points={
            (\xmin,\ymin,\zmax)  => (244,232-232)
            (\xmax,\ymin,\zmax)  => (0. ,232-150)
            (\xmin,\ymax,\zmax)  => (410,232-193)
            (\xmax,\ymax,\zmin)  => (165,232-0. )}
            ]{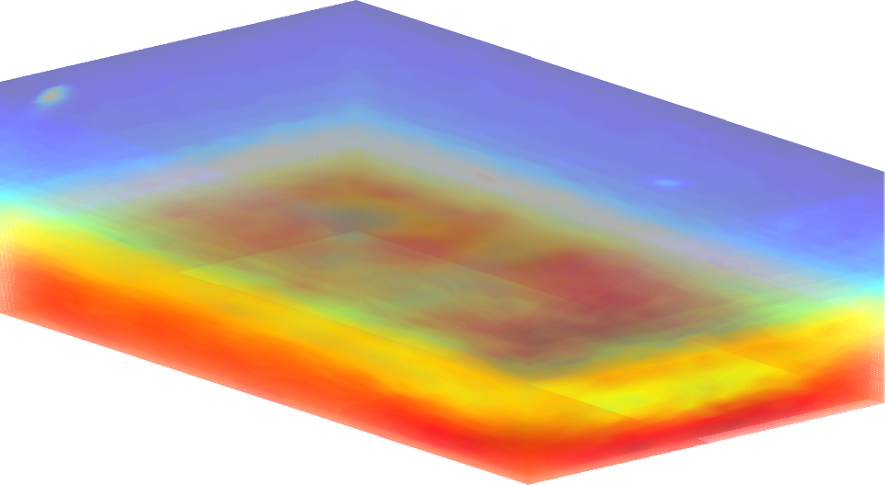};
  
  \addplot3[dashed,black,line width=1pt] 
      coordinates {(\xmin,\ycut,\zmin) (\xmin,\ycut,\zmax)
                   (\xmax,\ycut,\zmax) (\xmax,\ycut,\zmin) (\xmin,\ycut,\zmin)};
  \addplot3[dashed,black,line width=1pt] 
      coordinates {(\xmin,\ymin,\zcut) (\xmin,\ymax,\zcut) 
                   (\xmax,\ymax,\zcut) (\xmax,\ymin,\zcut) (\xmin,\ymin,\zcut)};
\end{axis}

&

\begin{axis}[%
anchor=center, width=\modelwidth, height=\modelheight,
xmin=\xmin, xmax=\xmax,ymin=\zmin, ymax=\zmax,hide axis,
tick label style={font=\small},y dir=reverse]
\addplot [forget plot] graphics 
         [xmin=\xmin,xmax=\xmax,ymin=\zmin,ymax=\zmax] {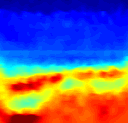};

\addplot[dashed,black,line width=1pt] 
      coordinates {(\xmin+.01,\zmin+.01) (\xmin+.01,\zmax-.01) 
                   (\xmax-.01,\zmax-.01) (\xmax-.01,\zmin+.01) (\xmin+.01,\zmin+.01)};
\end{axis}

\\

\begin{axis}[%
anchor=center, width=\modelwidth, height=\modelheight,
xmin=\xmin, xmax=\xmax,ymin=\ymin, ymax=\ymax,hide axis,
tick label style={font=\small}, y dir=reverse]
\addplot [forget plot] graphics 
         [xmin=\xmin,xmax=\xmax,ymin=\ymin,ymax=\ymax] {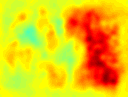};

\addplot[dashed,black,line width=1pt] 
      coordinates {(\xmin+.01,\ymin+.01) (\xmin+.01,\ymax-.01) 
                   (\xmax-.01,\ymax-.01) (\xmax-.01,\ymin+.01) (\xmin+.01,\ymin+.01)};
\end{axis}

&

\begin{axis}[anchor=south,yshift=.5cm,hide axis,height=.5cm,colorbar/width=.2cm,
     colormap/jet,colorbar horizontal,
     colorbar style={separate axis lines,
                     tick label style={font=\small},
                     title={\small{velocity (\si{\km\per\second})}}},
                     point meta min=\vmin, point meta max=\vmax,
                     height=.5\modelwidth]
\end{axis}
\\};

\draw[black,-stealth,->,dashed,line width=1pt] (-0.2,1.55)to[out=0,in=-180] ( 1.0 ,1.55);
\draw[black,-stealth,->,dashed,line width=1pt] (-4.5,1.0) to[out=-90,in=90] (-4.5,-0.5);

\end{tikzpicture}
  \caption{velocity reconstruction from the minimization of 
           $\misfitG$ in~\cref{eq:misfit_green}, starting 
           with the model of \cref{fig:statoil:start}.
           The reconstruction uses $30$ iterations per frequency
           between $5$ and $15$ \si{\Hz}, with pressure and normal 
           velocity data. Here, the numerical acquisition for 
           full reciprocity-gap waveform inversion follows the observational
           one (number of sources and position).
           For visualization, we picture sections at a fixed 
           depth $z = 800$ \si{\meter} (bottom) and for 
           $y = 1125$ \si{\meter} (right).}
  \label{fig:statoil:rwi_15hz}
\end{figure}

We first compare the performance of both misfit functionals,
$\misfitL$ and $\misfitG$, in the \emph{same context}: the 
numerical acquisition is taken to be the same as the 
observational one (which is anyway mandatory for $\misfitL$).
Therefore, we take $\nsrcSim = \nsrcObs$
and the same set for $f$ and $g$ in~\cref{eq:misfit_green}.
We use \cref{algo:FWI}, using sequential frequency
progression from $5$ to $15$ \si{\Hz}, more precisely, we use
$\{5,\, 6,\, 7,\, 8,\, 9,\, 10,\, 12,\, 15 \}$ \si{\Hz} data,
and \num{30} minimization iterations per frequency.
In \cref{fig:statoil:fwi_15hz,fig:statoil:rwi_15hz},
we show the final reconstruction, after $15$ \si{\Hz} iterations
when minimizing $\misfitL$ and $\misfitG$ respectively.
For the discretization, we employ a mesh of about $75$ thousands
tetrahedra, and polynomials of order three to five (depending on
the frequency) for accuracy (note the mesh differs from the one 
employed to generate the time-domain data).

\begin{remark} \label{rk:source-reconstruction}
  For the minimization of $\misfitL$~\cref{eq:misfit_classic}, 
  one has to use the \emph{same} source wavelet for the 
  observations and simulations. 
  Because the observational source wavelet is not precisely known,
  we need to reconstruct the source during the iterative process
  as well. 
  We employ the update formula given by~\cite{Pratt1999a} for 
  the iterative point-source reconstruction. 
  This can however induce additional difficulty in the case where 
  the source characteristic is not well recovered or, more important,
  when the source positions are not precisely known.
\end{remark}

In this experiment, we use the same acquisition context for 
the two choices of misfit functional, to observe intrinsic 
differences between the two methods. We observe the following:
\begin{itemize}
  \item both approaches provide a good reconstruction, where
        the layers of velocity  appear (see the vertical 
        section), and  the correct speed values are retrieved.
  \item The full reciprocity-gap waveform inversion provides slightly 
        better results, in particular for the parts that are near
        the boundaries (see the vertical and horizontal 
        sections of \cref{fig:statoil:fwi_15hz,fig:statoil:rwi_15hz}).
        This can be explained as FRgWI
        formulation~\cref{eq:misfit_green} tests \emph{every}
        simulation source with \emph{each} observational one
        (the two sums in~\cref{eq:misfit_green}) and 
        it somehow compensates for the limited illumination of 
        the boundary zones.
  \item Regarding the computational time, 
        both approaches are similar (the only difference
        is the two sums in the misfit 
        functional~\cref{eq:misfit_green}, which is 
        a computationally cheap operation compared to 
        the matrix factorization and linear system resolution).
\end{itemize}

Therefore, in this first test-case, we have demonstrated 
the efficiency of FRgWI, which produces better 
reconstruction compared to the traditional approach, without
incurring any increase of computational time. 
But more important, the reciprocity-gap offers flexibility for the 
choice of computational acquisition, which we now study.

\subsection{Comparison of multiple-point sources FRgWI with shot-stacking}

The main feature of FRgWI is to enable the
use of arbitrary probing sources for the numerical 
simulations ($g$ in~\cref{eq:misfit_green}). 
Here we investigate the use of multiple-point sources 
in order to reduce the computational cost. 
The observational acquisition is composed of 
$160$ \emph{point-sources}, i.e., 
each source function $f_i(\bx)$ in~\cref{eq:misfit_green} 
and~\cref{eq:misfit_classic} corresponds 
with a delta-Dirac function in $\bx_i^{(f)}$, 
according to~\cref{eq:single_ploint_source}.
For the computational acquisition ($g$ in~\cref{eq:misfit_green}),
we now consider $\nstack$ \emph{multiple-point sources}, 
each composed of $\npt$ points, cf.~\cref{eq:multi_point_source}.

We assume that the multiple-point sources all have the same 
number of points. 
The positions of the multiple-points are taken to coincide 
with the observational acquisition, and we consider a group
of sources in a structured partition of the original 
configuration, as illustrated in 
\cref{fig:multi-source-splitting}. We have
\begin{equation} \label{eq:multi_point_source_def}
  \bx_{j,\bullet}^{(g)} = \big\{ \bx_i^{(f)} \big\}_{i=i_1}^{i_2} \, , \qquad
  \text{ with } i_1 = (j-1) \npt + 1
  \quad \text{ and } \quad i_2 = i_1 + \npt - 1,
\end{equation}
where $\bx_j$ corresponds with the multiple-point source,
composed of the point-sources located in $\bx_i$.
The FWI counterpart is obtained using shot-stacking, 
following~\cref{eq:misfit_stack}, i.e., it requires
the summation of the observed data. 

\begin{figure}[ht!] \centering
\setlength{\modelwidth} {7.25cm}
\setlength{\modelheight}{4.00cm}
  \renewcommand{\modelfile}{cp_p-v-stack-5src_lbfgs_15hz_gauss2}
\begin{tikzpicture}
\pgfmathsetmacro{\xmin} {0.}
\pgfmathsetmacro{\xmax} {2.540}
\pgfmathsetmacro{\ymin} {0.}
\pgfmathsetmacro{\ymax} {1.440}
\pgfmathsetmacro{\zmin} {0.}
\pgfmathsetmacro{\zmax} {1.220}

\pgfmathsetmacro{\ycut} {1.14}
\pgfmathsetmacro{\zcut} {0.8}
\pgfmathsetmacro{\vmin} {1.5}
\pgfmathsetmacro{\vmax} {5.2}

\matrix[column sep=7mm, row sep=0mm] {
\begin{axis}[
  anchor=center,
  tick label style={font=\small},
  grid=both,minor tick num=1,
  xlabel={\small{$x$ (\si{\km})}},ylabel={\small{$y$ (\si{\km})}},
  zlabel={\small{depth (\si{\km})}}, 
  ztick pos=left,
  3d box,width=\modelwidth, 
  xmin=\xmin,ymin=\ymin,zmin=\zmin,xmax=\xmax,ymax=\ymax,zmax=\zmax,
  every axis x label/.style={at={(0.35, 0.00)},anchor=north},
  every axis y label/.style={at={(0.95, 0.02)},anchor=north},
  line width=.25pt]
  \addplot3[fill=white] graphics[points={
            (\xmin,\ymin,\zmax)  => (244,232-232)
            (\xmax,\ymin,\zmax)  => (0. ,232-150)
            (\xmin,\ymax,\zmax)  => (410,232-193)
            (\xmax,\ymax,\zmin)  => (165,232-0. )}
            ]{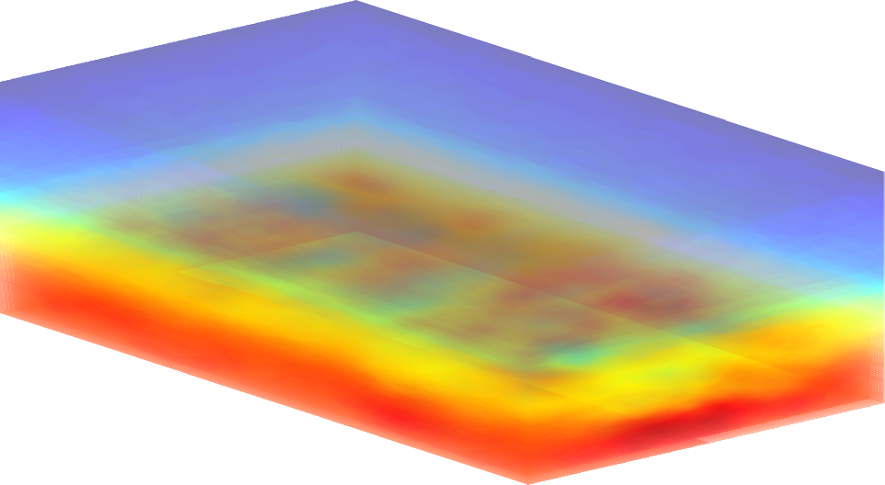};
  
  \addplot3[dashed,black,line width=1pt] 
      coordinates {(\xmin,\ycut,\zmin) (\xmin,\ycut,\zmax)
                   (\xmax,\ycut,\zmax) (\xmax,\ycut,\zmin) (\xmin,\ycut,\zmin)};
  \addplot3[dashed,black,line width=1pt] 
      coordinates {(\xmin,\ymin,\zcut) (\xmin,\ymax,\zcut) 
                   (\xmax,\ymax,\zcut) (\xmax,\ymin,\zcut) (\xmin,\ymin,\zcut)};
\end{axis}

&

\begin{axis}[%
anchor=center, width=\modelwidth, height=\modelheight,
xmin=\xmin, xmax=\xmax,ymin=\zmin, ymax=\zmax,hide axis,
tick label style={font=\small},y dir=reverse]
\addplot [forget plot] graphics 
         [xmin=\xmin,xmax=\xmax,ymin=\zmin,ymax=\zmax] {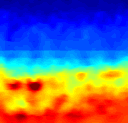};

\addplot[dashed,black,line width=1pt] 
      coordinates {(\xmin+.01,\zmin+.01) (\xmin+.01,\zmax-.01) 
                   (\xmax-.01,\zmax-.01) (\xmax-.01,\zmin+.01) (\xmin+.01,\zmin+.01)};
\end{axis}

\\

\begin{axis}[%
anchor=center, width=\modelwidth, height=\modelheight,
xmin=\xmin, xmax=\xmax,ymin=\ymin, ymax=\ymax,hide axis,
tick label style={font=\small}, y dir=reverse]
\addplot [forget plot] graphics 
         [xmin=\xmin,xmax=\xmax,ymin=\ymin,ymax=\ymax] {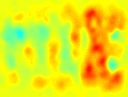};

\addplot[dashed,black,line width=1pt] 
      coordinates {(\xmin+.01,\ymin+.01) (\xmin+.01,\ymax-.01) 
                   (\xmax-.01,\ymax-.01) (\xmax-.01,\ymin+.01) (\xmin+.01,\ymin+.01)};
\end{axis}

&

\begin{axis}[anchor=south,yshift=.5cm,hide axis,height=.5cm,colorbar/width=.2cm,
     colormap/jet,colorbar horizontal,
     colorbar style={separate axis lines,
                     tick label style={font=\small},
                     title={\small{velocity (\si{\km\per\second})}}},
                     point meta min=\vmin, point meta max=\vmax,
                     height=.5\modelwidth]
\end{axis}
\\};

\draw[black,-stealth,->,dashed,line width=1pt] (-0.2,1.55)to[out=0,in=-180] ( 1.0 ,1.55);
\draw[black,-stealth,->,dashed,line width=1pt] (-4.5,1.0) to[out=-90,in=90] (-4.5,-0.5);

\end{tikzpicture}
  \caption{velocity reconstruction from the minimization of 
           $\misfitL^{\text{stack}}$ in~\cref{eq:misfit_stack}, 
           starting with the model of \cref{fig:statoil:start}.
           It uses $\nstack = 5$ multiple-point sources composed of $\npt = 32$, 
           cf.~\cref{eq:misfit_stack}.
           The reconstruction uses $30$ iterations per frequency
           between $5$ and $15$ \si{\Hz}. 
           For visualization, we picture sections at a fixed 
           depth $z = 800$ \si{\meter} (bottom) and for 
           $y = 1125$ \si{\meter} (right).}
  \label{fig:statoil:fwi_5stack_15hz}
\end{figure}

\begin{figure}[ht!] \centering
\setlength{\modelwidth} {7.25cm}
\setlength{\modelheight}{4.00cm}
  \renewcommand{\modelfile}{cp_reciprocity-5-src_nlcg_15hz_gauss2}
\begin{tikzpicture}
\pgfmathsetmacro{\xmin} {0.}
\pgfmathsetmacro{\xmax} {2.540}
\pgfmathsetmacro{\ymin} {0.}
\pgfmathsetmacro{\ymax} {1.440}
\pgfmathsetmacro{\zmin} {0.}
\pgfmathsetmacro{\zmax} {1.220}

\pgfmathsetmacro{\ycut} {1.14}
\pgfmathsetmacro{\zcut} {0.8}
\pgfmathsetmacro{\vmin} {1.5}
\pgfmathsetmacro{\vmax} {5.2}

\matrix[column sep=7mm, row sep=0mm] {
\begin{axis}[
  anchor=center,
  tick label style={font=\small},
  grid=both,minor tick num=1,
  xlabel={\small{$x$ (\si{\km})}},ylabel={\small{$y$ (\si{\km})}},
  zlabel={\small{depth (\si{\km})}}, 
  ztick pos=left,
  3d box,width=\modelwidth, 
  xmin=\xmin,ymin=\ymin,zmin=\zmin,xmax=\xmax,ymax=\ymax,zmax=\zmax,
  every axis x label/.style={at={(0.35, 0.00)},anchor=north},
  every axis y label/.style={at={(0.95, 0.02)},anchor=north},
  line width=.25pt]
  \addplot3[fill=white] graphics[points={
            (\xmin,\ymin,\zmax)  => (244,232-232)
            (\xmax,\ymin,\zmax)  => (0. ,232-150)
            (\xmin,\ymax,\zmax)  => (410,232-193)
            (\xmax,\ymax,\zmin)  => (165,232-0. )}
            ]{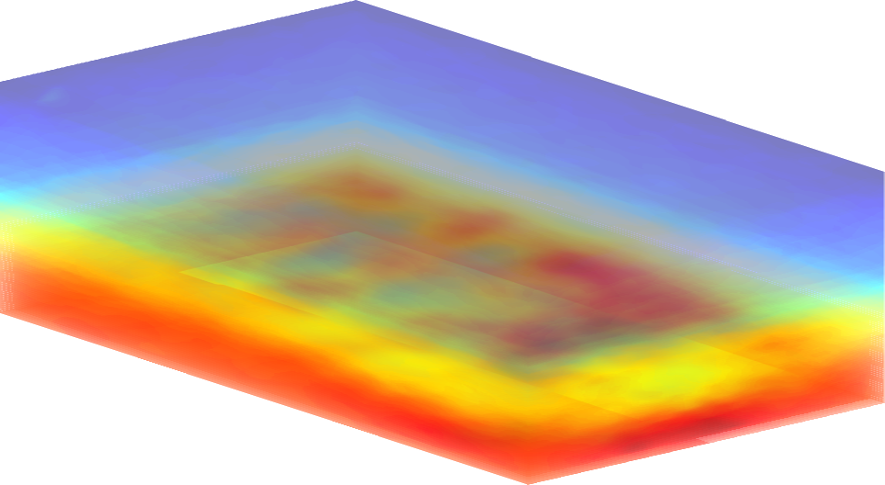};
  
  \addplot3[dashed,black,line width=1pt] 
      coordinates {(\xmin,\ycut,\zmin) (\xmin,\ycut,\zmax)
                   (\xmax,\ycut,\zmax) (\xmax,\ycut,\zmin) (\xmin,\ycut,\zmin)};
  \addplot3[dashed,black,line width=1pt] 
      coordinates {(\xmin,\ymin,\zcut) (\xmin,\ymax,\zcut) 
                   (\xmax,\ymax,\zcut) (\xmax,\ymin,\zcut) (\xmin,\ymin,\zcut)};
\end{axis}

&

\begin{axis}[%
anchor=center, width=\modelwidth, height=\modelheight,
xmin=\xmin, xmax=\xmax,ymin=\zmin, ymax=\zmax,hide axis,
tick label style={font=\small},y dir=reverse]
\addplot [forget plot] graphics 
         [xmin=\xmin,xmax=\xmax,ymin=\zmin,ymax=\zmax] {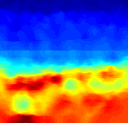};

\addplot[dashed,black,line width=1pt] 
      coordinates {(\xmin+.01,\zmin+.01) (\xmin+.01,\zmax-.01) 
                   (\xmax-.01,\zmax-.01) (\xmax-.01,\zmin+.01) (\xmin+.01,\zmin+.01)};
\end{axis}

\\

\begin{axis}[%
anchor=center, width=\modelwidth, height=\modelheight,
xmin=\xmin, xmax=\xmax,ymin=\ymin, ymax=\ymax,hide axis,
tick label style={font=\small}, y dir=reverse]
\addplot [forget plot] graphics 
         [xmin=\xmin,xmax=\xmax,ymin=\ymin,ymax=\ymax] {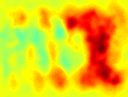};

\addplot[dashed,black,line width=1pt] 
      coordinates {(\xmin+.01,\ymin+.01) (\xmin+.01,\ymax-.01) 
                   (\xmax-.01,\ymax-.01) (\xmax-.01,\ymin+.01) (\xmin+.01,\ymin+.01)};
\end{axis}

&

\begin{axis}[anchor=south,yshift=.5cm,hide axis,height=.5cm,colorbar/width=.2cm,
     colormap/jet,colorbar horizontal,
     colorbar style={separate axis lines,
                     tick label style={font=\small},
                     title={\small{velocity (\si{\km\per\second})}}},
                     point meta min=\vmin, point meta max=\vmax,
                     height=.5\modelwidth]
\end{axis}
\\};

\draw[black,-stealth,->,dashed,line width=1pt] (-0.2,1.55)to[out=0,in=-180] ( 1.0 ,1.55);
\draw[black,-stealth,->,dashed,line width=1pt] (-4.5,1.0) to[out=-90,in=90] (-4.5,-0.5);

\end{tikzpicture}
  \caption{velocity reconstruction from the minimization 
           of $\misfitG$, starting with the model of 
           \cref{fig:statoil:start}.
           The numerical acquisition ($g$ in \cref{eq:misfit_green}) 
           uses $\nstack = 5$ sources composed of $\npt = 32$, 
           cf.~\cref{eq:multi_point_source}.
           The reconstruction uses $30$ iterations per frequency
           between $5$ and $15$ \si{\Hz}. 
           For visualization, we picture sections at a fixed 
           depth $z = 800$ \si{\meter} (bottom) and for 
           $y = 1125$ \si{\meter} (right).}
  \label{fig:statoil:rwi_5src_15hz}
\end{figure}

We perform the iterative reconstruction using 
$\nstack = 5$ multiple-point sources, each of them composed
of $\npt = 32$ points. The reconstruction using the shot-stacking
version of FWI~\cref{eq:misfit_stack} is shown
\cref{fig:statoil:fwi_5stack_15hz} and the result
using FRgWI (i.e., where only the computational
acquisition is changed) is shown 
\cref{fig:statoil:rwi_5src_15hz}.

We observe that FRgWI performs better than the shot-stacking in FWI,
\update{in this experiment of relatively small scale}:
\begin{itemize}
  \item the vertical and horizontal sections shown in 
        \cref{fig:statoil:rwi_5src_15hz} are actually 
        very close from the reconstruction obtained using 
        the same computational and observational acquisition, 
        see \cref{fig:statoil:rwi_15hz}. 
        The pattern of increasing and decreasing velocity 
        is well captured, with the appropriate values. 
  \item On the other hand, FWI with shot-stacking looses 
        accuracy, in particular the vertical section in
        \cref{fig:statoil:fwi_5stack_15hz} where 
        the non-monotone variation is not even recovered.
\end{itemize}
FRgWI, by testing each of the observational sources 
\emph{independently} with the arbitrary computational 
ones, is robust with multiple-point sources, 
and allows a sharp decrease in the number of numerical 
sources for the simulations.

\begin{remark}
  The results using the shot-stacking with FWI could 
  be improved by using more advanced strategy of shot blending 
  as mentioned in the introduction (e.g., \cite{Krebs2009,Li2012}).
  We illustrate here that even a naive approach (i.e. simple
  to implement numerically) of shot-stacking is sufficient for 
  the full reciprocity-gap waveform inversion to produce satisfactory
  results. 
  Also, note that we have tried to use a random selection of 
  positions for the multiple-point but it was not performing as
  well as the structured criterion. This would however require 
  more testing to confirm.
\end{remark}

\begin{figure}[ht!] \centering
\setlength{\modelwidth} {7.25cm}
\setlength{\modelheight}{4.00cm}
  \graphicspath{{figures/statoil/reconstruction_start1d-bkg/}}
  \renewcommand{\modelfile}{cp_reciprocity-2-src_nlcg_15hz_gauss2}
\begin{tikzpicture}
\pgfmathsetmacro{\xmin} {0.}
\pgfmathsetmacro{\xmax} {2.540}
\pgfmathsetmacro{\ymin} {0.}
\pgfmathsetmacro{\ymax} {1.440}
\pgfmathsetmacro{\zmin} {0.}
\pgfmathsetmacro{\zmax} {1.220}

\pgfmathsetmacro{\ycut} {1.14}
\pgfmathsetmacro{\zcut} {0.8}
\pgfmathsetmacro{\vmin} {1.5}
\pgfmathsetmacro{\vmax} {5.2}

\matrix[column sep=7mm, row sep=0mm] {
\begin{axis}[
  anchor=center,
  tick label style={font=\small},
  grid=both,minor tick num=1,
  xlabel={\small{$x$ (\si{\km})}},ylabel={\small{$y$ (\si{\km})}},
  zlabel={\small{depth (\si{\km})}}, 
  ztick pos=left,
  3d box,width=\modelwidth, 
  xmin=\xmin,ymin=\ymin,zmin=\zmin,xmax=\xmax,ymax=\ymax,zmax=\zmax,
  every axis x label/.style={at={(0.35, 0.00)},anchor=north},
  every axis y label/.style={at={(0.95, 0.02)},anchor=north},
  line width=.25pt]
  \addplot3[fill=white] graphics[points={
            (\xmin,\ymin,\zmax)  => (244,232-232)
            (\xmax,\ymin,\zmax)  => (0. ,232-150)
            (\xmin,\ymax,\zmax)  => (410,232-193)
            (\xmax,\ymax,\zmin)  => (165,232-0. )}
            ]{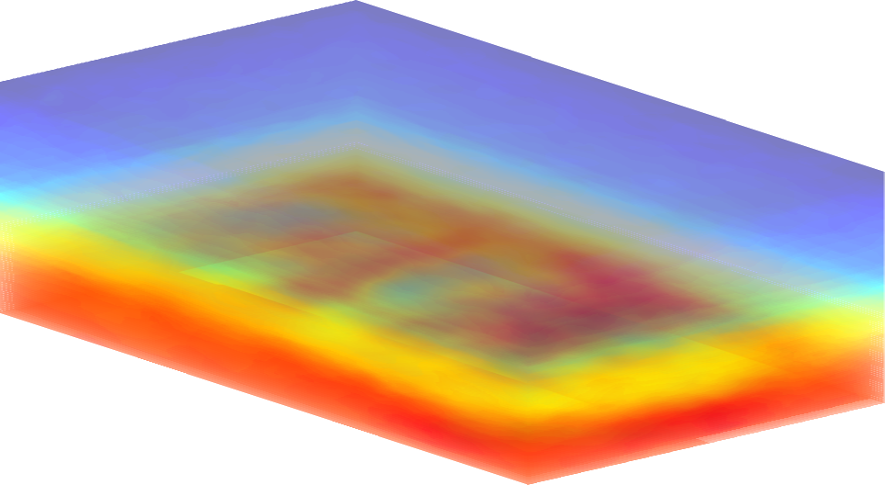};
  
  \addplot3[dashed,black,line width=1pt] 
      coordinates {(\xmin,\ycut,\zmin) (\xmin,\ycut,\zmax)
                   (\xmax,\ycut,\zmax) (\xmax,\ycut,\zmin) (\xmin,\ycut,\zmin)};
  \addplot3[dashed,black,line width=1pt] 
      coordinates {(\xmin,\ymin,\zcut) (\xmin,\ymax,\zcut) 
                   (\xmax,\ymax,\zcut) (\xmax,\ymin,\zcut) (\xmin,\ymin,\zcut)};
\end{axis}

&

\begin{axis}[%
anchor=center, width=\modelwidth, height=\modelheight,
xmin=\xmin, xmax=\xmax,ymin=\zmin, ymax=\zmax,hide axis,
tick label style={font=\small},y dir=reverse]
\addplot [forget plot] graphics 
         [xmin=\xmin,xmax=\xmax,ymin=\zmin,ymax=\zmax] {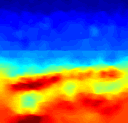};

\addplot[dashed,black,line width=1pt] 
      coordinates {(\xmin+.01,\zmin+.01) (\xmin+.01,\zmax-.01) 
                   (\xmax-.01,\zmax-.01) (\xmax-.01,\zmin+.01) (\xmin+.01,\zmin+.01)};
\end{axis}

\\

\begin{axis}[%
anchor=center, width=\modelwidth, height=\modelheight,
xmin=\xmin, xmax=\xmax,ymin=\ymin, ymax=\ymax,hide axis,
tick label style={font=\small}, y dir=reverse]
\addplot [forget plot] graphics 
         [xmin=\xmin,xmax=\xmax,ymin=\ymin,ymax=\ymax] {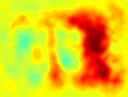};

\addplot[dashed,black,line width=1pt] 
      coordinates {(\xmin+.01,\ymin+.01) (\xmin+.01,\ymax-.01) 
                   (\xmax-.01,\ymax-.01) (\xmax-.01,\ymin+.01) (\xmin+.01,\ymin+.01)};
\end{axis}

&

\begin{axis}[anchor=south,yshift=.5cm,hide axis,height=.5cm,colorbar/width=.2cm,
     colormap/jet,colorbar horizontal,
     colorbar style={separate axis lines,
                     tick label style={font=\small},
                     title={\small{velocity (\si{\km\per\second})}}},
                     point meta min=\vmin, point meta max=\vmax,
                     height=.5\modelwidth]
\end{axis}
\\};

\draw[black,-stealth,->,dashed,line width=1pt] (-0.2,1.55)to[out=0,in=-180] ( 1.0 ,1.55);
\draw[black,-stealth,->,dashed,line width=1pt] (-4.5,1.0) to[out=-90,in=90] (-4.5,-0.5);

\end{tikzpicture}
  \caption{velocity reconstruction from the minimization 
           of $\misfitG$, starting with the model of 
           \cref{fig:statoil:start}.
           The numerical acquisition ($g$ in \cref{eq:misfit_green}) 
           uses $\nstack = 2$ sources composed of $\npt = 80$, 
           cf.~\cref{eq:multi_point_source}.
           The reconstruction uses $30$ iterations per frequency
           between $5$ and $15$ \si{\Hz}. 
           For visualization, we picture sections at a fixed 
           depth $z = 800$ \si{\meter} (bottom) and for 
           $y = 1125$ \si{\meter} (right).}
  \label{fig:statoil:rwi_2src_15hz}
\end{figure}
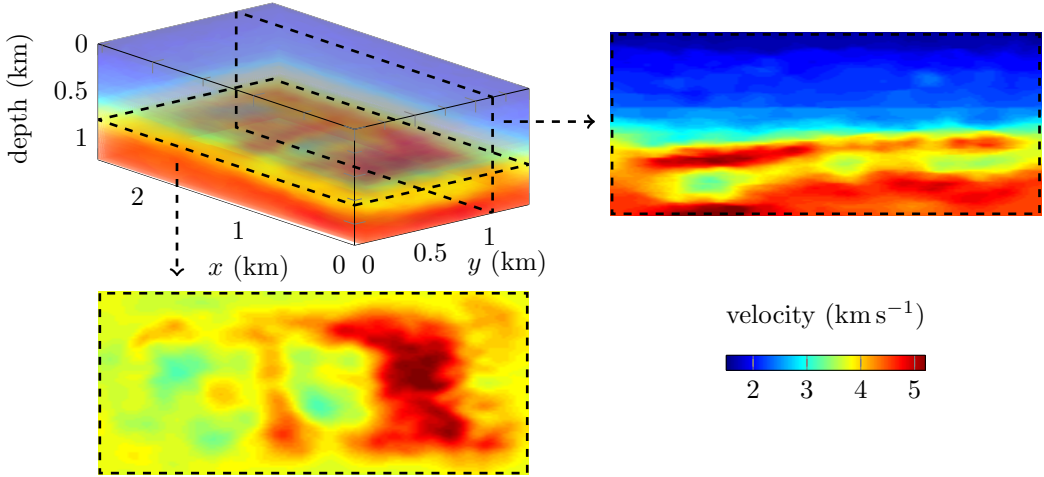

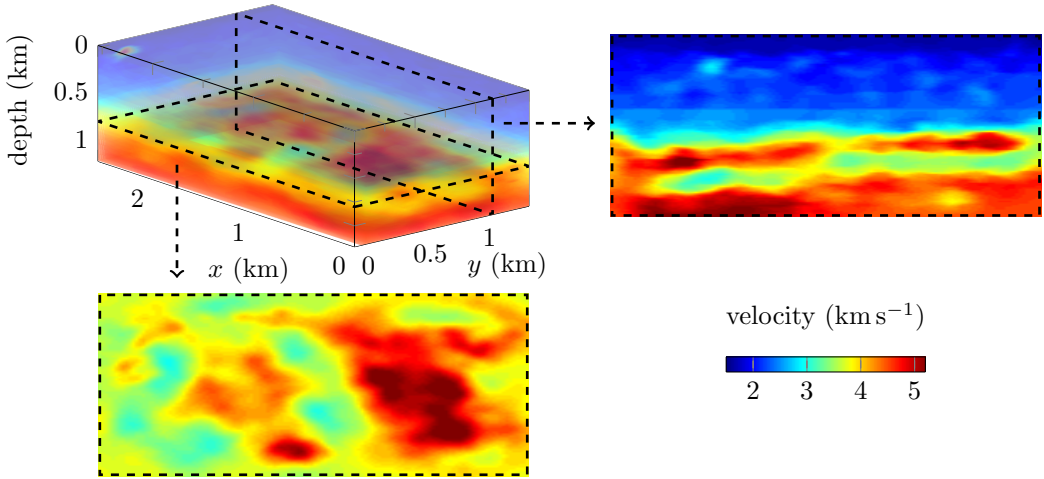
\begin{figure}[ht!] \centering
\setlength{\modelwidth} {7.25cm}
\setlength{\modelheight}{4.00cm}
  \renewcommand{\modelfile}{cp_reciprocity-1-src_nlcg_15hz_gauss2}
\begin{tikzpicture}
\pgfmathsetmacro{\xmin} {0.}
\pgfmathsetmacro{\xmax} {2.540}
\pgfmathsetmacro{\ymin} {0.}
\pgfmathsetmacro{\ymax} {1.440}
\pgfmathsetmacro{\zmin} {0.}
\pgfmathsetmacro{\zmax} {1.220}

\pgfmathsetmacro{\ycut} {1.14}
\pgfmathsetmacro{\zcut} {0.8}
\pgfmathsetmacro{\vmin} {1.5}
\pgfmathsetmacro{\vmax} {5.2}

\matrix[column sep=7mm, row sep=0mm] {
\begin{axis}[
  anchor=center,
  tick label style={font=\small},
  grid=both,minor tick num=1,
  xlabel={\small{$x$ (\si{\km})}},ylabel={\small{$y$ (\si{\km})}},
  zlabel={\small{depth (\si{\km})}}, 
  ztick pos=left,
  3d box,width=\modelwidth, 
  xmin=\xmin,ymin=\ymin,zmin=\zmin,xmax=\xmax,ymax=\ymax,zmax=\zmax,
  every axis x label/.style={at={(0.35, 0.00)},anchor=north},
  every axis y label/.style={at={(0.95, 0.02)},anchor=north},
  line width=.25pt]
  \addplot3[fill=white] graphics[points={
            (\xmin,\ymin,\zmax)  => (244,232-232)
            (\xmax,\ymin,\zmax)  => (0. ,232-150)
            (\xmin,\ymax,\zmax)  => (410,232-193)
            (\xmax,\ymax,\zmin)  => (165,232-0. )}
            ]{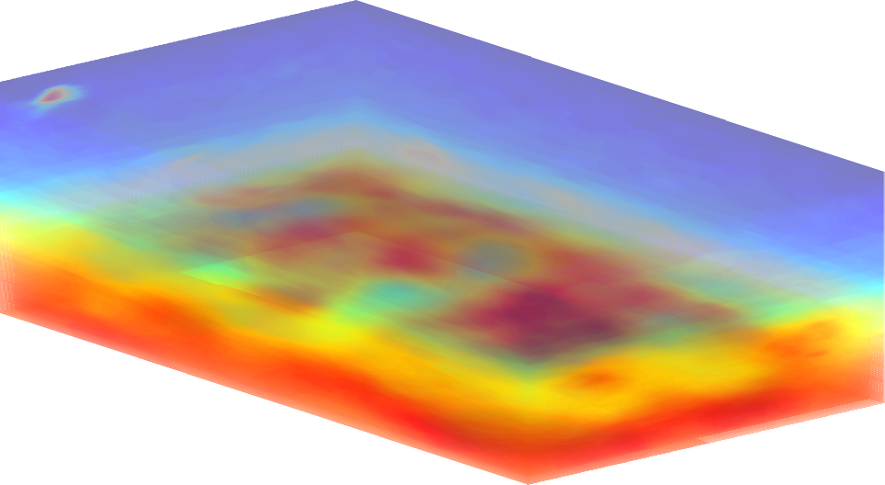};
  
  \addplot3[dashed,black,line width=1pt] 
      coordinates {(\xmin,\ycut,\zmin) (\xmin,\ycut,\zmax)
                   (\xmax,\ycut,\zmax) (\xmax,\ycut,\zmin) (\xmin,\ycut,\zmin)};
  \addplot3[dashed,black,line width=1pt] 
      coordinates {(\xmin,\ymin,\zcut) (\xmin,\ymax,\zcut) 
                   (\xmax,\ymax,\zcut) (\xmax,\ymin,\zcut) (\xmin,\ymin,\zcut)};
\end{axis}

&

\begin{axis}[%
anchor=center, width=\modelwidth, height=\modelheight,
xmin=\xmin, xmax=\xmax,ymin=\zmin, ymax=\zmax,hide axis,
tick label style={font=\small},y dir=reverse]
\addplot [forget plot] graphics 
         [xmin=\xmin,xmax=\xmax,ymin=\zmin,ymax=\zmax] {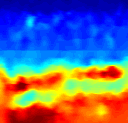};

\addplot[dashed,black,line width=1pt] 
      coordinates {(\xmin+.01,\zmin+.01) (\xmin+.01,\zmax-.01) 
                   (\xmax-.01,\zmax-.01) (\xmax-.01,\zmin+.01) (\xmin+.01,\zmin+.01)};
\end{axis}

\\

\begin{axis}[%
anchor=center, width=\modelwidth, height=\modelheight,
xmin=\xmin, xmax=\xmax,ymin=\ymin, ymax=\ymax,hide axis,
tick label style={font=\small}, y dir=reverse]
\addplot [forget plot] graphics 
         [xmin=\xmin,xmax=\xmax,ymin=\ymin,ymax=\ymax] {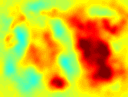};

\addplot[dashed,black,line width=1pt] 
      coordinates {(\xmin+.01,\ymin+.01) (\xmin+.01,\ymax-.01) 
                   (\xmax-.01,\ymax-.01) (\xmax-.01,\ymin+.01) (\xmin+.01,\ymin+.01)};
\end{axis}

&

\begin{axis}[anchor=south,yshift=.5cm,hide axis,height=.5cm,colorbar/width=.2cm,
     colormap/jet,colorbar horizontal,
     colorbar style={separate axis lines,
                     tick label style={font=\small},
                     title={\small{velocity (\si{\km\per\second})}}},
                     point meta min=\vmin, point meta max=\vmax,
                     height=.5\modelwidth]
\end{axis}
\\};

\draw[black,-stealth,->,dashed,line width=1pt] (-0.2,1.55)to[out=0,in=-180] ( 1.0 ,1.55);
\draw[black,-stealth,->,dashed,line width=1pt] (-4.5,1.0) to[out=-90,in=90] (-4.5,-0.5);

\end{tikzpicture}
  \caption{velocity reconstruction from the minimization 
           of $\misfitG$, starting with the model of 
           \cref{fig:statoil:start}.
           The numerical acquisition ($g$ in \cref{eq:misfit_green}) 
           uses $\nstack = 1$ source composed of $\npt = 160$, 
           cf.~\cref{eq:multi_point_source}.
           The reconstruction uses $30$ iterations per frequency
           between $5$ and $15$ \si{\Hz}. 
           For visualization, we picture sections at a fixed 
           depth $z = 800$ \si{\meter} (bottom) and for 
           $y = 1125$ \si{\meter} (right).}
  \label{fig:statoil:rwi_1src_15hz}
\end{figure}

We can further reduce the number of computational 
sources. The reconstruction using FRgWI with 
$\nstack = 2$, $\npt = 80$ is shown in
\cref{fig:statoil:rwi_2src_15hz}. 
In \cref{fig:statoil:rwi_1src_15hz},
we show the reconstruction using only one 
source: $\nstack = 1$, $\npt = 160$.
We see that FRgWI still behaves quite well,
in particular for two sources the reconstruction
carries similar accuracy than before: the vertical 
structures are accurately discovered, with 
pertinent variations of velocity. 
For the reconstruction using only one source, 
\cref{fig:statoil:rwi_1src_15hz}, we see
some deterioration in the recovered layers but
it remains more precise than the shot-stacking FWI
using $5$ sources (\cref{fig:statoil:fwi_5stack_15hz}).

The essence of FRgWI is that it \emph{does not}
modify the observational acquisition and instead
it probes every measured seismogram independently. 
It increases the robustness of the reconstruction
procedure, and allows
for the design of arbitrary computational acquisitions 
to reduce the computational cost. 
In this experiment, we have used a naive approach 
of shot summation, which requires no effort for its 
implementation, and it already shows accurate 
reconstructions.

\subsection{Sparse observational acquisition}

We now investigate a different situation: when the 
measurements are obtained from only a few set of 
multiple-point sources. 
This happens, for example, in marine seismic when 
several air guns are excited at the same time, i.e.
the source boat carries an array of air guns. 
It consequently reduces the amount of data obtained, 
and the cost of the field acquisition.

\begin{figure}[ht!] \centering
\setlength{\modelwidth} {7.25cm}
\setlength{\modelheight}{4.00cm}
  \renewcommand{\modelfile}{cp_aq5stack_reciprocity-160src_nlcg_10hz_gauss2}
\begin{tikzpicture}
\pgfmathsetmacro{\xmin} {0.}
\pgfmathsetmacro{\xmax} {2.540}
\pgfmathsetmacro{\ymin} {0.}
\pgfmathsetmacro{\ymax} {1.440}
\pgfmathsetmacro{\zmin} {0.}
\pgfmathsetmacro{\zmax} {1.220}

\pgfmathsetmacro{\ycut} {1.14}
\pgfmathsetmacro{\zcut} {0.8}
\pgfmathsetmacro{\vmin} {1.5}
\pgfmathsetmacro{\vmax} {5.2}

\matrix[column sep=7mm, row sep=0mm] {
\begin{axis}[
  anchor=center,
  tick label style={font=\small},
  grid=both,minor tick num=1,
  xlabel={\small{$x$ (\si{\km})}},ylabel={\small{$y$ (\si{\km})}},
  zlabel={\small{depth (\si{\km})}}, 
  ztick pos=left,
  3d box,width=\modelwidth, 
  xmin=\xmin,ymin=\ymin,zmin=\zmin,xmax=\xmax,ymax=\ymax,zmax=\zmax,
  every axis x label/.style={at={(0.35, 0.00)},anchor=north},
  every axis y label/.style={at={(0.95, 0.02)},anchor=north},
  line width=.25pt]
  \addplot3[fill=white] graphics[points={
            (\xmin,\ymin,\zmax)  => (244,232-232)
            (\xmax,\ymin,\zmax)  => (0. ,232-150)
            (\xmin,\ymax,\zmax)  => (410,232-193)
            (\xmax,\ymax,\zmin)  => (165,232-0. )}
            ]{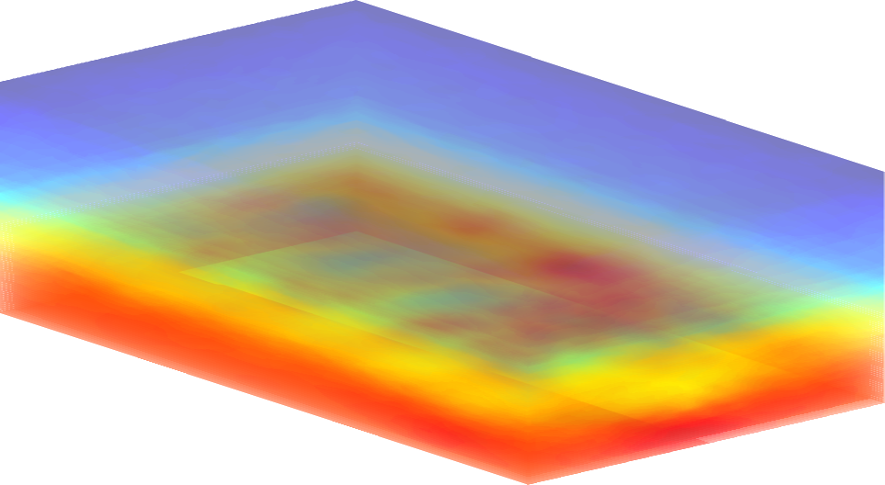};
  
  \addplot3[dashed,black,line width=1pt] 
      coordinates {(\xmin,\ycut,\zmin) (\xmin,\ycut,\zmax)
                   (\xmax,\ycut,\zmax) (\xmax,\ycut,\zmin) (\xmin,\ycut,\zmin)};
  \addplot3[dashed,black,line width=1pt] 
      coordinates {(\xmin,\ymin,\zcut) (\xmin,\ymax,\zcut) 
                   (\xmax,\ymax,\zcut) (\xmax,\ymin,\zcut) (\xmin,\ymin,\zcut)};
\end{axis}

&

\begin{axis}[%
anchor=center, width=\modelwidth, height=\modelheight,
xmin=\xmin, xmax=\xmax,ymin=\zmin, ymax=\zmax,hide axis,
tick label style={font=\small},y dir=reverse]
\addplot [forget plot] graphics 
         [xmin=\xmin,xmax=\xmax,ymin=\zmin,ymax=\zmax] {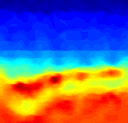};

\addplot[dashed,black,line width=1pt] 
      coordinates {(\xmin+.01,\zmin+.01) (\xmin+.01,\zmax-.01) 
                   (\xmax-.01,\zmax-.01) (\xmax-.01,\zmin+.01) (\xmin+.01,\zmin+.01)};
\end{axis}

\\

\begin{axis}[%
anchor=center, width=\modelwidth, height=\modelheight,
xmin=\xmin, xmax=\xmax,ymin=\ymin, ymax=\ymax,hide axis,
tick label style={font=\small}, y dir=reverse]
\addplot [forget plot] graphics 
         [xmin=\xmin,xmax=\xmax,ymin=\ymin,ymax=\ymax] {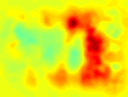};

\addplot[dashed,black,line width=1pt] 
      coordinates {(\xmin+.01,\ymin+.01) (\xmin+.01,\ymax-.01) 
                   (\xmax-.01,\ymax-.01) (\xmax-.01,\ymin+.01) (\xmin+.01,\ymin+.01)};
\end{axis}

&

\begin{axis}[anchor=south,yshift=.5cm,hide axis,height=.5cm,colorbar/width=.2cm,
     colormap/jet,colorbar horizontal,
     colorbar style={separate axis lines,
                     tick label style={font=\small},
                     title={\small{velocity (\si{\km\per\second})}}},
                     point meta min=\vmin, point meta max=\vmax,
                     height=.5\modelwidth]
\end{axis}
\\};

\draw[black,-stealth,->,dashed,line width=1pt] (-0.2,1.55)to[out=0,in=-180] ( 1.0 ,1.55);
\draw[black,-stealth,->,dashed,line width=1pt] (-4.5,1.0) to[out=-90,in=90] (-4.5,-0.5);

\end{tikzpicture}
  \caption{velocity reconstruction from the minimization 
           of $\misfitG$, starting with the model of 
           \cref{fig:statoil:start}.
           The observational acquisition ($f$ in \cref{eq:misfit_green}) 
           uses $\nstack = 5$ sources composed of $\npt = 32$, 
           cf.~\cref{eq:multi_point_source}
           while the computational acquisition 
           ($g$ in \cref{eq:misfit_green}) uses $160$ point-sources.
           The reconstruction uses $30$ iterations per frequency
           between $5$ and $15$ \si{\Hz}. 
           For visualization, we picture sections at a fixed 
           depth $z = 800$ \si{\meter} (bottom) and for 
           $y = 1125$ \si{\meter} (right).}
  \label{fig:statoil:rwi_aq5src_15hz}
\end{figure}

\begin{figure}[ht!] \centering
\setlength{\modelwidth} {7.25cm}
\setlength{\modelheight}{4.00cm}
  \renewcommand{\modelfile}{cp_aq1stack_reciprocity-160src_nlcg_12hz_gauss2}
\begin{tikzpicture}
\pgfmathsetmacro{\xmin} {0.}
\pgfmathsetmacro{\xmax} {2.540}
\pgfmathsetmacro{\ymin} {0.}
\pgfmathsetmacro{\ymax} {1.440}
\pgfmathsetmacro{\zmin} {0.}
\pgfmathsetmacro{\zmax} {1.220}

\pgfmathsetmacro{\ycut} {1.14}
\pgfmathsetmacro{\zcut} {0.8}
\pgfmathsetmacro{\vmin} {1.5}
\pgfmathsetmacro{\vmax} {5.2}

\matrix[column sep=7mm, row sep=0mm] {
\begin{axis}[
  anchor=center,
  tick label style={font=\small},
  grid=both,minor tick num=1,
  xlabel={\small{$x$ (\si{\km})}},ylabel={\small{$y$ (\si{\km})}},
  zlabel={\small{depth (\si{\km})}}, 
  ztick pos=left,
  3d box,width=\modelwidth, 
  xmin=\xmin,ymin=\ymin,zmin=\zmin,xmax=\xmax,ymax=\ymax,zmax=\zmax,
  every axis x label/.style={at={(0.35, 0.00)},anchor=north},
  every axis y label/.style={at={(0.95, 0.02)},anchor=north},
  line width=.25pt]
  \addplot3[fill=white] graphics[points={
            (\xmin,\ymin,\zmax)  => (244,232-232)
            (\xmax,\ymin,\zmax)  => (0. ,232-150)
            (\xmin,\ymax,\zmax)  => (410,232-193)
            (\xmax,\ymax,\zmin)  => (165,232-0. )}
            ]{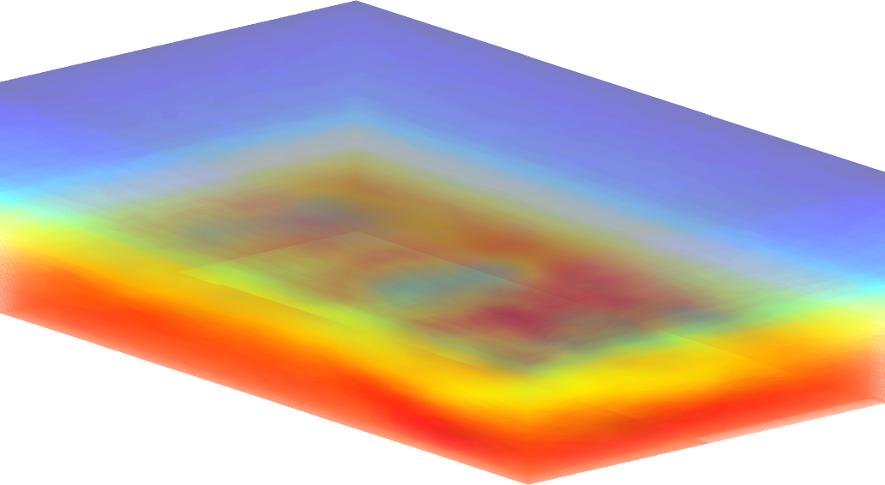};
  
  \addplot3[dashed,black,line width=1pt] 
      coordinates {(\xmin,\ycut,\zmin) (\xmin,\ycut,\zmax)
                   (\xmax,\ycut,\zmax) (\xmax,\ycut,\zmin) (\xmin,\ycut,\zmin)};
  \addplot3[dashed,black,line width=1pt] 
      coordinates {(\xmin,\ymin,\zcut) (\xmin,\ymax,\zcut) 
                   (\xmax,\ymax,\zcut) (\xmax,\ymin,\zcut) (\xmin,\ymin,\zcut)};
\end{axis}

&

\begin{axis}[%
anchor=center, width=\modelwidth, height=\modelheight,
xmin=\xmin, xmax=\xmax,ymin=\zmin, ymax=\zmax,hide axis,
tick label style={font=\small},y dir=reverse]
\addplot [forget plot] graphics 
         [xmin=\xmin,xmax=\xmax,ymin=\zmin,ymax=\zmax] {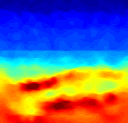};

\addplot[dashed,black,line width=1pt] 
      coordinates {(\xmin+.01,\zmin+.01) (\xmin+.01,\zmax-.01) 
                   (\xmax-.01,\zmax-.01) (\xmax-.01,\zmin+.01) (\xmin+.01,\zmin+.01)};
\end{axis}

\\

\begin{axis}[%
anchor=center, width=\modelwidth, height=\modelheight,
xmin=\xmin, xmax=\xmax,ymin=\ymin, ymax=\ymax,hide axis,
tick label style={font=\small}, y dir=reverse]
\addplot [forget plot] graphics 
         [xmin=\xmin,xmax=\xmax,ymin=\ymin,ymax=\ymax] {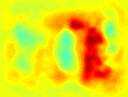};

\addplot[dashed,black,line width=1pt] 
      coordinates {(\xmin+.01,\ymin+.01) (\xmin+.01,\ymax-.01) 
                   (\xmax-.01,\ymax-.01) (\xmax-.01,\ymin+.01) (\xmin+.01,\ymin+.01)};
\end{axis}

&

\begin{axis}[anchor=south,yshift=.5cm,hide axis,height=.5cm,colorbar/width=.2cm,
     colormap/jet,colorbar horizontal,
     colorbar style={separate axis lines,
                     tick label style={font=\small},
                     title={\small{velocity (\si{\km\per\second})}}},
                     point meta min=\vmin, point meta max=\vmax,
                     height=.5\modelwidth]
\end{axis}
\\};

\draw[black,-stealth,->,dashed,line width=1pt] (-0.2,1.55)to[out=0,in=-180] ( 1.0 ,1.55);
\draw[black,-stealth,->,dashed,line width=1pt] (-4.5,1.0) to[out=-90,in=90] (-4.5,-0.5);

\end{tikzpicture}
  \caption{velocity reconstruction from the minimization 
           of $\misfitG$, starting with the model of 
           \cref{fig:statoil:start}.
           The observational acquisition ($f$ in \cref{eq:misfit_green}) 
           uses $\nstack = 1$ source composed of $\npt = 160$, 
           cf.~\cref{eq:multi_point_source}
           while the computational acquisition 
           ($g$ in \cref{eq:misfit_green}) uses $160$ point-sources.
           The reconstruction uses $30$ iterations per frequency
           between $5$ and $15$ \si{\Hz}. 
           For visualization, we picture sections at a fixed 
           depth $z = 800$ \si{\meter} (bottom) and for 
           $y = 1125$ \si{\meter} (right).}
  \label{fig:statoil:rwi_aq1src_15hz}
\end{figure}

Hence, we now consider that the measurements are obtained 
from $5$ multiple-point sources. 
we follow the configuration of the previous subsection but 
inverting the computational and observational acquisitions:
it is now $f$ in~\cref{eq:misfit_green} that is represented 
with a multiple-point sources~\cref{eq:multi_point_source_def}
while the computational sources $g$ consist of $160$ single 
point sources. 
It means that the observed measurements are acquired 
from the physical phenomenon corresponding with multiple-point sources.
Note that from this type of measurements, the FWI reconstruction
coincides with the shot-stacking result of \cref{fig:statoil:fwi_5stack_15hz}.
In \cref{fig:statoil:rwi_aq5src_15hz}, we show
the reconstruction where the observations are obtained
from $\nstack=5$ multiple-point sources and the computational 
acquisition consists of $160$ single-point sources.
In \cref{fig:statoil:rwi_aq1src_15hz}, we show 
the reconstruction where the measured data result 
from one multiple-point source, composed of $160$ points
(i.e., all sources are excited at the same time)
while the numerical acquisition remains with 
$160$ point-sources.

We observe that the reconstructions obtained 
from a drastically reduced set of measurements, 
resulting from multiple-point sources, retrieve
the appropriate velocity variation.
It displays similar accuracy compared to the use 
of single-point source measurements with
multiple-point source simulations 
(respectively \cref{fig:statoil:rwi_5src_15hz,fig:statoil:rwi_1src_15hz} 
for $5$ and $1$ 
computational sources).
Namely, it seems that FRgWI is insensitive to 
which acquisition is made sparse (i.e., with 
multiple-point sources). 

\section{Numerical experiment 2: salt-body SEAM model}
\label{section:numerical_experiments_2}

\begin{figure}[ht!] \centering
\setlength{\modelwidth} {7.0cm}
\setlength{\modelheight}{4.0cm}
\setlength{\jumpvert}   {2.8cm}
  \renewcommand{\modelfile}{cp_true}

\begin{tikzpicture}
\pgfmathsetmacro{\xmin} {0.}
\pgfmathsetmacro{\xmax} {7.}
\pgfmathsetmacro{\ymin} {0.}
\pgfmathsetmacro{\ymax} {6.5}
\pgfmathsetmacro{\zmin} {0.}
\pgfmathsetmacro{\zmax} {2.1}

\pgfmathsetmacro{\ycut} {3.25}
\pgfmathsetmacro{\xcut} {3.50}
\pgfmathsetmacro{\zcut} {1.47}
\pgfmathsetmacro{\vmin} {1.5}
\pgfmathsetmacro{\vmax} {4.8}

\matrix[column sep=6mm, row sep=-2mm] {
\begin{axis}[yshift=0.7*\jumpvert,
  anchor=center,
  tick label style={font=\small},
  grid=both,minor tick num=1,
  xlabel={\small{$x$ (\si{\km})}},ylabel={\small{$y$ (\si{\km})}},
  zlabel={\small{depth (\si{\km})}}, 
  ztick pos=left,
  3d box,width=.9\modelwidth, 
  xmin=\xmin,ymin=\ymin,zmin=\zmin,xmax=\xmax,ymax=\ymax,zmax=\zmax,
  every axis x label/.style={at={(0.30, 0.07)},anchor=north},
  every axis y label/.style={at={(0.80, 0.1)},anchor=north},
  line width=.25pt]
  \addplot3[fill=white] graphics[points={
            (\xmin,\ymin,\zmax)  => (154,203-203)
            (\xmax,\ymin,\zmax)  => (0. ,203-125)
            (\xmin,\ymax,\zmax)  => (324,203-144)
            (\xmax,\ymax,\zmin)  => (172,203-0. )}
            ]{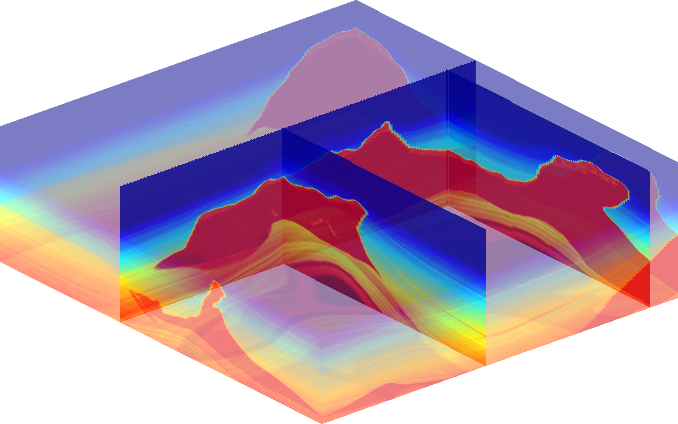};
  
  \addplot3[dashed,black,line width=1pt] 
      coordinates {(\xmin,\ycut,\zmin) (\xmin,\ycut,\zmax)
                   (\xmax,\ycut,\zmax) (\xmax,\ycut,\zmin) (\xmin,\ycut,\zmin)};
  \addplot3[dashed,black,line width=1pt] 
      coordinates {(\xmin,\ymin,\zcut) (\xmin,\ymax,\zcut) 
                   (\xmax,\ymax,\zcut) (\xmax,\ymin,\zcut) (\xmin,\ymin,\zcut)};

  \addplot3[dashed,black,line width=1pt] 
      coordinates {(\xcut,\ymin,\zmin) (\xcut,\ymin,\zmax)
                   (\xcut,\ymax,\zmax) (\xcut,\ymax,\zmin) (\xcut,\ymin,\zmin)};

\end{axis}

&

\begin{axis}[anchor=center,  width=\modelwidth, height=\modelheight,
             xmin=\xmin, xmax=\xmax,ymin=\zmin, ymax=\zmax,hide axis,
             tick label style={font=\small},y dir=reverse]
         \addplot [forget plot] graphics 
         [xmin=\xmin,xmax=\xmax,ymin=\zmin,ymax=\zmax] {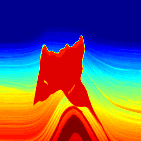};
         \addplot[dashed,black,line width=1pt] 
       coordinates {(\xmin+.01,\zmin+.01) (\xmin+.01,\zmax-.01) 
                    (\xmax-.01,\zmax-.01) (\xmax-.01,\zmin+.01) (\xmin+.01,\zmin+.01)};
\end{axis}

\begin{axis}[yshift=\jumpvert,
             anchor=center,  width=\modelwidth, height=\modelheight,
             xmin=\xmin, xmax=\xmax,ymin=\zmin, ymax=\zmax,hide axis,
             tick label style={font=\small},y dir=reverse]
         \addplot [forget plot] graphics 
         [xmin=\xmin,xmax=\xmax,ymin=\zmin,ymax=\zmax] {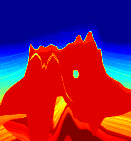};
         \addplot[dashed,black,line width=1pt] 
       coordinates {(\xmin+.01,\zmin+.01) (\xmin+.01,\zmax-.01) 
                    (\xmax-.01,\zmax-.01) (\xmax-.01,\zmin+.01) (\xmin+.01,\zmin+.01)};
\end{axis}

\\[-0.3cm]

\begin{axis}[anchor=center, width=\modelwidth, height=.60\modelwidth,
             xmin=\xmin, xmax=\xmax,ymin=\ymin, ymax=\ymax,hide axis,
             tick label style={font=\small}, y dir=reverse]
            \addplot [forget plot] graphics 
                     [xmin=\xmin,xmax=\xmax,ymin=\ymin,ymax=\ymax] {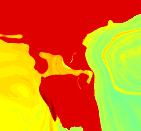};
            
            \addplot[dashed,black,line width=1pt] 
                  coordinates {(\xmin+.01,\ymin+.01) (\xmin+.01,\ymax-.01) 
                               (\xmax-.01,\ymax-.01) (\xmax-.01,\ymin+.01) (\xmin+.01,\ymin+.01)};
\end{axis}

&

\begin{axis}[yshift=-0.8cm,
     anchor=south,yshift=.5cm,hide axis,height=.5cm,colorbar/width=.2cm,
     colormap/jet,colorbar horizontal,
     colorbar style={separate axis lines,
                     tick label style={font=\small},
                     title={\small{velocity (\si{\km\per\second})}}},
                     point meta min=\vmin, point meta max=\vmax,
                     height=.5\modelwidth]
\end{axis}
\\};

\draw[black,-stealth,->,dashed,line width=1pt] (-0.7, 2.8) to[out=0,in=-180] ( 1.0, 2.8);
\draw[black,-stealth,->,dashed,line width=1pt] (-0.7, 0.3) to[out=0,in=-180] ( 1.0, 0.3);
\draw[black,-stealth,->,dashed,line width=1pt] (-4.7, 1.0) to[out=-90,in=90] (-4.7,-0.8);
%

\end{tikzpicture}
  \caption{velocity model of size 
           $7 \times 6.5 \times 2.1$ \si{\km\cubed}. 
           For visualization, we extract sections 
           at a fixed  depth $z = 1.5$  \si{\km} 
           (bottom), for     $x = 3.5$  \si{\km} 
           (top right) and   $y = 3.25$ \si{\km}
           (bottom right).}           
  \label{fig:seam:cp_true}
\end{figure}
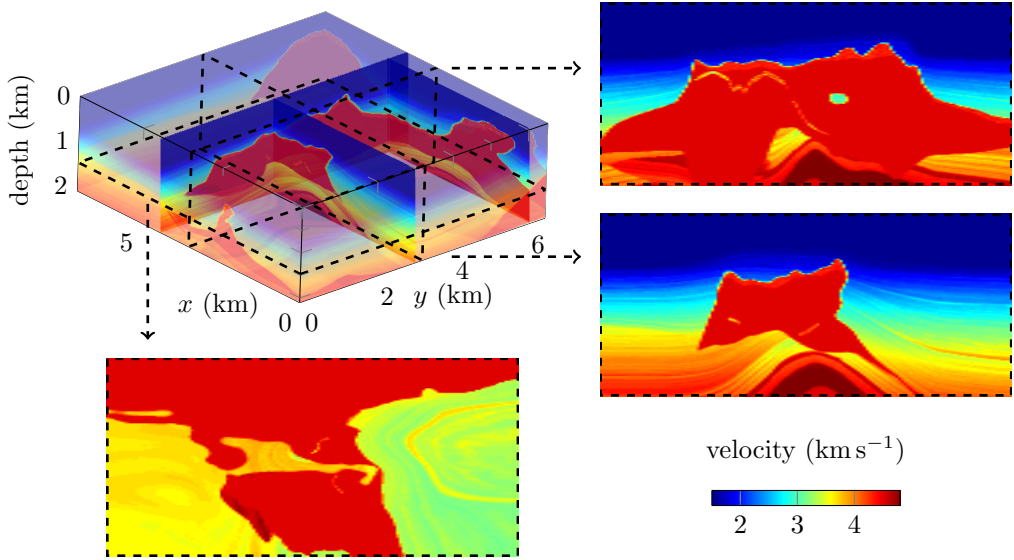

\begin{figure}[ht!] \centering
\setlength{\modelwidth} {7.0cm}
\setlength{\modelheight}{4.0cm}
\setlength{\jumpvert}   {2.8cm}
  \renewcommand{\modelfile}{rho_true}

\begin{tikzpicture}
\pgfmathsetmacro{\xmin} {0.}
\pgfmathsetmacro{\xmax} {7.}
\pgfmathsetmacro{\ymin} {0.}
\pgfmathsetmacro{\ymax} {6.5}
\pgfmathsetmacro{\zmin} {0.}
\pgfmathsetmacro{\zmax} {2.1}

\pgfmathsetmacro{\ycut} {3.25}
\pgfmathsetmacro{\xcut} {3.50}
\pgfmathsetmacro{\zcut} {1.47}
\pgfmathsetmacro{\vmin} {1000}
\pgfmathsetmacro{\vmax} {1700}

\matrix[column sep=6mm, row sep=-2mm] {
\begin{axis}[yshift=0.7*\jumpvert,
  anchor=center,
  tick label style={font=\small},
  grid=both,minor tick num=1,
  xlabel={\small{$x$ (\si{\km})}},ylabel={\small{$y$ (\si{\km})}},
  zlabel={\small{depth (\si{\km})}}, 
  ztick pos=left,
  3d box,width=.9\modelwidth, 
  xmin=\xmin,ymin=\ymin,zmin=\zmin,xmax=\xmax,ymax=\ymax,zmax=\zmax,
  every axis x label/.style={at={(0.30, 0.07)},anchor=north},
  every axis y label/.style={at={(0.80, 0.1)},anchor=north},
  line width=.25pt]
  \addplot3[fill=white] graphics[points={
            (\xmin,\ymin,\zmax)  => (154,203-203)
            (\xmax,\ymin,\zmax)  => (0. ,203-125)
            (\xmin,\ymax,\zmax)  => (324,203-144)
            (\xmax,\ymax,\zmin)  => (172,203-0. )}
            ]{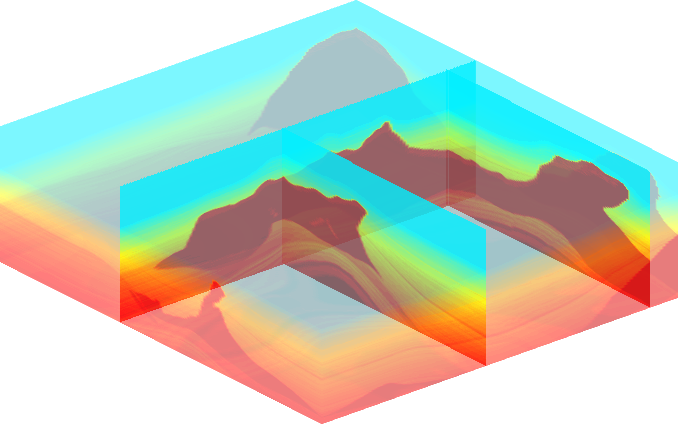};
  
  \addplot3[dashed,black,line width=1pt] 
      coordinates {(\xmin,\ycut,\zmin) (\xmin,\ycut,\zmax)
                   (\xmax,\ycut,\zmax) (\xmax,\ycut,\zmin) (\xmin,\ycut,\zmin)};
  \addplot3[dashed,black,line width=1pt] 
      coordinates {(\xmin,\ymin,\zcut) (\xmin,\ymax,\zcut) 
                   (\xmax,\ymax,\zcut) (\xmax,\ymin,\zcut) (\xmin,\ymin,\zcut)};

  \addplot3[dashed,black,line width=1pt] 
      coordinates {(\xcut,\ymin,\zmin) (\xcut,\ymin,\zmax)
                   (\xcut,\ymax,\zmax) (\xcut,\ymax,\zmin) (\xcut,\ymin,\zmin)};

\end{axis}

&

\begin{axis}[anchor=center,  width=\modelwidth, height=\modelheight,
             xmin=\xmin, xmax=\xmax,ymin=\zmin, ymax=\zmax,hide axis,
             tick label style={font=\small},y dir=reverse]
         \addplot [forget plot] graphics 
         [xmin=\xmin,xmax=\xmax,ymin=\zmin,ymax=\zmax] {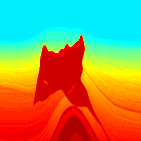};
         \addplot[dashed,black,line width=1pt] 
       coordinates {(\xmin+.01,\zmin+.01) (\xmin+.01,\zmax-.01) 
                    (\xmax-.01,\zmax-.01) (\xmax-.01,\zmin+.01) (\xmin+.01,\zmin+.01)};
\end{axis}

\begin{axis}[yshift=\jumpvert,
             anchor=center,  width=\modelwidth, height=\modelheight,
             xmin=\xmin, xmax=\xmax,ymin=\zmin, ymax=\zmax,hide axis,
             tick label style={font=\small},y dir=reverse]
         \addplot [forget plot] graphics 
         [xmin=\xmin,xmax=\xmax,ymin=\zmin,ymax=\zmax] {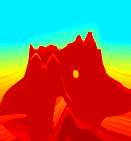};
         \addplot[dashed,black,line width=1pt] 
       coordinates {(\xmin+.01,\zmin+.01) (\xmin+.01,\zmax-.01) 
                    (\xmax-.01,\zmax-.01) (\xmax-.01,\zmin+.01) (\xmin+.01,\zmin+.01)};
\end{axis}

\\[-0.3cm]

\begin{axis}[anchor=center, width=\modelwidth, height=.60\modelwidth,
             xmin=\xmin, xmax=\xmax,ymin=\ymin, ymax=\ymax,hide axis,
             tick label style={font=\small}, y dir=reverse]
            \addplot [forget plot] graphics 
                     [xmin=\xmin,xmax=\xmax,ymin=\ymin,ymax=\ymax] {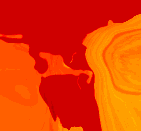};
            
            \addplot[dashed,black,line width=1pt] 
                  coordinates {(\xmin+.01,\ymin+.01) (\xmin+.01,\ymax-.01) 
                               (\xmax-.01,\ymax-.01) (\xmax-.01,\ymin+.01) (\xmin+.01,\ymin+.01)};
\end{axis}

&

\begin{axis}[yshift=-0.8cm,
     anchor=south,yshift=.5cm,hide axis,height=.5cm,colorbar/width=.2cm,
     colormap/jet,colorbar horizontal,
     colorbar style={separate axis lines,
                     tick label style={font=\small,},
                     xtick={1000,1700},
                     title={\small{density (\si{\kg\per\meter\cubed})}}},
                     point meta min=\vmin, point meta max=\vmax,
                     height=.5\modelwidth]
\end{axis}
\\};

\draw[black,-stealth,->,dashed,line width=1pt] (-0.7, 2.8) to[out=0,in=-180] ( 1.0, 2.8);
\draw[black,-stealth,->,dashed,line width=1pt] (-0.7, 0.3) to[out=0,in=-180] ( 1.0, 0.3);
\draw[black,-stealth,->,dashed,line width=1pt] (-4.7, 1.0) to[out=-90,in=90] (-4.7,-0.8);

\end{tikzpicture}
  \caption{Density model of size 
           $7 \times 6.5 \times 2.1$ \si{\km\cubed}. 
           For visualization, we extract sections 
           at a fixed  depth $z = 1.5$  \si{\km} 
           (bottom), for     $x = 3.5$  \si{\km} 
           (top right) and   $y = 3.25$ \si{\km}
           (bottom right).}  
  \label{fig:seam:rho_true}
\end{figure}
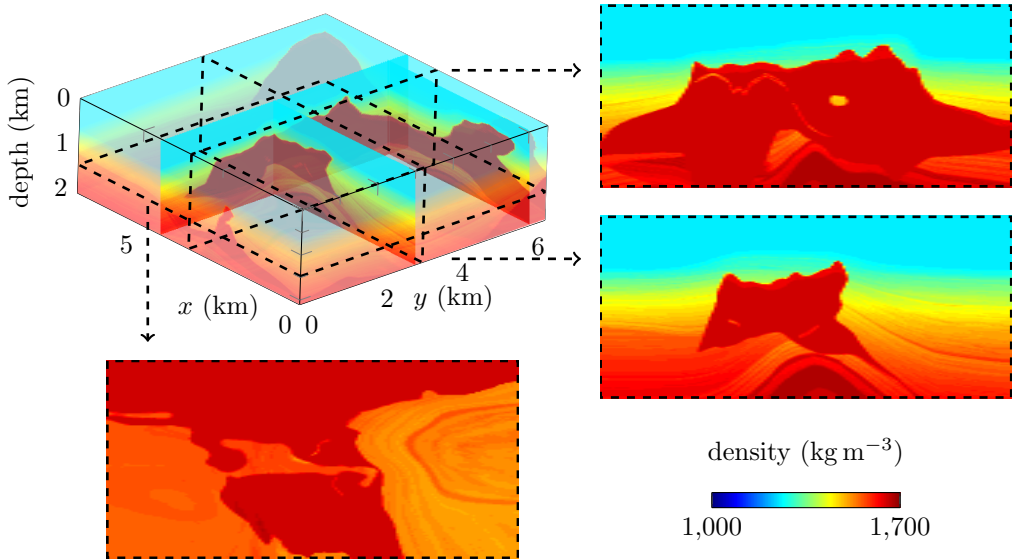

In this experiment, we consider a three-dimensional 
velocity model encompassing salt-domes. 
The velocity is extracted from the seismic
SEAM\footnote{SEG Advanced Modeling Corporation, 
              see $\text{https://seg.org/News-Resources/Research-and-Data/SEAM}$}
benchmark, and is of size 
$7 \times 6.5 \times 2.1$ \si{\km\cubed}. 
The velocity model is depicted in \cref{fig:seam:cp_true}, 
and varies from \num{1500} to \num{4800} \si{\meter\per\second}. 
In this experiment, the density is heterogeneous, and 
pictured in \cref{fig:seam:rho_true}.
Compared to previous test-case, it is of different nature
(salt-domes), it is larger and it has a heterogeneous density.
Therefore, we expect this numerical experiment to be more challenging.

\subsection{Time-domain data}

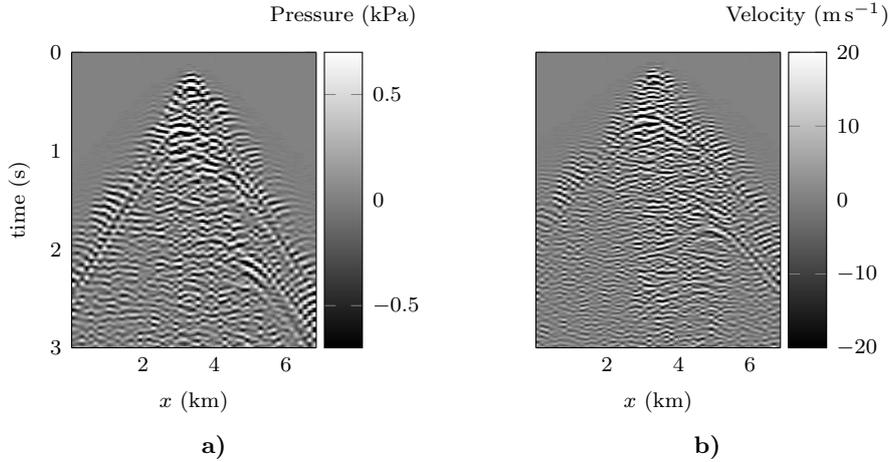
\begin{figure}[ht!] \centering
\setlength{\modelwidth} {4.80cm}
\setlength{\modelheight}{5.50cm}
  \graphicspath{{figures/seam/data/}}
  \renewcommand{\titlebar}{Pressure (\si{\kilo\pascal})}
  \pgfmathsetmacro{\xmin} {0.01} \pgfmathsetmacro{\xmax}{6.85}
  \pgfmathsetmacro{\cmin} {-0.7}   \pgfmathsetmacro{\cmax}{0.7}
  \pgfmathsetmacro{\zmax}{3.0}
  \renewcommand{\modelfile}{trace-rcvY25_pressure-noise_-7_7e2}
  \subfloat[][]{\begin{tikzpicture}

\pgfmathsetmacro{\zmin}    {0.000}
\pgfmathsetmacro{\zmaxall} {5.000}

\begin{axis}[%
width=\modelwidth,
height=\modelheight,
every node/.append style={font=\scriptsize},
        tick label style={font=\scriptsize},
             label style={font=\scriptsize},
axis on top, separate axis lines,
xmin=\xmin, xmax=\xmax, xlabel={$x$  (\si{\kilo\meter})},
ymin=\zmin, ymax=\zmax, ylabel={time (\si{\second})}, y dir=reverse,
colormap/jet,colorbar,
colormap={blackwhite}{gray(0cm)=(0);gray(1cm)=(1)},
colorbar style={title={{\scriptsize{\titlebar}}},
xshift=-.2cm},point meta min=\cmin,point meta max=\cmax
]
\addplot [forget plot] graphics [xmin=\xmin,xmax=\xmax,ymin=\zmin,ymax=\zmaxall] {{\modelfile}.png};
\end{axis}
\end{tikzpicture}%
                \label{fig:seam:data-2d_A}} \hspace*{.5cm}
  \pgfmathsetmacro{\cmin} {-20} \pgfmathsetmacro{\cmax} {20}
  \renewcommand{\modelfile}{trace-rcvY25_dpressure-noise_-2_2e1}
  \renewcommand{\titlebar}{Velocity (\si{\meter\per\second})}
  \subfloat[][]{\begin{tikzpicture}

\pgfmathsetmacro{\zmin} {0.000}
\pgfmathsetmacro{\zmaxall} {4.000}

\begin{axis}[%
width=\modelwidth,
height=\modelheight,
every node/.append style={font=\scriptsize},
        tick label style={font=\scriptsize},
             label style={font=\scriptsize},
axis on top, separate axis lines,
xmin=\xmin, xmax=\xmax, xlabel={$x$  (\si{\kilo\meter})},
ymin=\zmin, ymax=\zmax, 
y dir=reverse, yticklabels={,,},
colormap/jet,colorbar,
colormap={blackwhite}{gray(0cm)=(0);gray(1cm)=(1)},
colorbar style={title={{\scriptsize{\titlebar}}},
xshift=-.2cm},point meta min=\cmin,point meta max=\cmax
]
\addplot [forget plot] graphics [xmin=\xmin,xmax=\xmax,ymin=\zmin,ymax=\zmaxall] {{\modelfile}.png};
\end{axis}
\end{tikzpicture}%
                \label{fig:seam:data-2d_B}} 
  \caption{Time-domain \protect\subref{fig:seam:data-2d_A}
           pressure trace and 
           \protect\subref{fig:seam:data-2d_B}
           normal velocity trace with 20 \si{\deci\bel} 
           signal-to-noise ratio. 
           These correspond to a line of receivers at a fixed 
           $y = \num{3.1}$ \si{\km}, for a single-point source 
           located in $(x_{s},\, y_{s},\, z_{s}) 
           = (\num{3.4},\, \num{3.4},\, \num{0.01})~\si{\km}$
           for the velocity and density models given 
           in \cref{fig:seam:cp_true,fig:seam:rho_true}. 
           For our 
           experiment, we apply $20$ \si{\deci\bel} 
           signal-to-noise ratio to the synthetic data
           and employ the Fourier transform of the noisy
           traces. The available frequency ranges from 
           $2$ to $7$ \si{\Hz}.}
  \label{fig:seam:data-2d}
\end{figure}

Similarly to the previous experiment, we work with 
noisy time-domain data and compute their Fourier 
transform to generate the harmonic data used in the 
iterative reconstruction.
There is a total of \num{272} point-sources in the 
observational acquisition, which are located at \num{10}
\si{\m} depth. 
The sources are placed on a two-dimensional plane 
with $400$ \si{\m} between each source along the $x$
and $y$-axes. 
For multiple-point acquisition, we follow the structured
combination illustrated in \cref{fig:multi-source-splitting_B}.
In this experiment, we generate the time-domain data
and we use $20$ \si{\deci\bel} signal-to-noise
ratio. The resulting seismograms are illustrated 
in \cref{fig:seam:data-2d} for a single source,
where we show the pressure and vertical velocity
measurements.
The data are acquired by a total of \num{2805} receivers. 
For the reconstruction, we perform a Fourier transform 
of the time-domain noisy data, and use frequency content from 
$2$ to $7$ \si{\Hz}.

\subsection{Reconstruction using FRgWI}

\begin{figure}[ht!] \centering
\setlength{\modelwidth} {7.0cm}
\setlength{\modelheight}{4.0cm}
\setlength{\jumpvert}   {2.6cm}
  \renewcommand{\modelfile}{cp_start}

\begin{tikzpicture}
\pgfmathsetmacro{\xmin} {0.}
\pgfmathsetmacro{\xmax} {7.}
\pgfmathsetmacro{\ymin} {0.}
\pgfmathsetmacro{\ymax} {6.5}
\pgfmathsetmacro{\zmin} {0.}
\pgfmathsetmacro{\zmax} {2.1}

\pgfmathsetmacro{\ycut} {3.25}
\pgfmathsetmacro{\xcut} {3.50}
\pgfmathsetmacro{\zcut} {1.47}
\pgfmathsetmacro{\vmin} {1.5}
\pgfmathsetmacro{\vmax} {4.8}

\matrix[column sep=6mm, row sep=-2mm] {
\begin{axis}[yshift=0.5*\jumpvert,
  anchor=center,
  tick label style={font=\small},
  grid=both,minor tick num=1,
  xlabel={\small{$x$ (\si{\km})}},ylabel={\small{$y$ (\si{\km})}},
  zlabel={\small{depth (\si{\km})}}, 
  ztick pos=left,
  3d box,width=.9\modelwidth, 
  xmin=\xmin,ymin=\ymin,zmin=\zmin,xmax=\xmax,ymax=\ymax,zmax=\zmax,
  every axis x label/.style={at={(0.25, 0.05)},anchor=north},
  every axis y label/.style={at={(0.80, 0.1)},anchor=north},
  line width=.25pt]
  \addplot3[fill=white] graphics[points={
            (\xmin,\ymin,\zmax)  => (154,203-203)
            (\xmax,\ymin,\zmax)  => (0. ,203-125)
            (\xmin,\ymax,\zmax)  => (324,203-144)
            (\xmax,\ymax,\zmin)  => (172,203-0. )}
            ]{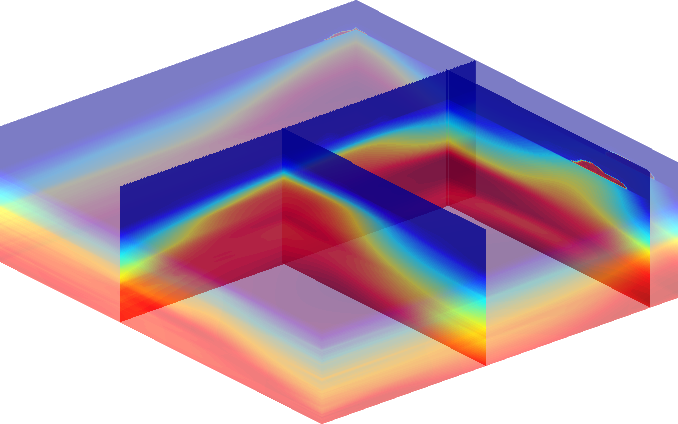};
  
  \addplot3[dashed,black,line width=1pt] 
      coordinates {(\xmin,\ycut,\zmin) (\xmin,\ycut,\zmax)
                   (\xmax,\ycut,\zmax) (\xmax,\ycut,\zmin) (\xmin,\ycut,\zmin)};

  \addplot3[dashed,black,line width=1pt] 
      coordinates {(\xcut,\ymin,\zmin) (\xcut,\ymin,\zmax)
                   (\xcut,\ymax,\zmax) (\xcut,\ymax,\zmin) (\xcut,\ymin,\zmin)};

\end{axis}

&

\begin{axis}[anchor=center,  width=\modelwidth, height=\modelheight,
             xmin=\xmin, xmax=\xmax,ymin=\zmin, ymax=\zmax,hide axis,
             tick label style={font=\small},y dir=reverse]
         \addplot [forget plot] graphics 
         [xmin=\xmin,xmax=\xmax,ymin=\zmin,ymax=\zmax] {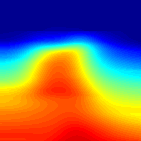};
         \addplot[dashed,black,line width=1pt] 
       coordinates {(\xmin+.01,\zmin+.01) (\xmin+.01,\zmax-.01) 
                    (\xmax-.01,\zmax-.01) (\xmax-.01,\zmin+.01) (\xmin+.01,\zmin+.01)};
\end{axis}

\begin{axis}[yshift=\jumpvert,
             anchor=center,  width=\modelwidth, height=\modelheight,
             xmin=\xmin, xmax=\xmax,ymin=\zmin, ymax=\zmax,hide axis,
             tick label style={font=\small},y dir=reverse]
         \addplot [forget plot] graphics 
         [xmin=\xmin,xmax=\xmax,ymin=\zmin,ymax=\zmax] {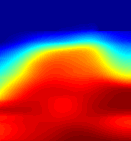};
         \addplot[dashed,black,line width=1pt] 
       coordinates {(\xmin+.01,\zmin+.01) (\xmin+.01,\zmax-.01) 
                    (\xmax-.01,\zmax-.01) (\xmax-.01,\zmin+.01) (\xmin+.01,\zmin+.01)};
\end{axis}

&

\hspace*{-.5cm}\begin{axis}[xshift=-1cm,
             height=1.2\modelheight,
             tick label style={font=\small},
             hide axis,colorbar,colormap/jet,
             colorbar style={yshift=0cm,
             ylabel={\small{velocity (\si{\km\per\second})}},
             point meta min=\vmin, point meta max=\vmax,
             rotate=0,
             yticklabel pos=right,
             },]
{}; \end{axis}

\\};

\draw[black,-stealth,->,dashed,line width=1pt] (-1.8, 1.10) to[out=0,in=-180] (-0.05, 1.10);
\draw[black,-stealth,->,dashed,line width=1pt] (-1.8,-1.35) to[out=0,in=-180] (-0.05,-1.35);

\end{tikzpicture}
  \caption{Starting velocity model for the reconstruction 
           of the model of \cref{fig:seam:cp_true}.
           We picture vertical sections at a fixed $x = 3.5$  \si{\km} 
           (top right) and $y = 3.25$ \si{\km} (bottom right).}
  \label{fig:seam:cp_start}
\end{figure}
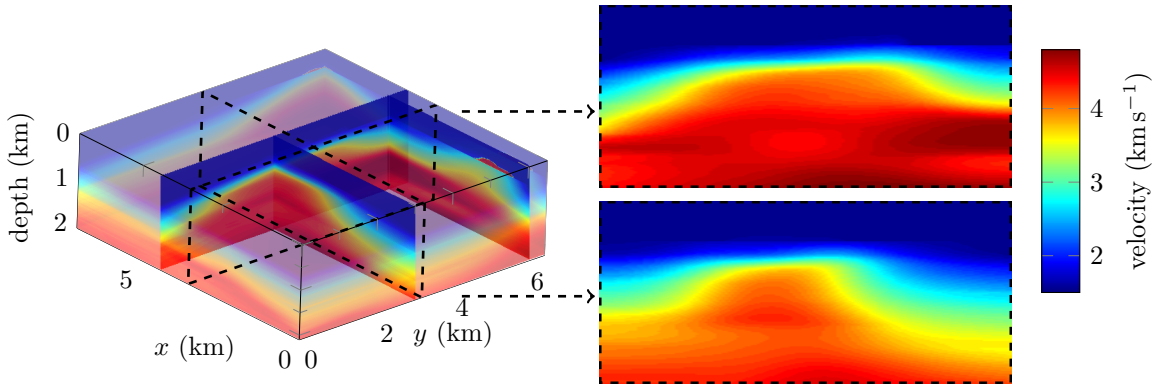

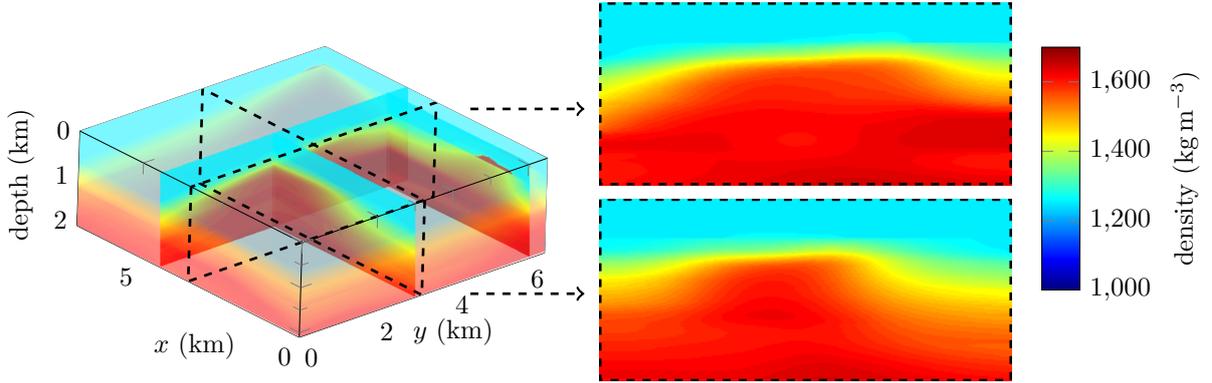
\begin{figure}[ht!] \centering
\setlength{\modelwidth} {7.0cm}
\setlength{\modelheight}{4.0cm}
\setlength{\jumpvert}   {2.6cm}
  \renewcommand{\modelfile}{rho_start}

\begin{tikzpicture}
\pgfmathsetmacro{\xmin} {0.}
\pgfmathsetmacro{\xmax} {7.}
\pgfmathsetmacro{\ymin} {0.}
\pgfmathsetmacro{\ymax} {6.5}
\pgfmathsetmacro{\zmin} {0.}
\pgfmathsetmacro{\zmax} {2.1}

\pgfmathsetmacro{\ycut} {3.25}
\pgfmathsetmacro{\xcut} {3.50}
\pgfmathsetmacro{\zcut} {1.47}
\pgfmathsetmacro{\vmin} {1000}
\pgfmathsetmacro{\vmax} {1700}

\matrix[column sep=6mm, row sep=-2mm] {
\begin{axis}[yshift=0.5*\jumpvert,
  anchor=center,
  tick label style={font=\small},
  grid=both,minor tick num=1,
  xlabel={\small{$x$ (\si{\km})}},ylabel={\small{$y$ (\si{\km})}},
  zlabel={\small{depth (\si{\km})}}, 
  ztick pos=left,
  3d box,width=.9\modelwidth, 
  xmin=\xmin,ymin=\ymin,zmin=\zmin,xmax=\xmax,ymax=\ymax,zmax=\zmax,
  every axis x label/.style={at={(0.25, 0.05)},anchor=north},
  every axis y label/.style={at={(0.80, 0.1)},anchor=north},
  line width=.25pt]
  \addplot3[fill=white] graphics[points={
            (\xmin,\ymin,\zmax)  => (154,203-203)
            (\xmax,\ymin,\zmax)  => (0. ,203-125)
            (\xmin,\ymax,\zmax)  => (324,203-144)
            (\xmax,\ymax,\zmin)  => (172,203-0. )}
            ]{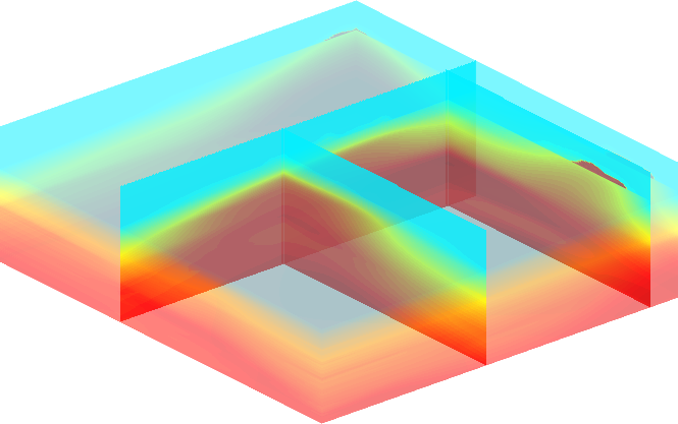};
  
  \addplot3[dashed,black,line width=1pt] 
      coordinates {(\xmin,\ycut,\zmin) (\xmin,\ycut,\zmax)
                   (\xmax,\ycut,\zmax) (\xmax,\ycut,\zmin) (\xmin,\ycut,\zmin)};

  \addplot3[dashed,black,line width=1pt] 
      coordinates {(\xcut,\ymin,\zmin) (\xcut,\ymin,\zmax)
                   (\xcut,\ymax,\zmax) (\xcut,\ymax,\zmin) (\xcut,\ymin,\zmin)};

\end{axis}

&

\begin{axis}[anchor=center,  width=\modelwidth, height=\modelheight,
             xmin=\xmin, xmax=\xmax,ymin=\zmin, ymax=\zmax,hide axis,
             tick label style={font=\small},y dir=reverse]
         \addplot [forget plot] graphics 
         [xmin=\xmin,xmax=\xmax,ymin=\zmin,ymax=\zmax] {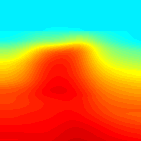};
         \addplot[dashed,black,line width=1pt] 
       coordinates {(\xmin+.01,\zmin+.01) (\xmin+.01,\zmax-.01) 
                    (\xmax-.01,\zmax-.01) (\xmax-.01,\zmin+.01) (\xmin+.01,\zmin+.01)};
\end{axis}

\begin{axis}[yshift=\jumpvert,
             anchor=center,  width=\modelwidth, height=\modelheight,
             xmin=\xmin, xmax=\xmax,ymin=\zmin, ymax=\zmax,hide axis,
             tick label style={font=\small},y dir=reverse]
         \addplot [forget plot] graphics 
         [xmin=\xmin,xmax=\xmax,ymin=\zmin,ymax=\zmax] {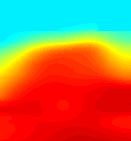};
         \addplot[dashed,black,line width=1pt] 
       coordinates {(\xmin+.01,\zmin+.01) (\xmin+.01,\zmax-.01) 
                    (\xmax-.01,\zmax-.01) (\xmax-.01,\zmin+.01) (\xmin+.01,\zmin+.01)};
\end{axis}

&

\hspace*{-.5cm}\begin{axis}[xshift=-1cm,
             height=1.2\modelheight,
             tick label style={font=\small},
             hide axis,colorbar,colormap/jet,
             colorbar style={yshift=0cm,
             ylabel={\small{density (\si{\kg\per\meter\cubed})}},
             point meta min=\vmin, point meta max=\vmax,
             rotate=0,
             yticklabel pos=right,
             },]
{}; \end{axis}

\\};

\draw[black,-stealth,->,dashed,line width=1pt] (-2.0, 1.10) to[out=0,in=-180] (-0.5, 1.10);
\draw[black,-stealth,->,dashed,line width=1pt] (-2.0,-1.35) to[out=0,in=-180] (-0.5,-1.35);

\end{tikzpicture}
  \caption{Starting density model for the reconstruction 
           of the model of \cref{fig:seam:cp_true,fig:seam:rho_true}.
           We picture vertical sections at a fixed $x = 3.5$  \si{\km} 
           (top right) and $y = 3.25$ \si{\km} (bottom right).}
  \label{fig:seam:rho_start}
\end{figure}

For the reconstruction, we start with initial guesses that 
correspond with smooth version of the true models, they are
shown in \cref{fig:seam:cp_start,fig:seam:rho_start} for 
the starting velocity and density respectively. 
We actually only focuses on the reconstruction of 
the velocity model, and keep the density as its 
initial representation of \cref{fig:seam:rho_start}. 
The density is known to be more complicated to recover than
the velocity because of lack of sensitivity in the data, 
\cite{Jeong2012}, but it should not prevent us from recovering 
the velocity, see, e.g.,~\cite{Faucher2017}.

\begin{figure}[ht!] \centering
\setlength{\modelwidth} {7.0cm}
\setlength{\modelheight}{4.0cm}
\setlength{\jumpvert}   {2.8cm}
  \renewcommand{\modelfile}{cp_recycle-3hz-reciprocity-34src_gauss1}

\begin{tikzpicture}
\pgfmathsetmacro{\xmin} {0.}
\pgfmathsetmacro{\xmax} {7.}
\pgfmathsetmacro{\ymin} {0.}
\pgfmathsetmacro{\ymax} {6.5}
\pgfmathsetmacro{\zmin} {0.}
\pgfmathsetmacro{\zmax} {2.1}

\pgfmathsetmacro{\ycut} {3.25}
\pgfmathsetmacro{\xcut} {3.50}
\pgfmathsetmacro{\zcut} {1.47}
\pgfmathsetmacro{\vmin} {1.5}
\pgfmathsetmacro{\vmax} {4.8}

\matrix[column sep=6mm, row sep=-2mm] {
\begin{axis}[yshift=0.7*\jumpvert,
  anchor=center,
  tick label style={font=\small},
  grid=both,minor tick num=1,
  xlabel={\small{$x$ (\si{\km})}},ylabel={\small{$y$ (\si{\km})}},
  zlabel={\small{depth (\si{\km})}}, 
  ztick pos=left,
  3d box,width=.9\modelwidth, 
  xmin=\xmin,ymin=\ymin,zmin=\zmin,xmax=\xmax,ymax=\ymax,zmax=\zmax,
  every axis x label/.style={at={(0.30, 0.07)},anchor=north},
  every axis y label/.style={at={(0.80, 0.1)},anchor=north},
  line width=.25pt]
  \addplot3[fill=white] graphics[points={
            (\xmin,\ymin,\zmax)  => (154,203-203)
            (\xmax,\ymin,\zmax)  => (0. ,203-125)
            (\xmin,\ymax,\zmax)  => (324,203-144)
            (\xmax,\ymax,\zmin)  => (172,203-0. )}
            ]{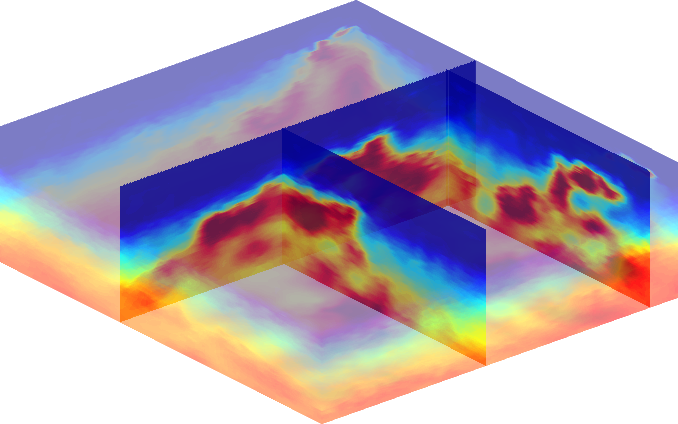};
  
  \addplot3[dashed,black,line width=1pt] 
      coordinates {(\xmin,\ycut,\zmin) (\xmin,\ycut,\zmax)
                   (\xmax,\ycut,\zmax) (\xmax,\ycut,\zmin) (\xmin,\ycut,\zmin)};
  \addplot3[dashed,black,line width=1pt] 
      coordinates {(\xmin,\ymin,\zcut) (\xmin,\ymax,\zcut) 
                   (\xmax,\ymax,\zcut) (\xmax,\ymin,\zcut) (\xmin,\ymin,\zcut)};

  \addplot3[dashed,black,line width=1pt] 
      coordinates {(\xcut,\ymin,\zmin) (\xcut,\ymin,\zmax)
                   (\xcut,\ymax,\zmax) (\xcut,\ymax,\zmin) (\xcut,\ymin,\zmin)};

\end{axis}

&

\begin{axis}[anchor=center,  width=\modelwidth, height=\modelheight,
             xmin=\xmin, xmax=\xmax,ymin=\zmin, ymax=\zmax,hide axis,
             tick label style={font=\small},y dir=reverse]
         \addplot [forget plot] graphics 
         [xmin=\xmin,xmax=\xmax,ymin=\zmin,ymax=\zmax] {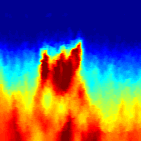};
         \addplot[dashed,black,line width=1pt] 
       coordinates {(\xmin+.01,\zmin+.01) (\xmin+.01,\zmax-.01) 
                    (\xmax-.01,\zmax-.01) (\xmax-.01,\zmin+.01) (\xmin+.01,\zmin+.01)};
\end{axis}

\begin{axis}[yshift=\jumpvert,
             anchor=center,  width=\modelwidth, height=\modelheight,
             xmin=\xmin, xmax=\xmax,ymin=\zmin, ymax=\zmax,hide axis,
             tick label style={font=\small},y dir=reverse]
         \addplot [forget plot] graphics 
         [xmin=\xmin,xmax=\xmax,ymin=\zmin,ymax=\zmax] {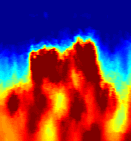};
         \addplot[dashed,black,line width=1pt] 
       coordinates {(\xmin+.01,\zmin+.01) (\xmin+.01,\zmax-.01) 
                    (\xmax-.01,\zmax-.01) (\xmax-.01,\zmin+.01) (\xmin+.01,\zmin+.01)};
\end{axis}

\\[-0.3cm]

\begin{axis}[anchor=center, width=\modelwidth, height=.60\modelwidth,
             xmin=\xmin, xmax=\xmax,ymin=\ymin, ymax=\ymax,hide axis,
             tick label style={font=\small}, y dir=reverse]
            \addplot [forget plot] graphics 
                     [xmin=\xmin,xmax=\xmax,ymin=\ymin,ymax=\ymax] {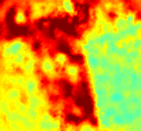};
            
            \addplot[dashed,black,line width=1pt] 
                  coordinates {(\xmin+.01,\ymin+.01) (\xmin+.01,\ymax-.01) 
                               (\xmax-.01,\ymax-.01) (\xmax-.01,\ymin+.01) (\xmin+.01,\ymin+.01)};
\end{axis}

&

\begin{axis}[yshift=-0.8cm,
     anchor=south,yshift=.5cm,hide axis,height=.5cm,colorbar/width=.2cm,
     colormap/jet,colorbar horizontal,
     colorbar style={separate axis lines,
                     tick label style={font=\small},
                     title={\small{velocity (\si{\km\per\second})}}},
                     point meta min=\vmin, point meta max=\vmax,
                     height=.5\modelwidth]
\end{axis}
\\};

\draw[black,-stealth,->,dashed,line width=1pt] (-0.7, 2.8) to[out=0,in=-180] ( 1.0, 2.8);
\draw[black,-stealth,->,dashed,line width=1pt] (-0.7, 0.3) to[out=0,in=-180] ( 1.0, 0.3);
\draw[black,-stealth,->,dashed,line width=1pt] (-4.7, 1.0) to[out=-90,in=90] (-4.7,-0.8);
%

\end{tikzpicture}
  \caption{velocity reconstruction of the SEAM model 
           from the minimization of $\misfitG$, starting 
           with the models of \cref{fig:seam:cp_start}.
           It uses a \emph{dense} observational acquisition of 
           $272$ point-sources, while the numerical 
           acquisition ($g$ in \cref{eq:misfit_green}) 
           employs $\nstack = 34$ sources composed of 
           $\npt = 8$ points.
           The reconstruction follows $30$ iterations per 
           frequency between $2$ and $7$ \si{\Hz}. 
           For visualization, we picture sections 
           at a fixed  depth $z = 1.5$  \si{\km} 
           (bottom), for     $x = 3.5$  \si{\km} 
           (top right) and   $y = 3.25$ \si{\km}
           (bottom right).}  
  \label{fig:seam:rwi}
\end{figure}

\begin{figure}[ht!] \centering
\setlength{\modelwidth} {7.0cm}
\setlength{\modelheight}{4.0cm}
\setlength{\jumpvert}   {2.8cm}
  \renewcommand{\modelfile}{cp_aq34src-7hz-reciprocity-allsrc_gauss1}

\begin{tikzpicture}
\pgfmathsetmacro{\xmin} {0.}
\pgfmathsetmacro{\xmax} {7.}
\pgfmathsetmacro{\ymin} {0.}
\pgfmathsetmacro{\ymax} {6.5}
\pgfmathsetmacro{\zmin} {0.}
\pgfmathsetmacro{\zmax} {2.1}

\pgfmathsetmacro{\ycut} {3.25}
\pgfmathsetmacro{\xcut} {3.50}
\pgfmathsetmacro{\zcut} {1.47}
\pgfmathsetmacro{\vmin} {1.5}
\pgfmathsetmacro{\vmax} {4.8}

\matrix[column sep=6mm, row sep=-2mm] {
\begin{axis}[yshift=0.7*\jumpvert,
  anchor=center,
  tick label style={font=\small},
  grid=both,minor tick num=1,
  xlabel={\small{$x$ (\si{\km})}},ylabel={\small{$y$ (\si{\km})}},
  zlabel={\small{depth (\si{\km})}}, 
  ztick pos=left,
  3d box,width=.9\modelwidth, 
  xmin=\xmin,ymin=\ymin,zmin=\zmin,xmax=\xmax,ymax=\ymax,zmax=\zmax,
  every axis x label/.style={at={(0.30, 0.07)},anchor=north},
  every axis y label/.style={at={(0.80, 0.1)},anchor=north},
  line width=.25pt]
  \addplot3[fill=white] graphics[points={
            (\xmin,\ymin,\zmax)  => (154,203-203)
            (\xmax,\ymin,\zmax)  => (0. ,203-125)
            (\xmin,\ymax,\zmax)  => (324,203-144)
            (\xmax,\ymax,\zmin)  => (172,203-0. )}
            ]{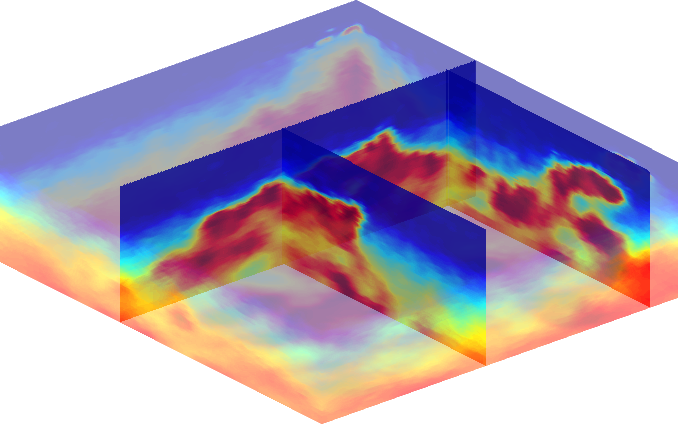};
  
  \addplot3[dashed,black,line width=1pt] 
      coordinates {(\xmin,\ycut,\zmin) (\xmin,\ycut,\zmax)
                   (\xmax,\ycut,\zmax) (\xmax,\ycut,\zmin) (\xmin,\ycut,\zmin)};
  \addplot3[dashed,black,line width=1pt] 
      coordinates {(\xmin,\ymin,\zcut) (\xmin,\ymax,\zcut) 
                   (\xmax,\ymax,\zcut) (\xmax,\ymin,\zcut) (\xmin,\ymin,\zcut)};

  \addplot3[dashed,black,line width=1pt] 
      coordinates {(\xcut,\ymin,\zmin) (\xcut,\ymin,\zmax)
                   (\xcut,\ymax,\zmax) (\xcut,\ymax,\zmin) (\xcut,\ymin,\zmin)};

\end{axis}

&

\begin{axis}[anchor=center,  width=\modelwidth, height=\modelheight,
             xmin=\xmin, xmax=\xmax,ymin=\zmin, ymax=\zmax,hide axis,
             tick label style={font=\small},y dir=reverse]
         \addplot [forget plot] graphics 
         [xmin=\xmin,xmax=\xmax,ymin=\zmin,ymax=\zmax] {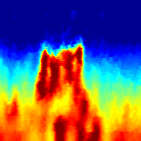};
         \addplot[dashed,black,line width=1pt] 
       coordinates {(\xmin+.01,\zmin+.01) (\xmin+.01,\zmax-.01) 
                    (\xmax-.01,\zmax-.01) (\xmax-.01,\zmin+.01) (\xmin+.01,\zmin+.01)};
\end{axis}

\begin{axis}[yshift=\jumpvert,
             anchor=center,  width=\modelwidth, height=\modelheight,
             xmin=\xmin, xmax=\xmax,ymin=\zmin, ymax=\zmax,hide axis,
             tick label style={font=\small},y dir=reverse]
         \addplot [forget plot] graphics 
         [xmin=\xmin,xmax=\xmax,ymin=\zmin,ymax=\zmax] {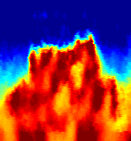};
         \addplot[dashed,black,line width=1pt] 
       coordinates {(\xmin+.01,\zmin+.01) (\xmin+.01,\zmax-.01) 
                    (\xmax-.01,\zmax-.01) (\xmax-.01,\zmin+.01) (\xmin+.01,\zmin+.01)};
\end{axis}

\\[-0.3cm]

\begin{axis}[anchor=center, width=\modelwidth, height=.60\modelwidth,
             xmin=\xmin, xmax=\xmax,ymin=\ymin, ymax=\ymax,hide axis,
             tick label style={font=\small}, y dir=reverse]
            \addplot [forget plot] graphics 
                     [xmin=\xmin,xmax=\xmax,ymin=\ymin,ymax=\ymax] {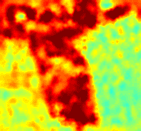};
            
            \addplot[dashed,black,line width=1pt] 
                  coordinates {(\xmin+.01,\ymin+.01) (\xmin+.01,\ymax-.01) 
                               (\xmax-.01,\ymax-.01) (\xmax-.01,\ymin+.01) (\xmin+.01,\ymin+.01)};
\end{axis}

&

\begin{axis}[yshift=-0.8cm,
     anchor=south,yshift=.5cm,hide axis,height=.5cm,colorbar/width=.2cm,
     colormap/jet,colorbar horizontal,
     colorbar style={separate axis lines,
                     tick label style={font=\small},
                     title={\small{velocity (\si{\km\per\second})}}},
                     point meta min=\vmin, point meta max=\vmax,
                     height=.5\modelwidth]
\end{axis}
\\};

\draw[black,-stealth,->,dashed,line width=1pt] (-0.7, 2.8) to[out=0,in=-180] ( 1.0, 2.8);
\draw[black,-stealth,->,dashed,line width=1pt] (-0.7, 0.3) to[out=0,in=-180] ( 1.0, 0.3);
\draw[black,-stealth,->,dashed,line width=1pt] (-4.7, 1.0) to[out=-90,in=90] (-4.7,-0.8);
%

\end{tikzpicture}
  \caption{velocity reconstruction of the SEAM model 
           from the minimization of $\misfitG$, starting 
           with the models of \cref{fig:seam:cp_start}.
           It uses a \emph{sparse} observational acquisition of 
           $\nstack = 34$ sources composed of $\npt = 8$ points, 
           while the numerical acquisition ($g$ in \cref{eq:misfit_green}) 
           employs a dense lattice of $272$ point-sources.
           The reconstruction follows $30$ iterations per 
           frequency between $2$ and $7$ \si{\Hz}. 
           For visualization, we picture sections 
           at a fixed  depth $z = 1.5$  \si{\km} 
           (bottom), for     $x = 3.5$  \si{\km} 
           (top right) and   $y = 3.25$ \si{\km}
           (bottom right).}  
  \label{fig:seam:rwi_sparse-obs}
\end{figure}

We use the FRgWI algorithm with the minimization 
of~\cref{eq:misfit_green} using sequential 
frequency content from $2$ to $7$ \si{\Hz}. 
We perform $30$ iterations per frequency which
leads to a total of $180$ iterations.
We analyze the two following configurations.
\begin{itemize}
  \item We first use the dense observational acquisition 
        composed of $272$ point-sources, and 
        multiple-point sources for the computational
        acquisition ($g$ in~\cref{eq:misfit_green}). 
        The multiple-point sources use 
        $\npt = 8$ points, resulting in $\nstack = 34$ 
        sources. The reconstructed velocity 
        after the iterations at $7$ \si{\Hz} 
        is shown in \cref{fig:seam:rwi}.
  \item Next, we consider the opposite situation: 
        we assume a sparse observational acquisition
        of $\nstack = 34$ sources each composed
        of $\npt = 8$ points. Here, the computational
        acquisition is made dense, with 
        $272$ point-sources. The recovered velocity
        is shown in \cref{fig:seam:rwi_sparse-obs}.
\end{itemize}


The minimization performs well in both situations:
the salt dome is appearing with appropriate values 
(about \num{4500} \si{\meter\per\second}). 
The upper boundary of the salt is accurately 
recovered (cf. the right of \cref{fig:seam:rwi,fig:seam:rwi_sparse-obs}) 
while
the deepest parts are more difficult to obtain.
It is possible that the high velocity contrast 
prevents us from imaging below the salt.
The horizontal section (bottom of 
\cref{fig:seam:rwi,fig:seam:rwi_sparse-obs})
also shows the appropriate pattern, in particular with the 
decrease of velocity in the middle, and only the parts near 
the boundary are slightly less accurate.
It confirms the results of our first experiment that FRgWI 
is relatively insensitive to whether the observational or 
computational acquisitions is taken sparse or dense: 
the reconstructions display similar resolution for the two cases.
In terms of numerical cost, sparse computational acquisition 
reduces the computational time but, in the frequency domain
with the use of direct solver, this gain remains marginal.
In order to further improve the reconstruction, we shall 
include a regularization parameter in the minimization, 
e.g. with Total Variation approach, which can indeed be 
supported by our misfit functional; see also 
\cite{Faucher2020EV} and the references therein.

\section{Analysis of results and perspectives}

We have implemented the iterative minimization method 
based upon the reciprocity-gap functional in the frequency 
domain, with the hybridizable discontinuous 
Galerkin (HDG) discretization to efficiently address  
the first-order system.
To probe the method, we have carried out two three-dimensional 
experiments, one with a velocity made of layers, and another 
with a medium containing a salt dome, extracted from the 
SEAM benchmark.
Firstly, our experiments have shown that the Full 
Reciprocity-gap Waveform Inversion (FRgWI) method performs 
better than FWI in the \emph{same configuration}, that is,
when we keep the observational acquisition for the numerical 
one, in particular by improving the illumination on the sides. 
\update{
These promising results motivate the implementation of the 
method for larger, field benchmarks, in order to probe 
the scalability of FRgWI. 
In this case, the time-domain formulation should be considered
to continue the performance analysis we have started.
}

\subsection{Flexibility in the computational acquisition}

The main feature of the FRgWI is its robustness regarding the 
absence of information on the observational acquisition,
and the consequent freedom in the computational one.
This is due to the misfit functional~\cref{eq:misfit_green}
that works with a product of a datum with a simulation,
and that relates with the correlation-based family of methods.
For instance, the wavelet used for the computational
source does not have to match the observational one,
as long as there is an overlap in their frequency contents,
and we have not observed any other requirement on the 
choice of the source function.
The use of less traditional sources such as plane waves
is theoretically possible, and should be investigated in
the future.

Our main result comes from the positions of the sources 
used for the simulations that can be arbitrarily chosen. 
This flexibility is in the core of the misfit functional
which does not compare directly an observation with a 
simulation, but tests the \emph{product} of \emph{any} 
observation with \emph{any} simulation, hence enforcing 
the domain illumination.
It is crucial to note that the measurements are not 
modified and are independently tested one by one 
against each of the (chosen) computations.
In our work, we remain with illumination from one side 
(i.e., simulations and observations are generated near 
the surface), motivated by \cite{Alessandrini2019} 
which prove the stability of the method.
This choice is also motivated from the derivation of 
the variational formulation we give in appendix, to 
avoid the appearance of additional integrals 
(see the discussion at the end of the first appendix).
The use of probing sources from other part of the domain 
(e.g., deeper in the domain) has to be carefully 
investigated, and is part of our ongoing research.

\subsection{Using sparse acquisitions}

The use of multiple-point sources allows for the reduction
of the computational time \emph{or} of the field acquisition
process.
Here, the key is that \emph{only one} of the two acquisitions 
is reduced, while the other remains dense. 
Once again, as every datum in the dense acquisition is 
tested with respect to each of the sparse one, it appears
to overall produce the proper illumination of the domain. 
This is possible due to the misfit formula which offers
high flexibility, contrary to more traditional approaches 
where both acquisitions must be the same.

We have experimented with the two configurations (sparse 
computational acquisition with dense observational one or
the opposite) 
and it results in similar accuracy in the reconstruction. 
Because FRgWI tests independently all combinations of an observation
with a simulation, \update{it further appears robust with respect to shot-stacking}.
With FRgWI, we do not exactly observe the cross-talk effect, 
as the measurements are not modified before entering the misfit, instead, we observe 
some inherent difficulties when the acquisition is too sparse. 
This is unavoidable, for instance, if the original field data 
have been obtained from a few sources only: we cannot 
artificially enhance the information contained in the measurements.
On the other hand, the difficulties that come from the summation 
of data (i.e., having multiple-point sources in a sparse acquisition) 
and that results in cross-talk are diminished by the misfit 
functional which takes the product with each datum in the 
dense acquisition.

In the context of shot-stacking, there is a ``reference'' 
solution which is the use of the non-summed fields. 
Here, we raise the following question for future applications 
of the FRgWI method: \emph{how to select the numerical 
acquisition to make the best use of the observed measurements?}

\subsection{Data coverage}

It is important to remind that the reciprocity-gap
misfit functional relates, in terms of receivers, to 
the discretization of a surface integral, as depicted 
in the first appendix.
In our experiments, we have a dense array of receivers,
according to the standards in exploration geophysics.
In the case where the surface is not properly covered, 
we certainly need to appropriately weight the records, 
using, for instance, existing techniques from 
Earth seismology, such as the one prescribed by \cite{Montagner2012}.
Similarly, the use of specific weight (e.g., from quadrature
rules) could give a first indication on how to select the 
computational acquisition in order to ensure the equal illumination 
of the domain with minimal points.

\update{
\subsection{Towards practical applications}

Our reconstruction algorithm has been carried out in the 
frequency domain, in which it is difficult to address the 
scales encountered in field applications, due to the operations 
of linear algebra (mainly the memory required for the matrix factorization). 
While the efficiency of the frequency domain is improved by 
the use of HDG discretization (to reduce the size of linear systems) 
and by new techniques developed for the solver \citep{XLiu2018}, time or hybrid 
domain inversion should be implemented to further probe the 
method in field studies.
In this case, the multiplication of fields in the misfit functional \cref{eq:misfit_green}
turns into convolution products, maintaining the same features
regarding the flexibility of the numerical acquisition.
We do not expect a difference in the performance of the method when 
implemented in the time-domain for the situation we have studied here, 
especially if one follows the (usual) progression of increasing 
frequency-content in the data, as introduced by \cite{Bunks1995}.
On the other hand, it is crucial to experiment on larger, field 
configurations, to further evaluate the method compared to the 
existing techniques. Here, the impact of the (limited) data coverage 
has to be carefully investigated. 
}


\section{Conclusion}

In this paper, we have performed waveform inversion 
using a reciprocity-gap misfit functional which is 
adapted to the dual-sensors devices which capture 
both the pressure field and the normal velocity of the waves. 
In this case, the measurements can be seen as Cauchy 
data, motivating our choice of misfit from the
point of view of inverse scattering theory. 
Our method enables the use of different acquisitions 
for the observations and the simulations, by relying on a 
product of data. 
We have investigated its performance with sparse acquisitions
made of multiple-point sources and have demonstrated its
efficiency compared to the traditional FWI.
We have used a sparse acquisition either for the observational one 
(to reduce the cost of the field acquisition) or for the 
computational one (to reduce the numerical cost of 
the iterative minimization), while not altering the resolution 
of the reconstruction. 

\update{
We have implemented the method in the frequency domain
and carried out two experiments of different natures.
These represent the first steps of validation, while
time and hybrid-domain inversion should now be considered.
The implementation of the misfit with time signals 
does not result in any technicality and should not lead to 
a difference in performance for the scales we have studied 
with our frequency-domain setup. 
However, it is necessary to further probe the behaviour of 
FRgWI in larger, practical test-cases, and to analyze
how it competes with state-of-the art methods.
}

The FRgWI method allows for arbitrary computational 
source positions that now needs to be further analyzed. 
For instance, the definition of criteria for the 
design of the computational acquisition, and the study 
of the source positions (e.g., not only close to the 
near surface area, with quadrature rules to ensure equal 
illumination) is the subject our future research. 
Note also that the concept of full reciprocity-gap 
waveform inversion extends readily to elasticity. 


\section*{Acknowledgments}

The authors thank Jean Virieux and Andreas Fichtner for 
thoughtful and encouraging discussions.
FF is funded by the Austrian Science Fund (FWF) under 
the Lise Meitner fellowship M 2791-N.
The work of GA and ES was performed under the PRIN 
grant No. 201758MTR2-007.
The research of HB has received funding from the 
European Union's Horizon 2020 research and innovation 
program under the Marie Sklodowska-Curie grant 
agreement No. 777778.
The research of HB and FF is also supported by the 
Inria--TOTAL strategic action DIP.
MVdH was supported by the Simons Foundation under the 
MATH + X program, the National Science Foundation under 
grant DMS-1815143, and by the members of the 
Geo-Mathematical Imaging Group at Rice University.
The work of RG was supported by the Higher Education Authority 
(HEA) Government of Ireland International Academic Mobility Program 2019.
ES has also been supported by the Individual Funding 
for Basic Research (FFABR) granted by MIUR.
The numerical experiments have been performed on 
the CEA's cluster Irene, as part of the GENCI 
allocation AP010411013; and on the cluster PlaFRIM
(Plateforme F\'ed\'erative pour la Recherche en Informatique 
et Math\'ematiques, $\text{https://www.plafrim.fr/fr}$).
The code used for the experiments, \texttt{hawen}, is developed by the
first author and available at https://gitlab.inria.fr/ffaucher/hawen.

\appendix
\section{Reciprocity-gap formula with Euler's equations in seismic}
\label{appendix:justification_reciprocity}

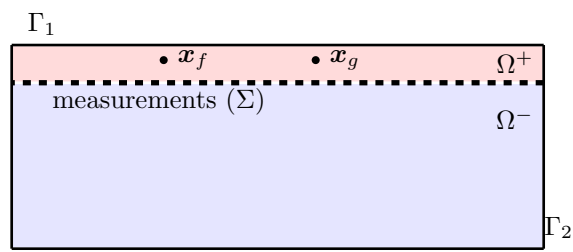
\begin{figure}[ht!]
\centering
  \begin{tikzpicture}[every node/.append style={font=\small}]
  \coordinate (d1) at ( 0.,0.);   \coordinate (d2) at ( 7.,0.);
  \coordinate (d3) at ( 7.,2.7);  \coordinate (d4) at ( 0.,2.7);
  \coordinate (d3r)at ( 7.,2.2);  \coordinate (d4r)at ( 0.,2.2);

  \draw[line width=1,color=white,fill=blue!50!white,opacity=0.2] (d1)to(d2)to(d3r)to(d4r)to(d1);
  \draw[line width=1,color=white,fill=red!70!white,opacity=0.2] (d4r)to(d3r)to(d3)to(d4)to(d4r);

  \coordinate (src)at ( 4.,2.50);
  \draw[mark=*,mark size=1,mark options={color=black,line width=1}] 
       plot[] coordinates{(src)};
  \draw (src)node[right=0.05 of src,xshift=0mm,black,anchor=west]
                 {$\bx_g$};  

  \coordinate (src)at ( 2.,2.50);
  \draw[mark=*,mark size=1,mark options={color=black,line width=1}] plot[] 
       coordinates{(src)};
  \draw (src)node[right=0.05 of src,xshift=0mm,black,anchor=west]
                 {$\bx_f$};  
  
  \draw (d4) node[right=0.4 of d4,xshift=0mm,black,anchor=south]
                 {$\Gamma_1$};  
  \draw (d4) node[right=0.2 of d2,xshift=0mm,black,anchor=south]
                 {$\Gamma_2$};

  \draw[line width=1,color=black] (d1) to[out=  0,in=180] (d2);
  \draw[line width=1,color=black] (d2) to[out= 90,in=-90] (d3);
  \draw[line width=1,color=black] (d3) to[out=180,in=  0] (d4);
  \draw[line width=1,color=black] (d4) to[out=-90,in= 90] (d1);
  \draw[line width=2,color=black,dashed] (d3r) to[] (d4r);
  \draw (d4r)node[below=0.25 of d4r,xshift=4mm,black,anchor=west]
                 {measurements ($\Sigma$)};
  \draw (d3r) node[above=0.0 of d3r ,   anchor=south east]{$\Omega^+$} ;
  \draw (d3r) node[below=0.0 of d3r ,   anchor=north east, yshift=-2mm]{$\Omega^-$} ;


\end{tikzpicture}
\caption{Illustration of the domain of interest for the Euler's 
         equation~\cref{eq:euler-continuity-conditions}, the 
         line of measurements cuts the domain into 
         $\Omega = \Omega^+ \cup \Omega^-$ which are colored 
         in red and blue respectively. We also consider the 
         sources ($x_f$ and $x_g$) to be located in $\Omega^+$.}
\label{fig:domain}
\end{figure}

In this appendix, we motivate the reciprocity-gap misfit,
constructed from the variational formulation of 
Problem~\ref{eq:euler_main}.
In the context of seismic, we assume that the data (i.e. pressure 
and normal velocity) are acquired on a line $\Sigma$ which is 
slightly underneath the surface $\Gamma_1$.
The domain is decomposed into above and below area from the line 
of receivers, such that $\Omega = \Omega^+ \cup \Omega^-$, we 
illustrate in \cref{fig:domain}.

Let us first rewrite Problem~\ref{eq:euler_main}
introducing continuity conditions at $\Sigma$, 
using exponent $^+$  and $^-$ for the fields that 
are in $\Omega^+$ and $\Omega^-$ respectively,
such that 
\begin{equation}
  \pressure \mid_{\Omega^+} = \pressure^+ \, , \qquad 
  \pressure \mid_{\Omega^-} = \pressure^- \, ,
\end{equation}
and similarly for the velocity. 
Problem~\ref{eq:euler_main} is 
equivalent to,
\begin{subequations} \label{eq:euler-continuity-conditions}
\begin{empheq}[left={\empheqlbrace}]{alignat=2} 
   -\ii \omega \rho(\bx) \velocity(\bx) &= -\nabla \pressure(\bx), 
        &\text{in $\Omega^{+} \cup \Omega^{-}$,}        \\
   -\ii \omega \kappa(\bx)^{-1} \pressure(\bx)  &= -\divergence \velocity(\bx) + f(\bx),
        &\text{in $\Omega^{+} \cup \Omega^{-}$,}        \\
    \pressure(\bx) & = 0,  &\text{on $\Gamma_1$,}       \\
    \partial_{\n} \pressure(\bx) - \dfrac{\ii \omega}{c(\bx)} \pressure(\bx) 
                   & = 0, &\text{on $\Gamma_2$,}        \\
    \pressure^{+} & = \pressure^{-} \, , \quad 
    \partial_\nu \pressure^{+} = \partial_\nu \pressure^{-} \, , &\text{on $\Sigma$,}   \\
    \velocity^{+} & = \velocity^{-} \, , \quad 
    \partial_\nu \velocity^{+} = \partial_\nu \velocity^{-} \, , &\text{on $\Sigma$.} 
\end{empheq} \end{subequations}

We consider two couples $(\pressure_1,  \, \velocity_1)$
and $(\pressure_2,  \, \velocity_2)$ solutions to
Problem~\cref{eq:euler-continuity-conditions}.
They are respectively associated to two sources, $f$ 
and $g$, and two different sets of physical parameters: 
$(\kappa_1, \, \rho_1)$ and $(\kappa_2, \, \rho_2)$
respectively.
We take $\pressure_k \in H^1(\Omega)$
ad $\velocity_k \in (H^1(\Omega))^3$, for $k=1$, $2$,
where $H$ refers to the Hilbert space.
For the derivation of the reciprocity-gap formula, we 
write the variational formulation on $\Omega^{-}$ only.
The variational formulation of~\cref{eq:euler-continuity-conditions}
for $(\pressure_1, \, \velocity_1)$, using for the 
test functions $(\pressure_2,  \, \velocity_2)$ gives
\begin{empheq}[left={\empheqlbrace}]{align}
\label{eq:appendix:variational_main}
\begin{split}
&   \int_{\Omega^-} -\ii \omega \rho_1 \velocity_1 \cdot \velocity_2 
                    +\nabla \pressure_1 \cdot \velocity_2 \, \, \dd \Omega^{-} = 0, \\
&   \int_{\Omega^-} -\ii \omega \kappa_1^{-1} \pressure_1 \, \pressure_2  
                    + (\divergence \velocity_1) \, \pressure_2 \, 
                    \, \dd{\Omega^-} = 0,
\end{split} \end{empheq}
where we omit the $^-$ exponent in the fields 
$\pressure$ and $\velocity$ for the sake of 
clarity.
Furthermore we assume that the source $f$ 
is supported \emph{outside} $\Omega^-$, e.g., 
by using delta-Dirac functions in some position
$\bx_f \in \Omega^+$, according to \cref{fig:domain}.

We use an integration by part for both equations, 
and subtract the first one from the second one to 
get 
\begin{empheq}[]{align}
\label{eq:omega-minus_variational}
\begin{split}
   \int_{\Omega^-} \ii \omega \rho_1 \velocity_1 \cdot \velocity_2   
               + \pressure_1 \, \divergence(\velocity_2) \, 
             & - \ii \omega \kappa_1^{-1} \pressure_1 \, \pressure_2  
               - \velocity_1 \cdot \nabla \pressure_2 \, \, \dd{\Omega^-} \\ 
&=\int_{\partial\Omega^-} \pressure_1 \, (\velocity_2 \cdot \n)
                        - (\velocity_1 \cdot \n) \, \pressure_2 \, \, \dd \partial\Omega^-  .
\end{split} \end{empheq}
In the volume integral, we replace 
$(\divergence \velocity_2)$ and 
$(\nabla \pressure_2)$ using the fact that 
$(\pressure_2, \, \velocity_2)$ solves~\cref{eq:euler-continuity-conditions}:
\begin{empheq}[]{align}\begin{split}
   \int_{\Omega^-} \ii \omega \rho_1 \velocity_1 \cdot  \velocity_2   
             & + \pressure_1 \cdot \divergence(\velocity_2) \, 
               - \ii \omega \kappa_1^{-1} \pressure_1 \, \pressure_2  
               - \velocity_1 \cdot \nabla \pressure_2 \, \, \dd{\Omega^-} \\ 
& =\int_{\Omega^-} \ii \omega \rho_1 \velocity_1 \cdot \velocity_2   
               + \pressure_1 \, (\ii \omega \kappa_2^{-1} \pressure_2) \, 
               - \ii \omega \kappa_1^{-1} \pressure_1 \, \pressure_2  
               - \velocity_1 \cdot (\ii \omega \rho_2 \velocity_2) \, \, \dd{\Omega^-} \\ 
& =\int_{\Omega^-} \ii \omega (\rho_1 - \rho_2) \velocity_1 \cdot \velocity_2   
               + \ii \omega \big( \kappa_2^{-1} - \kappa_1^{-1} \big) \, \pressure_1 \, \pressure_2
                     \, \, \dd{\Omega^-}.
\end{split} \end{empheq}

Next, the integral on the boundary 
in~\cref{eq:omega-minus_variational}
is decomposed between 
$\Gamma_2^- = \Gamma_2 \cap \partial\Omega^-$ 
and $\Sigma$ such that 
\begin{empheq}[]{align}\begin{split}
    \int_{\partial\Omega^-} \pressure_1 \, (\velocity_2 \cdot \n) &
                        - (\velocity_1 \cdot \n) \, \pressure_2 \, \, \dd \partial\Omega^- \\
 = &\int_{\Gamma_2^-} \pressure_1 \, (\velocity_2 \cdot \n) 
                 - (\velocity_1 \cdot \n) \, \pressure_2 \, \, \dd \Gamma_2^-
  \, + \,
  \int_{\Sigma} \pressure_1 \, (\velocity_2 \cdot \n) 
                 - (\velocity_1 \cdot \n) \, \pressure_2 \, \, \dd \Sigma \, .
\end{split} \end{empheq}
On $\Gamma_2^-$, we first use \cref{eq:euler_main_a}
to replace $(\velocity_1 \cdot \n)$ and $(\velocity_2 \cdot \n)$
and then the absorbing boundary condition~\cref{eq:euler_main_bc_abc}:
\begin{empheq}[]{align}\begin{split}
&\int_{\Gamma_2^-} \pressure_1 \, (\velocity_2 \cdot \n) 
                 - (\velocity_1 \cdot \n) \, \pressure_2 \, \, \dd \Gamma_2^- \\
& = 
  \int_{\Gamma_2^-} \pressure_1 \, 
                   \big((\ii \omega \rho_2)^{-1} \partial_{\n} \pressure_2 \big) 
               -   \big((\ii \omega \rho_1)^{-1} \partial_{\n} \pressure_1 \big)  
                   \, \pressure_2 \, \, \dd \Gamma_2^- \\
& = 
  \int_{\Gamma_2^-} \pressure_1 \, 
                   \big((c_2 \rho_2)^{-1} \pressure_2 \big) 
               -   \big((c_1 \rho_1)^{-1} \pressure_1 \big)  
                   \, \pressure_2 \, \, \dd \Gamma_2^- \\
& = 
  \int_{\Gamma_2^-} \big((c_2 \rho_2)^{-1} - (c_1 \rho_1)^{-1} \big) 
                    \, \pressure_1 \, \pressure_2 \, \, \dd \Gamma_2^-.
\end{split} \end{empheq}

Eventually, injecting the new formulas for the 
volume and integral equations 
in~\cref{eq:omega-minus_variational}, we obtain
\begin{empheq}[]{align} \label{eq:appendix:final_corres}
\begin{split}
& \int_{\Omega^-} \ii \omega (\rho_1 - \rho_2) \velocity_1 \, \velocity_2   
               + \ii \omega \big( \kappa_2^{-1} - \kappa_1^{-1} \big) \, \pressure_1 \, \pressure_2
                     \, \, \dd{\Omega^-} \\
& -   \int_{\Gamma_2^-} \big((c_2 \rho_2)^{-1} - (c_1 \rho_1)^{-1} \big) 
                    \, \pressure_1 \, \pressure_2 \, \, \dd \Gamma_2^- 
 = \int_{\Sigma} \pressure_1 \, (\velocity_2 \cdot \n) 
                 - (\velocity_1 \cdot \n) \, \pressure_2 \, \, \dd \Sigma \, .
\end{split} \end{empheq}
The right-hand side coincides with our choice misfit 
functional for full reciprocity-gap waveform inversion 
cf.~\cref{eq:misfit_green} (where we take the squared 
of the expression to have a positive functional), and 
it equates to zero \emph{if and only if} $\rho_1 = \rho_2$ 
and $\kappa_1 = \kappa_2$ (meaning that $c_1 = c_2$) over 
the \emph{whole domain $\Omega^-$}.
We further refer to \cite{Alessandrini2018,Alessandrini2019}
for stability results.
We see that the formula~\cref{eq:appendix:final_corres} does 
not involve the sources, because we have positioned them above
of the receivers line (see \cref{fig:domain}). In the case 
where sources are below, or if one wants to create a numerical
acquisition with sources inside the domain, it results in an 
additional integral in \cref{eq:appendix:variational_main}, 
which then has to be included in the misfit.


\section{Gradient formulation using adjoint-state method}
\label{appendix:gradient}

%

The gradient of the misfit functional is computed 
using the \emph{adjoint-state} method, which comes
from the work of Lions in optimal control, cf.~\cite{Lions1971},
with early applications in~\cite{Chavent1974}.
The method is popular in seismic as it avoids the computation 
of the Jacobian matrix, thus requiring limited numerical effort;
it is reviewed in \cite{Plessix2006}.

For generality, we consider the misfit functional $\misfitall$, 
which is either $\misfitL$ of $\misfitG$, and define the 
constrained minimization problem,
\begin{equation}
  \min_m \misfitall(m) \, \qquad \text{ subject to } \quad 
                                 \sum_{i} \woperator (\sol^{(f_i)}) = \srcall_i ,
  \quad \text{for $i \in \{1, \ldots, \nsrc\}$}, 
\end{equation}
where $\woperator$ is the linear wave operator corresponding
with the Euler's equations of Problem~\ref{eq:euler_main}. 
We denote by $f_i$ the sources and use the notation
$\sol^{(f_i)}=\{\velocity^{(f_i)}, \, \pressure^{(f_i)}\} = 
 \{v_x^{(f_i)}, \, v_y^{(f_i)}, \, v_z^{(f_i)}, \, \pressure^{(f_i)}\}$ 
for the solution associated with the 
volume source $F_i = \{0, \, 0, \, 0, \, f_i\}$,
in accordance with Problem~\ref{eq:euler_main}.
For the sake of notation, we first consider a single source 
and drop the index $f_i$, the formulation with Lagrangian gives 
\begin{equation}
  \lagrangian(m,\tildsol,\tildadj) = 
                \misfitall \, + \, \big< \woperator \tildsol - \srcall \, , \, \tildadj \big>_\Omega \, .
\end{equation}
Here, $< . \, , \, .>_\Omega$ denotes the complex inner product 
in $L^2(\Omega)$ such that 
$< a, \, b>_\Omega = \int_\Omega \overline{a} \, b \, \dd\Omega$,  
with $\overline{\phantom{a}}$ the complex conjugate.

The first step of the adjoint-state 
method is to take $\tildsol = \sol$ 
such that
\begin{equation} \label{eq:gradient_lagrange_general}
  \nabla_m \lagrangian(m, \tildsol = \sol, \tildadj) 
= \nabla_m \misfitall .
\end{equation}
Then, the adjoint-state $\adj$ is selected
such that the derivative of the Lagrangian 
with respect to $U$ equates zero, i.e., 
$\gamma$ is solution to
\begin{equation} \label{eq:adjoint_eq}
  \woperator^*  \adj  = -\partial_\sol \misfitall \, ,
\end{equation}
where $^*$ denotes the adjoint (transposed of the complex conjugate).
With this choice of adjoint-state, the gradient of the misfit
functional (which coincides with the one of the Lagrangian 
from~\cref{eq:gradient_lagrange_general}) is 
\begin{equation} \label{eq:gradient_lagrange}
  \nabla_m \misfitall =   \nabla_m \lagrangian(m, \tildsol = \sol, \tildadj = \adj) 
                      = \Real \bigg( \big<  \partial_m \woperator \, \sol \, , \, \adj \big>_\Omega \bigg).
\end{equation}
We further refer to the Appendix~A of \cite{Faucher2019IP}, 
\cite{Faucher2017} and \cite{Barucq2018} for more details on 
the complex-variable adjoint-state, and to 
\cite{Faucher2020adjoint} for the specificity with HDG
discretization.
\medskip

When we incorporate back the different sources, 
the adjoint problem~\cref{eq:adjoint_eq} 
for the misfit functional $\misfitL$ is,
for each source $f_i$ in the acquisition,
\begin{equation} \label{eq:adjoint_eq_l2}
  \woperator^* \begin{pmatrix}
                 \gamma_{v_x}^{(f_i)}       \\
                 \gamma_{v_y}^{(f_i)}       \\
                 \gamma_{v_z}^{(f_i)}       \\
                 \gamma_{\pressure}^{(f_i)} \\
               \end{pmatrix} = - \restrict^*
               \begin{pmatrix}
                   \eta \big(\restrict(  v_{\n}^{(f_i)} ) - d_{v}^{(f_i)}    \big) \n_x \\
                   \eta \big(\restrict(  v_{\n}^{(f_i)} ) - d_{v}^{(f_i)}    \big) \n_y \\
                   \eta \big(\restrict(  v_{\n}^{(f_i)} ) - d_{v}^{(f_i)}    \big) \n_z \\
                   \restrict(\pressure^{(f_i)}) - d_{\pressure}^{(f_i)}
               \end{pmatrix} 
\qquad \text{adjoint-state problem for $\misfitL$. }
\end{equation}
Here we use $\n$ to indicate the normal direction
(in the case of a flat surface, i.e. $x-y$ plane, 
$\n_x = \n_y = 0$ and $\n_z = 1$).
The gradient using all sources is 
\begin{equation} \label{eq:gradient_l2}
  \nabla_m \misfitL =  \Real \bigg( \sum_{i=1}^{\nsrcObs}
                       \big<  \partial_m \woperator \, \sol^{(f_i)} \, , 
                       \, \adj^{(f_i)} \big>_\Omega \bigg).
\end{equation}

For the misfit functional $\misfitG$ 
of~\cref{eq:misfit_green}, the 
adjoint-state solves, for each 
\emph{computational} source $g_j$,
\begin{equation} \label{eq:adjoint_eq_reciprocity}
  \woperator^* \begin{pmatrix}
                 \gamma_{v_x}^{(g_j)}       \\
                 \gamma_{v_y}^{(g_j)}       \\
                 \gamma_{v_z}^{(g_j)}       \\
                 \gamma_{\pressure}^{(g_j)} \\
               \end{pmatrix} = - \sum_{i=1}^{\nsrcObs} \xi_{i,j} \, \restrict^* 
               \begin{pmatrix}
                - \overline{d}_{\pressure}^{(f_i)} \, \n_x \\
                - \overline{d}_{\pressure}^{(f_i)} \, \n_y \\
                - \overline{d}_{\pressure}^{(f_i)} \, \n_z \\
                  \overline{d}_{v}^{(f_i)} 
               \end{pmatrix} 
\qquad \text{adjoint-state problem for $\misfitG$,}
\end{equation}
with the scalar $\xi_{i,j}$ given by
\begin{equation}
  \xi_{i,j} = \sum_{k=1}^{\nrcv} 
         \bigg( d_{v}^{(f_i)}(\bx_k) \pressure^{(g_j)}(\bx_k)
            -   d_{\pressure}^{(f_i)}(\bx_k) v_{\n}^{(g_j)}(\bx_k) \bigg).
\end{equation}
The gradient is 
\begin{equation} \label{eq:gradient_rwi}
  \nabla_m \misfitG =  \Real \bigg(
                       \sum_{j=1}^{\nsrcSim} \big<  \partial_m \woperator \, \sol^{(g_j)} \, , 
                       \, \adj^{(g_j)} \big>_\Omega \bigg).
\end{equation}
We refer to \cite{Alessandrini2019} for more
details on the adjoint-state method using 
Cauchy data.
Therefore, we see that the adjoint-state for 
a \emph{single} source in the computational 
acquisition for $\misfitG$ encodes information 
from \emph{all} measurements because of the
reciprocity-gap (the sum over 
$i$ in~\cref{eq:adjoint_eq_reciprocity}) while it 
only takes the current source with $\misfitL$,
see~\cref{eq:adjoint_eq_l2}.

%

\bibliographystyle{apalike}
\bibliography{bibliography}

\end{document}